\documentclass[12pt,a4paper]{article}
\usepackage{amssymb,enumitem,booktabs,subfig,xcolor,todonotes,microtype,setspace}
\usepackage[top=1.4in, bottom=1.4in, left=1.35in, right=1.35in]{geometry}
\usepackage{natbib}
\usepackage{url} 
\usepackage[pdftex,colorlinks=true,hypertexnames=false]{hyperref}
\definecolor{darkblue}{rgb}{0,0,.6}
\hypersetup{citecolor=darkblue,linkcolor=darkblue,urlcolor=darkblue}
\usepackage{amsmath}
\usepackage{amsfonts}
\usepackage{epsfig}
\usepackage{graphics}
\usepackage{palatino,eulervm}
\usepackage{graphicx}
\usepackage{lscape}
\usepackage{rotating}
\usepackage{float}
\usepackage{tablefootnote}
\allowdisplaybreaks[4]

\DeclareMathOperator*{\argmin}{arg\,min}

\usepackage{amsthm,thmtools}
\usepackage{mathrsfs}
\declaretheorem{theorem}
\declaretheorem{lemma}
\makeatletter
\def\th@newremark{\th@remark\thm@headfont{\bfseries}}
\makeatletter
\theoremstyle{newremark}
\newtheorem{remark}{Remark}
\newtheorem{prop}{Proposition}

\newtheorem{assumption}{Assumption}
\declaretheoremstyle[
  spaceabove=6pt, spacebelow=6pt,
  headfont=\bfseries,
  notefont=\mdseries, notebraces={(}{)},
bodyfont=\normalfont,
  postheadspace=0.5em,
]{mystyle}

\providecommand{\U}[1]{\protect\rule{.1in}{.1in}}

\graphicspath{{plots/}}

\begin{document}

\title{\LARGE Estimating Factor-Based Spot Volatility Matrices with Noisy and Asynchronous High-Frequency Data}

\author{{\normalsize Degui Li\thanks{Department of Mathematics, University of York, UK. Email: degui.li@york.ac.uk.},\ \ \ Oliver Linton\thanks{Faculty of Economics, University of Cambridge, UK. Email: obl20@cam.ac.uk. },\ \ \ Haoxuan Zhang\thanks{Department of Mathematics, University of York, UK. Email: hz885@york.ac.uk.}}\\
{\normalsize\em University of York and University of Cambridge}}
\date{\normalsize This version: \today}

\maketitle

\centerline{\bf Abstract}

\medskip

We propose a new estimator of  high-dimensional spot volatility matrices satisfying a low-rank plus sparse structure from noisy and asynchronous high-frequency data collected for an ultra-large number of assets. The noise processes are allowed to be temporally correlated, heteroskedastic, asymptotically vanishing and dependent on the efficient prices. We define a kernel-weighted pre-averaging method to jointly tackle the microstructure noise and asynchronicity issues, and we obtain uniformly consistent estimates for latent prices. We impose a continuous-time factor model with time-varying factor loadings on the price processes, and estimate the common factors and loadings via a local principal component analysis. Assuming a uniform sparsity condition on the idiosyncratic volatility structure, we combine the POET and kernel-smoothing techniques to estimate the spot volatility matrices for both the latent prices and idiosyncratic errors. Under some mild restrictions, the estimated spot volatility matrices are shown to be uniformly consistent under various matrix norms. We provide Monte-Carlo simulation and empirical studies to examine the numerical performance of the developed estimation methodology.  

\medskip


\newpage

\section{Introduction}\label{sec1}
\renewcommand{\theequation}{1.\arabic{equation}}
\setcounter{equation}{0}

In high-frequency financial econometrics, the so-called realised volatility has been commonly used to measure the integrated volatility of asset returns over a fixed time window \citep[e.g.,][]{BS02, BS04, ABDL03, S05, AJ14}. However, this results in a question of how to choose the time window, in particular when the financial market is volatile. In practice, it is often important to recover the actual spot/instantaneous volatility structure, which plays an important role in various applications such as testing price jumps \citep[e.g.,][]{LM07} and estimating stochastic volatility models \citep[e.g.,][]{KK16, BR18}. There have been many studies about spot volatility estimation. For the case of a single asset without market microstructure noise, \cite{FW08} and \cite{K10} use a nonparametric kernel smoothing method to estimate the spot volatility function and derive its in-fill asymptotic properties. For the more general high-frequency data setting with microstructure noise, \cite{ZB14} propose a local version of the two-scale realised volatility \citep{ZMA05} to estimate the spot volatility, whereas \cite{KK16} combine classic kernel smoothing with the pre-averaging method \citep{JLMPV09, CKP10}. 

\smallskip

Nowadays, practitioners often have to work with high-frequency financial data collected for a large number of assets. The aforementioned spot volatility estimation methods developed for a single or finite number of assets do not generally work well in the high-dimensional and high-frequency data setting. Under a uniform sparsity condition, \cite{BLLW23} estimate high-dimensional spot volatility matrices when the number of assets is ultra large, and derive the uniform convergence properties via the joint in-fill and increasing dimensionality asymptotics. However, the sparsity assumption imposed on large volatility matrices is too restrictive, since the price processes are often highly correlated between a large number of assets (in particular those from the same sector). It is well known that there may exist co-movements between these highly-correlated asset prices, and these co-movements may be captured by some latent risk factors. Hence, to relax the restrictive sparsity assumption and estimate meaningful volatility structures, the following continuous-time factor model is often employed for a $p$-dimensional vector of asset prices:
\begin{equation}\label{eq1.1}
X_t=\Lambda F_t+ U_t,
\end{equation}
where $\Lambda$ is a $p\times k$ matrix of constant factor loadings, $F_t$ and $ U_t$ are $k$-dimensional and $p$-dimensional continuous semimartingales, respectively (see Section \ref{sec2} for the definition). Model (\ref{eq1.1}) is the approximate factor model, which has been extensively studied for low-frequency data \citep[e.g.,][]{CR83, BN02}. By imposing a sparse structural assumption on the volatility of $ U_t$, it follows from (\ref{eq1.1}) that $X_t$ has a low-rank plus sparse volatility structure, which is often called conditional sparsity \citep{FLM13}. \cite{FFX16} estimate the large (integrated) volatility matrix of $X_t$ when the factors $F_t$ are observable; whereas \cite{AX17} use the principal component analysis (PCA) to estimate the factor model (\ref{eq1.1}) and further construct the large volatility matrix estimation for $X_t$ when the factors are latent. \cite{P19} estimates the factor model allowing jumps in the latent factor process and derives the convergence rates and limit distribution theory for the PCA estimated factors and loadings. \cite{DLX19} estimate the conditionally sparse large covariance matrices and their inverse for the asynchronous high-frequency data which may be contaminated by the microstructure noise, combine the pre-averaging and generalised shrinkage in the estimation procedure and cover three different scenarios for the factor model specification. Other recent developments on estimation and testing of model (\ref{eq1.1}) can be found in \cite{KL18} and \cite{SX22}.

\smallskip

The continuous-time factor model (\ref{eq1.1}) is essentially static with constant factor loadings. This model assumption may be insufficient when the main interest lies in the spot volatility matrix estimation. In particular, it becomes invalid when there are smooth structural changes or breaks in the process governing the large data series. This motivates the following time-varying factor model for continuous-time processes:
\begin{equation}\label{eq1.2}
dX_t=\Lambda(t) dF_t+d U_t,
\end{equation}
where $\Lambda(t)$ is a $p\times k$ matrix of time-varying factor loading processes and the other components are the same as those in (\ref{eq1.1}). Model (\ref{eq1.2}) covers model (\ref{eq1.1}) as a special case when $\Lambda(t)=\Lambda$. It can be regarded as a natural extension of the time-varying factor model from the low-frequency data setting \citep[e.g.,][]{MHvS11, SW17} to the high-frequency data setting. The aim of this paper is to estimate the large spot volatility matrix for $X_t$ based on the time-varying factor model (\ref{eq1.2}). \cite{K18} generalises \cite{FLM13}'s POET (Principal Orthogonal complEment Thresholding) method to estimate the large spot volatility matrix of $X_t$ defined in (\ref{eq1.2}) and its decomposition into the systematic and idiosyncratic volatility components for noise-free and synchronised high-frequency data. In practice, it is not uncommon that the large high-frequency data are non-synchronised and may be contaminated by the market microstructure noise. A direct application of \cite{K18}'s method in the latter setting would result in biased volatility estimation. Hence, a substantial extension is required to address this problem and consistently estimate large spot volatility matrices. For the noisy high-frequency data but with observed (and possibly noise contaminated) factors, two recent papers by \cite{BLLW23} and \cite{CMZ23} estimate the spot volatility structure of $X_t$ under (\ref{eq1.2}). 

\smallskip

We assume the asset prices are observed with an additive noise structure to be defined in Section \ref{sec2.1}, where the microstructure noise is allowed to be temporally correlated with nonlinear heteroskedasticity, asymptotically vanishing and dependent on the latent prices. The empirical studies in \cite{JLZ17} and \cite{LL22a} reveal non-trivial temporal dependence for the microstructure noise of individual asset prices. The assumption of nonlinear heteroskedasticity allows the microstructure noise to have prominent intraday patterns such as the well-known U-shape. Some authors have argued that the microstructure noise is small in magnitude, and so, as in \cite{KL08}, we allow the microstructure noise to be shrinking as the data frequency increases. Such a ``small noise" model structure is also adopted by \cite{DX21} and \cite{LL22b}. Furthermore, we allow the noise to be correlated with the latent prices but do not impose any explicit correlation structure \citep[e.g.,][]{KL08}. The existence of endogenous noise may be due to rounding effects, price stickiness and asymmetric information \citep[e.g.,][]{HL06}.

\smallskip

In practice different assets are traded at different frequencies and so their transaction prices are observed at times that are not synchronised and can be quite differently spaced due to liquidity levels varying over assets. This often leads to volatility matrix estimation bias and possibly enhances the so-called Epps effect \citep[e.g.,][]{E79}. In order to jointly tackle the noise and asynchronicity issues, we extend the conventional pre-averaging method and introduce a kernel-weighted version, i.e., we take a weighted average of the observed prices over a local neighborhood of $t$ by kernel smoothing to obtain an approximation of the latent price at time $t$. This kernel-weighted pre-averaging has been used by \cite{KK16} in spot volatility estimation for a single asset, and recently has been adopted by \cite{BLLW23} in large spot volatility matrix estimation for synchronised high-frequency data. We show that the the first-step kernel filter of the asset prices is uniformly consistent with the approximation order determined by the bandwidth, the dimension and the sample size, extending the uniform consistency derived by \cite{KK16} and \cite{BLLW23} to a much more general setting.  

\smallskip

To estimate the latent structure in the time-varying factor model (\ref{eq1.2}), it is natural to apply a local version of PCA which relies on the local expansion of the smooth time-varying factor loading functions \citep[e.g.,][]{MHvS11, SW17, K18, WPLL21}. Since the latent prices are unavailable, we have to apply the local PCA to the estimated prices obtained via pre-averaging. This extension is non-trivial and leads to an extra challenge in proving the in-fill asymptotics for the local PCA estimates of common factors and factor loadings. The mean squared convergence rates for the estimated factors (taking the first-order difference) and the uniform convergence rates for the estimated time-varying factor loadings are slower than those in \cite{K18} who considers synchronised noise-free data. This is reasonable, since approximating latent prices results in extra estimation errors that consequently slow down the convergence rates for local PCA. 

\smallskip

It follows from (\ref{eq1.2}) that the spot volatility matrix of $X_t$ is decomposed into the common and idiosyncratic spot volatility matrices. The latter is assumed to satisfy a uniform sparsity condition, which is similar to that used by \cite{CXW13}, \cite{CL16}, \cite{CLL19} and \cite{BLLW23}, and results in the low-rank plus sparse matrix structure. Accordingly, we combine the classic POET with kernel smoothing to estimate the spot volatility matrices for both the latent prices and idiosyncratic errors. In particular, a generalised shrinkage technique is applied to the idiosyncratic spot volatility matrix estimation. We derive the uniform convergence property for the estimated idiosyncratic spot volatility matrix in the (elementwise) max and spectral norms, and obtain the rates comparable to those derived in \cite{CMZ20} and \cite{BLLW23}. Due to the spiked volatility structure, we derive the uniform convergence rates for the estimated spot volatility matrix of $X_t$ not only in the max norm but also in the relative error measurement \citep[e.g.,][]{FLM13, WPLL21}.

\smallskip

The finite-sample Monte-Carlo simulation studies show that the developed large spot volatility estimation method outperforms the naive estimation (without applying shrinkage to the estimated spot idiosyncratic volatility matrix). An empirical application to the one-min intraday log-price of S\&P 500 index constituents reveals significant time-varying patterns of the spot volatility and covariance, and demonstrates rationality of the low-rank plus sparse spot volatility structure. In particular, the estimated factor number varies over time with fewer factors during the period of market collapse and more factors when the market is stable.
 
\smallskip

The rest of the paper is organised as follows. Section \ref{sec2} introduces the additive microstructure noise model, low-rank plus sparse spot volatility matrix structure and estimation technique for large spot volatility matrices. Section \ref{sec3} gives some regularity conditions and the in-fill asymptotic properties for the developed estimates together with some remarks. Section \ref{sec4} conducts the Monte-Carlo simulation study and Section \ref{sec5} reports the empirical application. Section \ref{sec6} concludes the paper. Proofs of the main theoretical results and some technical lemmas are available in the Supplementary Material \citep{LLZ2024}. Throughout the paper, we let $\Vert\cdot\Vert$ be the Euclidean norm of a vector; and for a $p\times p$ matrix $A=(A_{i_1i_2})_{p\times p}$, we let $\Vert A\Vert_s$ and $\Vert A\Vert_F$ be the matrix spectral norm and Frobenius norm, $\vert A\vert_1=\sum_{i_1=1}^p\sum_{i_2=1}^p |A_{i_1i_2}|$, $\Vert A\Vert_1=\max_{1\leq i_2\leq p}\sum_{i_1=1}^p |A_{i_1i_2}|$, $\Vert A\Vert_{\infty,q}=\max_{1\leq i_1\leq p}\sum_{i_2=1}^p |A_{i_1i_2}|^q$ and $\Vert A\Vert_{\max}=\max_{1\leq i_1\leq p}\max_{1\leq i_2\leq p} |A_{i_1i_2}|$.


\section{Model and methodology}\label{sec2}
\renewcommand{\theequation}{2.\arabic{equation}}
\setcounter{equation}{0}

In this section, we first introduce an additive noise structure for contaminated high-frequency data and the low-rank plus sparse spot volatility matrix structure, followed by description of the estimation methodology.

\subsection{Microstructure noise}\label{sec2.1}

For the $i$-th asset, suppose that the asset prices are observed with the additive noise structure:
\begin{equation}\label{eq2.1}
Y_{i,t_j^i}=X_{i,t_j^i}+\varepsilon_{i,t_j^i},\ \ \ i=1,\cdots,p,\ \ j=1,\cdots,n_i,
\end{equation}
where $X_{i,t}$ is the $i$-th component of $X_t$, $t_1^i,\cdots,t_{n_i}^i$ are the data collection time points (which may be non-equidistant) for the $i$-th asset, and 
\begin{equation}\label{eq2.2}
\varepsilon_{i,t_j^i}=n_i^{-\beta_i}\chi_i(t_j^i)\varepsilon_{i,j}^\ast,\ \ 0\leq \beta_i<1/2,
\end{equation}
is the noise with $\chi_i(\cdot)$ and $\varepsilon_{i,j}^\ast$ satisfying Assumption \ref{ass:2} in Section \ref{sec3.1} below. The general structure in (\ref{eq2.2}) shows that the microstructure noise not only has a nonlinear heteroskedastic structure $\chi_i(\cdot)$, but also is asymptotically vanishing as the sampling frequency increases if $0<\beta_i<1/2$. In addition, we allow $\varepsilon_{i,j}^\ast$ to be correlated over $i$ and $j$, and dependent on the latent prices $X_{i,t_j^i}$. Throughout the paper, we let $t_{0}^i\equiv0$ and $t_{n_i}^i=T_i$.

\smallskip

The latent vector process $X_t=(X_{1,t},\cdots,X_{p,t})^{^\intercal}$ satisfies the time-varying factor model structure (\ref{eq1.2}), where $F_t$ and $ U_t$ are $k$-dimensional and $p$-dimensional Brownian semimartingales,  respectively, solving
\begin{equation}\label{eq2.3}
dF_t=\mu_t^Fdt+\sigma_t^Fd W_t^F
\end{equation}
and 
\begin{equation}\label{eq2.4}
d U_t=\mu_t^Udt+\sigma_t^Ud W_t^U,
\end{equation}
$\mu_t^F$ and $\mu_t^U$ are vectors of drift with dimensions $k$ and $p$, respectively, $\sigma_t^F$ and $\sigma_t^U$ are $k\times k$ and $p\times p$ matrices of spot volatilities, $W_t^F$ and $W_t^U$ are $k$-dimensional and $p$-dimensional standard Brownian motions. The drifts and spot volatilities are progressively measurable processes. As in \cite{K18} and \cite{P19}, we assume that $\{W_t^F: t\geq0\}$ and $\{W_t^U: t\geq0\}$ are independent standard Brownian motions. This assumption may be relaxed to allow for weak correlation between the factor and idiosyncratic error processes. Without loss of generality, we may further assume that $\sigma_t^F=I_k$, a $k\times k$ identity matrix; otherwise, we can re-define the factor loading matrix as $\Lambda(t) \sigma_t^F$ in (\ref{eq1.2}) and the drift as $(\sigma_t^F)^{-1}\mu_t^F$ in (\ref{eq2.3}).

\subsection{Large spot volatility matrices with conditional sparsity}\label{sec2.2} 

Our main interest lies in estimating the factor-based spot volatility structure of the latent process $X_t$. For $0<\tau<T$ with $T$ being fixed, we let $\Sigma_X(\tau)$, $\Sigma_F(\tau)$ and $\Sigma_U(\tau)$ denote the spot volatility matrices for $X_t$, $F_t$ and $ U_t$, respectively. From (\ref{eq1.2}), (\ref{eq2.3}) and (\ref{eq2.4}), assuming $F_t$ and $U_t$ are independent, we readily have that 
\begin{equation}\label{eq2.5}
\Sigma_X(\tau)=\Lambda(\tau) \Lambda(\tau)^{^\intercal}+\sigma_\tau^U\left(\sigma_\tau^U\right)^{^\intercal}=\Sigma_C(\tau)+\Sigma_U(\tau),
\end{equation}
indicating that $\Sigma_X(\tau)$ can be decomposed into the common and idiosyncratic spot volatility matrices: $\Sigma_C(\tau)$ and $\Sigma_U(\tau)$. Throughout the paper, we assume the following uniform sparsity condition: $\left\{\Sigma_U(t):\ 0\leq t\leq T\right\}\in \mathcal{S}(q,\varpi_p)$ which is defined by
\begin{equation}\label{eq2.6}
\left\{\Sigma(t)=\left[\sigma_{i_1i_2}(t)\right]_{p\times p},\ t\in[0, T]\ \big|\ \Sigma(t)\succ0,\ \ \sup_{0\leq t \leq T}\Vert\Sigma(t)\Vert_{\infty,q}\le \Psi\cdot\varpi_p\right\},
\end{equation}
where ``$\succ0$" denotes positive definiteness, $0\le q<1$ and $\Psi$ is a positive random variable satisfying ${\mathsf E}(\Psi)\leq C_\Psi<\infty$. This is similar to the sparsity assumption used by \cite{CXW13}, \cite{CL16}, \cite{CLL19} and \cite{BLLW23} and is a natural extension of the approximate sparsity \citep[e.g.,][]{BL08}. 

\smallskip

With the sparsity restriction on $\Sigma_U(\cdot)$,  we obtain a low-rank plus sparse (or conditionally sparse) structure on $\Sigma_X(\cdot)$. In the low-frequency data setting, \cite{FLM13} introduce the POET method to estimate the large covariance matrix of $X_t$ which satisfies the discrete-time approximate factor model; and \cite{WPLL21} propose a local POET via kernel smoothing to estimate the large dynamic covariance matrix of $X_t$ which satisfies the state-varying or time-varying factor model. Extending their methods to estimate $\Sigma_X(\cdot)$ is non-trivial since $X_t$ is latent and $Y_{i,t}$ is non-synchronised. \cite{K18} generalises the POET method to estimate the spot volatilities $\Sigma_X(\cdot)$, $\Sigma_C(\cdot)$ and $\Sigma_U(\cdot)$ as well as their integrated versions for noise-free and synchronised high-frequency data. His method is not applicable to our high-frequency data setting and would result in biased volatility estimation due to presence of the microstructure noise and asynchronicity in model (\ref{eq2.1}).

\subsection{Estimation of the large spot volatility matrix}\label{sec2.3}

Partly motivated by \cite{KK16} and \cite{BLLW23}, we next introduce a kernel-weighted pre-averaging method to jointly tackle the microstructure noise and asynchronicity issues. The proposed technique is a local extension of the conventional pre-averaging which is introduced by \cite{JLMPV09} and \cite{CKP10} to estimate the integrated volatility for a single asset and has been further extended by \cite{KWZ16} and \cite{DLX19} to estimate large matrices of integrated volatility. 

\smallskip

Start with locally averaging the high-frequency observations $Y_{i,t_j^i}$ via a kernel filter:
\begin{equation}\label{eq2.7}
\widetilde{X}_{i,t}=\sum_{j=1}^{n_i} \left(t_j^i-t_{j-1}^i\right)L_b(t_j^i-t)Y_{i,t_j^i},\ \ i=1,\cdots,p,
\end{equation}
where $L(\cdot)$ is a kernel function, $b$ is a bandwidth and $L_b(\cdot)=b^{-1}L(\cdot/b)$. The $\widetilde{X}_{i,t}$ can be seen as a fitted value for the latent $X_{i,t}$, and Proposition \ref{prop:3.1} in Section \ref{sec3.2} below establishes the uniform consistency property for $\widetilde{X}_{i,t}$. We adopt the modified kernel weights $(t_j^i-t_{j-1}^i)L_b(t_j^i-t)$ in (\ref{eq2.7}) to address the issue of irregular/asynchronous sampling times. For the special case of equally-spaced time points in high-frequency data collection, i.e., $t_j^i-t_{j-1}^i\equiv\Delta$, $j=1,\cdots,n_i$, the kernel filter in (\ref{eq2.7}) can be simplified to 
\[\widetilde{X}_{i,t}=\Delta\sum_{j=1}^{n_i} L_b(t_j^i-t)Y_{i,t_j^i},\]
see \cite{KK16}.

\smallskip

Let $\widetilde X_t=(\widetilde{X}_{1,t},\cdots, \widetilde{X}_{p,t})^{^\intercal}$ and $t_j=j\Delta_\circ$,  $j=0,1,\cdots,N$, with $\Delta_\circ$ being a user-specified time difference and $N=\lfloor T/\Delta_\circ\rfloor$. We construct the kernel-weighted realised volatility matrix:
\begin{equation}\label{eq2.8}
\widetilde\Sigma_X(\tau)=\sum_{j=1}^NK_h(t_j-\tau)\Delta \widetilde X_j\Delta \widetilde X_j^{^\intercal},
\end{equation}
where
\[\Delta \widetilde X_{j}=\widetilde X_{t_j}-\widetilde X_{t_{j-1}}=\left(\Delta \widetilde X_{1,j}, \cdots, \Delta \widetilde X_{p,j}\right)^{^\intercal}.\] 
With eigen-analysis on $\widetilde\Sigma_X(\tau)$, we obtain $(\widetilde{\lambda}_{l}(\tau),\widetilde{\eta}_{l}(\tau))$, $l=1,\cdots,p$, as pairs of eigenvalues and normalised eigenvectors, where the eigenvalues are arranged in a descending order. From the matrix spectral decomposition, assuming the factor number is known a priori, we re-write
\begin{equation}\label{eq2.9}
\widetilde\Sigma_X(\tau)=\sum_{l=1}^k \widetilde{\lambda}_{l}(\tau)\widetilde{\eta}_{l}(\tau)\widetilde{\eta}_{l}(\tau)^{^\intercal}+\sum_{l=k+1}^p \widetilde{\lambda}_{l}(\tau)\widetilde{\eta}_{l}(\tau)\widetilde{\eta}_{l}(\tau)^{^\intercal}=:\widetilde\Sigma_C(\tau)+\widetilde\Sigma_U(\tau),
\end{equation}
where $\widetilde\Sigma_C(\tau)$ and $\widetilde\Sigma_U(\tau)$ are the estimated spot volatility matrices at time $\tau$ for the common and idiosyncratic error components, respectively. By the construction in (\ref{eq2.9}), $\widetilde\Sigma_C(\tau)$ is a low-rank covariance matrix. With the sparsity restriction on $\Sigma_U(\cdot)$, we may further apply a generalised shrinkage. Let $s_\rho(\cdot)$ be a shrinkage function satisfying that (i) $\vert s_\rho(u)\vert \leq \vert u\vert$ for $u\in{\cal R}$; (ii) $s_\rho(u)=0$ if $\vert u\vert \leq \rho$; and (iii) $\vert s_\rho(u)-u\vert\leq \rho$, where $\rho$ is a user-specified tuning parameter controlling the level of shrinkage. Letting $\widetilde{\sigma}_{U,i_1i_2}(\tau)$ be the $(i_1,i_2)$-entry of $\widetilde\Sigma_U(\tau)$, we define
\begin{equation}\label{eq2.10}
\overline\Sigma_U(\tau)=\left[\overline{\sigma}_{U,i_1i_2}(\tau)\right]_{p\times p}\ \ {\rm with}\ \ \overline{\sigma}_{U,i_1i_2}(\tau)=s_{\rho(\tau)}\left(\widetilde{\sigma}_{U,i_1i_2}(\tau)\right),
\end{equation}
where $\rho(\cdot)$ is a time-varying tuning parameter. Consequently, we obtain the kernel POET estimate of $\Sigma_X(\tau)$:
\begin{equation}\label{eq2.11}
\overline\Sigma_X(\tau)=\widetilde\Sigma_C(\tau)+\overline\Sigma_U(\tau).
\end{equation}

\smallskip

It follows from Theorem 1 in \cite{FLM13} and Proposition 1 in \cite{WPLL21} that the kernel POET estimate defined in (\ref{eq2.11}) is equivalent to the local PCA method to be introduced shortly. By (\ref{eq1.2}) with some smoothness condition on $\Lambda(\cdot)$ (see Assumption \ref{ass:1}(iii) in Section \ref{sec3.1}), we may obtain the following local approximation of the time-varying factor model using the increments of the stochastic processes:
\begin{equation}\label{eq2.12}
\Delta X_j K_h^{1/2}(t_j-\tau)\approx \Lambda(\tau)\Delta F_j K_h^{1/2}(t_j-\tau)+\Delta U_j K_h^{1/2}(t_j-\tau),
\end{equation}
where $\Delta X_{j}=X_{t_j}-X_{t_{j-1}}=(\Delta X_{1,j}, \cdots, \Delta X_{p,j})^{^\intercal}$, $\Delta F_j$ and $\Delta U_j$ are defined analogously. As in (\ref{eq2.8}), we replace $\Delta X_j$ by $\Delta \widetilde X_j$ in the following local PCA. Write $\Lambda(\tau)=[\Lambda_{1}(\tau),\cdots,\Lambda_{p}(\tau)]^{^\intercal}$ with $\Lambda_{i}(\tau)=[\Lambda_{i,1}(\tau),\cdots,\Lambda_{i,k}(\tau)]^{^\intercal}$,
\begin{eqnarray}
\Delta\widetilde X(\tau)&=&\left[\Delta\widetilde X_1 K_h^{1/2}(t_1-\tau),\cdots,\Delta\widetilde X_N K_h^{1/2}(t_N-\tau)\right],\notag\\
\Delta F(\tau)&=&\left[\Delta F_1 K_h^{1/2}(t_1-\tau),\cdots,\Delta F_N K_h^{1/2}(t_N-\tau)\right]^{^\intercal}.\notag
\end{eqnarray}
Define the kernel-weighted least squares objective function:
\begin{equation}\label{eq2.13}
\sum_{j=1}^N\left(\Delta\widetilde X_j-\overline\Lambda\cdot\overline F_{j}\right)^{^\intercal}\left(\Delta\widetilde X_j-\overline\Lambda\cdot\overline F_{j}\right)K_h(t_j-\tau)=\left\Vert \Delta\widetilde X(\tau)-\overline\Lambda\cdot\overline F(\tau)^{^\intercal}\right\Vert_F^2,
\end{equation}
where $\overline\Lambda$ and $\overline F_j$ are generic notation for the factor loading matrix and common factor in the local PCA estimation procedure, and $\overline F(\tau)$ is defined similarly to $\Delta F(\tau)$ but with $\Delta F_j$ replaced by $\overline F_{j}$.

\smallskip

Consider the following identification condition:
\begin{equation}\label{eq2.14}
\Delta F(\tau)^{^\intercal}\Delta F(\tau)=I_k\ \ \ {\rm and}\ \ \ \frac{1}{p}\Lambda(\tau)^{^\intercal}\Lambda(\tau)\ {\rm is\ diagonal},
\end{equation}
which are commonly used in PCA estimation of the factor models \citep[e.g.,][]{BN02,FLM13, CMZ20}. With the identification condition (\ref{eq2.14}), we replace $\overline\Lambda$ in (\ref{eq2.13}) by $\Delta\widetilde X(\tau)\overline F(\tau)$. Then, the kernel-weighted objective function in (\ref{eq2.13}) becomes
\[
\mathsf{trace}\left\{\Delta\widetilde X(\tau)^{^\intercal}\Delta\widetilde X(\tau)\right\}-\mathsf{trace}\left\{\overline F(\tau)^{^\intercal}\Delta\widetilde X(\tau)^{^\intercal}\Delta\widetilde X(\tau)\overline F(\tau)\right\},
\]
indicating that minimising (\ref{eq2.13}) subject to the restriction (\ref{eq2.14}) is equivalent to maximising the trace of $\overline F(\tau)^{^\intercal}\Delta\widetilde X(\tau)^{^\intercal}\Delta\widetilde X(\tau)\overline F(\tau)$ subject to $\overline F(\tau)^{^\intercal}\overline F(\tau)=I_k$. Hence, we conduct the eigen-analysis on the $N\times N$ matrix $\Delta\widetilde X (\tau)^{^\intercal}\Delta\widetilde X(\tau)$ and obtain the local PCA estimate of $\Delta F(\tau)$: 
\[\Delta\widetilde F(\tau)=\left[\Delta\widetilde F_1(\tau),\cdots,\Delta\widetilde F_N(\tau)\right]^{^\intercal}\] 
which is a matrix consisting of the eigenvectors corresponding to the $k$ largest eigenvalues. Furthermore, the time-varying factor loading matrix $\Lambda(\tau)$ is estimated as
\[\widetilde\Lambda(\tau)=\Delta\widetilde X(\tau)\Delta\widetilde F(\tau)=\left[\widetilde\Lambda_1(\tau),\cdots,\widetilde\Lambda_p(\tau)\right]^{^\intercal}.\]
The kernel-weighted residuals $\Delta U_j K_h^{1/2}(t_j-\tau)$ in the local approximation (\ref{eq2.12}) are then approximated by
\[\widetilde U_j(\tau)=\left[\widetilde{U}_{1,j}(\tau),\cdots,\widetilde{U}_{p,j}(\tau)\right]^{^\intercal}=\Delta\widetilde X_j K_h^{1/2}(t_j-\tau)-\widetilde\Lambda(\tau)\Delta\widetilde F_j(\tau),\]
which are subsequently used to estimate $\Sigma_U(\cdot)$. However, the conventional sample covariance matrix using $\widetilde U_j(\tau)$:
\begin{equation}\label{eq2.15}
\check\Sigma_U(\tau)=\left[\check{\sigma}_{U,i_1i_2}(\tau)\right]_{p\times p}=\sum_{j=1}^N\widetilde U_j(\tau)\widetilde U_j(\tau)^{^\intercal},
\end{equation}
usually performs poorly when the number of assets $p$ is ultra large. To address this problem, we again apply the generalised shrinkage and estimate $\Sigma_U(\tau)$ by
\begin{equation}\label{eq2.16}
\widehat\Sigma_U(\tau)=\left[\widehat{\sigma}_{U,i_1i_2}(\tau)\right]_{p\times p}\ \ {\rm with}\ \ \widehat{\sigma}_{U,i_1i_2}(\tau)=s_{\rho(\tau)}\left(\check{\sigma}_{U,i_1i_2}(\tau)\right).
\end{equation}
Finally, the estimate of $\Sigma_X(\tau)$ is obtained via 
\begin{equation}\label{eq2.17}
\widehat\Sigma_X(\tau)=\widetilde\Lambda(\tau)\widetilde\Lambda(\tau)^{^\intercal}+\widehat\Sigma_U(\tau).
\end{equation}

\smallskip

It is worth comparing our model assumption and methodology with those in \cite{CMZ20} before concluding this section. Although our main model framework is similar to that in \cite{CMZ20}, we impose more general assumptions on the microstructure noises, allowing them to be nonlinear heteroskedastic and asymptotically vanishing. In particular, the noises can be endogenous but no explicit correlation structure (between noises and latent prices) is required. The estimation methodology in \cite{CMZ20} is mainly built on the smoothed two-scale realised volatility introduced in \cite{CMZ19}, which combines the pre-averaging and two-scale realised volatility. This is further combined with local POET to estimate spot volatility matrices in the high dimension. In contrast, our proposed methodology is virtually simpler, using modified kernel weights in the first step of pre-averaging and kernel POET (or local PCA) in the second step.


\section{Main theoretical results}\label{sec3}
\renewcommand{\theequation}{3.\arabic{equation}}
\setcounter{equation}{0}

In this section, we give some regularity conditions with remarks and then present the theoretical properties for the developed large spot volatility matrix estimates via the in-fill asymptotics.

\subsection{Regularity conditions}\label{sec3.1}

Write 
\begin{eqnarray}
&&\mu_{t}^F=\left(\mu_{1,t}^F,\cdots,\mu_{k,t}^F\right)^{^\intercal},\quad \mu_{t}^U=\left(\mu_{1,t}^U,\cdots,\mu_{p,t}^U\right)^{^\intercal},\notag\\
&&\sigma_{t}^U=\left(\sigma_{i_1i_2,t}^U\right)_{p\times p},\quad \Sigma_U(t)=\left[\sigma_{U,i_1i_2}(t)\right]_{p\times p}=\sigma_t^U(\sigma_t^U)^{^\intercal}.\notag
\end{eqnarray}

\begin{assumption}\label{ass:1}          
        
{\em (i)\ Let $\left\{\mu_{i,t}^F: t\geq0\right\}$, $\left\{\mu_{i,t}^U: t\geq0\right\}$, and $\left\{\sigma_{i_1i_2,t}^U: t\geq0\right\}$ be adapted locally bounded processes with continuous sample path.}

{\em (ii)\ Let $\left\{\sigma_{U,i_1i_2}(t): t\geq0\right\}$, $1\leq i_1,i_2\leq p$, be adapted locally bounded and satisfy that
\[
\min_{1\leq i\leq p}\inf_{0\leq t\leq T}\sigma_{U,ii}(t)>0,\ \ \min_{1\leq i_1\neq i_2\leq p}\inf_{0\leq t\leq T} \left[\sigma_{U,i_1i_1}(t)+\sigma_{U,i_2i_2}(t)+2\sigma_{U,i_1i_2}(t)\right]>0,
\]
and
\begin{equation}\label{eq3.1}
\sup_{1\leq i_1,i_2\leq p}\left\vert \sigma_{U,i_1i_2}(t+\epsilon)-\sigma_{U,i_1i_2}(t)\right\vert\leq B_U(t,\epsilon)|\epsilon|^\delta+o(|\epsilon|^\delta),\ \ \epsilon\rightarrow0,
\end{equation}
with probability one, where $0<\delta<1/2$, and $B_U(t,\epsilon)$ is a positive random function which is continuous with respect to $t$ and slowly varying at $\epsilon=0$.} 

{\em (iii)\ Let $\left\{\Lambda_{i}(t): t\geq0\right\}$, $1\leq i\leq p$, be adapted locally bounded processes satisfying that
\begin{equation}\label{eq3.2}
\sup_{1\leq i\leq p}\left\Vert \Lambda_{i}(t+\epsilon)-\Lambda_{i}(t)\right\Vert\leq B_\Lambda(t,\epsilon)|\epsilon|^\delta+o(|\epsilon|^\delta),\ \ \epsilon\rightarrow0,
\end{equation}
with probability one, where $B_\Lambda(t,\epsilon)$ is defined similarly to $B_U(t,\epsilon)$ in (\ref{eq3.1}), and 
\begin{equation}\label{eq3.3}
\sup_{0\leq t\leq T}\left\Vert \frac{1}{p}\sum_{i=1}^p\Lambda_{i}(t)\Lambda_{i}(t)^{^\intercal}-\Sigma_\Lambda(t)\right\Vert=o_P(1),\quad p\rightarrow\infty, 
\end{equation}
where $\Sigma_\Lambda(t)$ is positive definite with uniformly bounded eigenvalues.  } 

\end{assumption}

\begin{assumption}\label{ass:2}

{\em (i)\ Let $\{\varepsilon_{i,j}^\ast:\ j=1,2,\cdots\}$ be a stationary and $\alpha$-mixing random sequence satisfying that the mixing coefficient $\alpha_i(t)=O(\gamma_0^t)$ uniformly over $i$, $0<\gamma_0<1$, $\mathsf{E}(\varepsilon_{i,j}^\ast)=0$, $\mathsf{Var}(\varepsilon_{i,j}^\ast)=1$ and}
\begin{equation}\label{eq3.4}
\max_{1\leq i\leq p}\mathsf{E}\left[\exp\left\{s_0\left|\varepsilon_{i,j}^\ast\right|\right\}\right]\leq C_{\varepsilon}<\infty, \ \ 0<s_0<\infty.
\end{equation}

{\em (ii)\ Let $0\leq \beta_i<1/2$ and the deterministic function $\chi_i(\cdot)$ satisfy that }
\[\max_{1\leq i\leq p}\sup_{0\leq t\leq T}|\chi_i(t)|\leq C_\chi<\infty.\]

\end{assumption}

\begin{assumption}\label{ass:3}

{\em (i)\ Let $t_j^i-t_{j-1}^i=c_{j}^in_i^{-1}$, where 
\[0<\underline c\leq\min_{1\leq i\leq p}\min_{1\leq j\leq n_i}c_{j}^i\leq \max_{1\leq i\leq p}\max_{1\leq j\leq n_i}c_{j}^i\leq \overline{c}<\infty.\]
In addition, there exists a $\kappa_0>0$ such that $N=O(\underline{n}^{\kappa_0})$, where $N$ is the number of pseudo-sampling time points in (\ref{eq2.8}) and $\underline{n}=\min_{1\leq i\leq p}n_i$.}

{\em (ii)\ The bandwidth $b$, the dimension $p$ and the sample sizes $n_i$ satisfy that $b\rightarrow0$,
\begin{equation}\label{eq3.5}
\min_{1\leq i\leq p}\frac{n_i^{1-4\kappa_1}b}{\log (p\vee n_i)}\rightarrow\infty,\ \ \ \sum_{i=1}^pn_i \exp\left\{-s_0n_i^{\kappa_1}\right\}\rightarrow0,
\end{equation}
where $0<\kappa_1<1/4$ and $s_0$ is defined in (\ref{eq3.4}).}

\end{assumption}

\begin{assumption}\label{ass:4}

{\em (i)\ The kernels $L(\cdot)$ and $K(\cdot)$ are bounded and Lipschitz continuous probability density functions with a compact support $[-1,1]$.}

{\em (ii)\ The bandwidth $h$, the dimension $p$ and the pseudo-sample size $N$ satisfy that 
\begin{equation}\label{eq3.6}
h^{2\delta}\log(p\vee N)\rightarrow0,\ \ \frac{Nh}{\log(p\vee N)}\rightarrow\infty,\ \ \frac{ph}{\log^2(p\vee N)\varpi_p}\rightarrow\infty,
\end{equation}
where $\varpi_p$ is defined in (\ref{eq2.6}). In addition,
\begin{equation}\label{eq3.7}
N^{1/2}\overline\nu\left(p,b, {\mathbf n}\right)\rightarrow0,\quad \overline\nu\left(p,b, {\mathbf n}\right)=\max_{1\leq i\leq p}\nu(p,b,n_i),
\end{equation}
where ${\mathbf n}=\{n_i\}_{i=1}^p$,
\[
\nu(p,b,n_i)=\sqrt{\log(p\vee n_i)}\left[(n_i^{2\beta_i+1} b)^{-1/2}+(n_ib)^{-1}+b^{1/2}\right].\]

}

{\em (iii)\ Let the time-varying tuning parameter $\rho(\cdot)$ in the generalised shrinkage be chosen as 
\begin{equation*}
\rho(\tau)=M(\tau)\left[\zeta_1(p,b,{\mathbf n},N)+\zeta_2(p,h,N)\right],
\end{equation*}
where $M(\cdot)$ is a positive function satisfying that} 
\begin{equation*}
0<\underline{C}_M\le \inf_{0\leq t\leq T}M(t)\leq\sup_{ 0\leq t\leq T}M(t)\leq\overline{C}_M<\infty,
\end{equation*}
{\em $\zeta_1(p,b,{\mathbf n},N)=N^{1/2}\overline\nu\left(p,b, {\mathbf n}\right)$ with $\overline\nu\left(p,b, {\mathbf n}\right)$ defined in (\ref{eq3.7}), and }
\[
\zeta_2(p,h,N)=\left(\frac{\varpi_p}{ph}\right)^{1/2}+\left(\frac{\log(p\vee N)}{Nh}\right)^{1/2}+h^{\delta}.
\]

\end{assumption}

\renewcommand{\theremark}{3.\arabic{remark}}\setcounter{remark}{0}

\begin{remark}\label{re:3.1}

(i) Assumption \ref{ass:1} contains some mild conditions on the drift, volatility and factor loading processes \citep[e.g.,][]{K18}. In particular, the local boundedness conditions can be strengthened to the uniform boundedness via the localisation technique \citep[e.g.,][]{JP12}: 
\begin{eqnarray}
&&\max_{1\leq i\leq k}\sup_{0\leq t\leq T} \left|\mu_{i,t}^F\right|+\max_{1\leq i\leq p}\sup_{0\leq t\leq T}\left|\mu_{i,t}^U\right|\leq C_\mu<\infty,\label{eq3.8}\\
&&\max_{1\leq i_1,i_2\leq p}\sup_{0\leq t\leq T} \left[\left\vert\sigma_{i_1i_2,t}^U\right\vert+\left\vert\sigma_{U,i_1i_2}(t)\right\vert \right]\leq C_\sigma<\infty,\label{eq3.9}\\
&&\max_{1\leq i\leq p}\sup_{0\leq t\leq T}\left\Vert \Lambda_i(t)\right\Vert \leq C_{\Lambda}<\infty,\label{eq3.10}
\end{eqnarray}
with probability one. The smoothness conditions (\ref{eq3.1}) and (\ref{eq3.2}) are similar to those in \cite{K10}, \cite{ZB14} and \cite{BLLW23}, which determine the bias order of the kernel-based estimation. The convergence assumption (\ref{eq3.3}) is similar to the condition (3.1) in \cite{WPLL21}, indicating that all the latent high frequency factors are pervasive if the minimum eigenvalue of $\Sigma_\Lambda(t)$ is bounded away from zero (uniformly over $t$). 

(ii) Assumption \ref{ass:2} shows that the microstructure noise vector is allowed to be serially correlated, nonlinear heteroskedastic and asymptotically vanishing. We do not impose any explicit correlation structure between the noises and latent prices. The moment condition in (\ref{eq3.4}) is weaker than the typical sub-Gaussian condition commonly adopted in high-dimensional covariance matrix estimation \citep[e.g.,][]{BL08}. The restriction of $0\leq\beta_i<1/2$ in Assumption \ref{ass:2}(ii) is the same as that in \cite{KL08}, and the uniform boundedness condition on $\chi_i(\cdot)$ parallels those in (\ref{eq3.8})--(\ref{eq3.10}). 

(iii) Assumption \ref{ass:3}(i) imposes some mild restrictions on the transaction time points which are allowed to be non-equidistant (for each asset). The liquidity level of the asynchronous high-frequency data may vary over assets. For synchronised data setting with $n_1=\cdots=n_p=n$, the conditions on $b$ and $p$ in Assumption \ref{ass:3}(ii) can be simplified to
\[
\frac{n^{1-4\kappa_1}b}{\log (p\vee n)}\rightarrow\infty\ \ \ {\rm and}\ \ \ \frac{pn}{\exp\left\{s_0n^{\kappa_1}\right\}}\rightarrow0.
\]
The first condition is comparable to the commonly-used bandwidth condition $n b/\log n$ if $\kappa_1$ tends to zero and $p$ diverges at a polynomial rate of $n$, whereas the second one indicates that the number of assets can be ultra large, diverging at an exponential rate of $n$. Although the transaction times $t_j^i$ are assumed to be non-random, we conjecture that the methodology and theory may be applicable when the transaction times are stochastic \citep[e.g.,][]{JLZ17, JLZ19, LL22a, LL22b} with modified Assumption \ref{ass:3}(i). For example, it is easy to verify the main theorems when the stochastic transaction times are independent of all the other random elements in the model and the conditions in Assumption \ref{ass:3}(i) hold with probability one. A more challenging case of endogenous transaction times will be explored in our future studies.

(iv) Assumption \ref{ass:4}(i) imposes some mild conditions for the kernel functions $K(\cdot)$ and $L(\cdot)$, which are satisfied for the commonly-used uniform and Epanechnikov kernels. Assumption \ref{ass:4}(ii) imposes some conditions on $h$, $p$ and the pseudo-sample size $N$ in the kernel POET and local PCA. The first condition in (\ref{eq3.6}) is to handle the kernel estimation bias order uniformly whereas the fourth condition ensures that the approximation error of the latent price by the kernel filter converges to zero. When $p$ is of order higher than $N$ and $\varpi_p$ is bounded, the third condition in (\ref{eq3.6}) is implied by the second one and thus can be removed. The restriction (\ref{eq3.7}) ensures that the accumulation of pre-averaging approximation errors is ``small" in the local PCA and kernel POET. Assumption \ref{ass:4}(iii) is similar to Assumption 4(iii) in \cite{BLLW23}. In addition, the developed methodology and theory continue to hold when the time-varying tuning parameter varies over entries of the spot volatility matrix. For instance, we may set $\rho_{i_1i_2}(\tau)=\rho(\tau)[\widehat{\sigma}_{U,i_1i_1}(\tau)\widehat{\sigma}_{U,i_2i_2}(\tau)]^{1/2}$ and shrink the estimate of the $(i_1,i_2)$-entry to zero if the spot {\em correlation} is smaller than $\rho(\tau)$. 

\end{remark} 


\subsection{Asymptotic theorems}\label{sec3.2}

We next present the in-fill asymptotic properties of the proposed estimates, letting $t_j^i-t_{j-1}^i\rightarrow0$ and thus $n_i\rightarrow\infty$ for each asset. Proposition \ref{prop:3.1} below gives the uniform approximation order of the latent price estimates by kernel-weighted pre-averaging.

\smallskip

\renewcommand{\theprop}{3.\arabic{prop}}\setcounter{prop}{0}
\begin{prop}\label{prop:3.1}

Suppose that Assumptions \ref{ass:1}--\ref{ass:3} and \ref{ass:4}(i) are satisfied. Define
\[
{\mathcal D}_i=\left\{\max_{1\leq j\leq N}\left\vert \widetilde{X}_{i,t_j}-X_{i,t_j}\right\vert>c_\dagger \nu(p,b,n_i)\right\},
\]
where $c_\dagger$ is a sufficiently large positive constant and $\nu(p,b,n_i)$ is defined in Assumption \ref{ass:4}(ii). Then 
\begin{equation}\label{eq3.11}
\mathsf{P}\left(\bigcup_{i=1}^p{\mathcal D}_i\right)\rightarrow0,\ \ {\it as}\ p,\underline{n}\rightarrow\infty,
\end{equation}
where $\underline{n}$ is defined in Assumption \ref{ass:3}(i).

\end{prop}

\smallskip

\begin{remark}\label{re:3.2}

(i) The approximation rate in Proposition \ref{prop:3.1} is determined by the orders of $p$, $n_i$, $b$ and the magnitude of $\beta_i$. When $\beta_i$ approaches zero, $(n_ib)^{-1}$ is dominated by $(n_i^{2\beta_i+1} b)^{-1/2}$ and thus $\nu(p,b,n_i)=[\log(p\vee n_i)]^{1/2}[(n_i^{2\beta_i+1} b)^{-1/2}+b^{1/2}]$. When $\beta_i$ approaches $1/2$, $(n_i^{2\beta_i+1} b)^{-1/2}$ is of order smaller than $(n_ib)^{-1}$ and $\nu(p,b,n_i)=[\log(p\vee n_i)]^{1/2}[(n_i b)^{-1}+b^{1/2}]$. It follows from Proposition \ref{prop:3.1} that 
\begin{equation}\label{eq3.12}
\max_{1\leq i\leq p}\max_{1\leq j\leq N}\left\vert \widetilde{X}_{i,t_j}-X_{i,t_j}\right\vert=O_P\left(\overline\nu(p, b, {\mathbf n})\right),
\end{equation}
where $\overline\nu(p, b, {\mathbf n})$ is defined in (\ref{eq3.7}).

(ii) We next discuss explicit uniform approximation rates for some special cases. When the liquidity level remains the same over assets and the homogeneity restriction is imposed on $\beta_i$, i.e., $n_1=\cdots=n_p=n$ and $\beta_1=\cdots=\beta_p=\beta$, by (\ref{eq3.12}), we readily have that 
\[
\max_{1\leq i\leq p}\max_{1\leq j\leq N}\left\vert \widetilde{X}_{i,t_j}-X_{i,t_j}\right\vert=O_P\left(\left[\log(p\vee n)\right]^{1/2}\left[(n^{2\beta+1} b)^{-1/2}+(n b)^{-1}+b^{1/2}\right]\right).
\]
This approximation rate may be faster than that obtained in Lemma B.1 of \cite{BLLW23} as $\beta$ may be positive. When the noise is not shrinking, i.e., $\beta\equiv0$, we obtain  
\[
\max_{1\leq i\leq p}\max_{1\leq j\leq N}\left\vert \widetilde{X}_{i,t_j}-X_{i,t_j}\right\vert=O_P\left(\left[\log(p\vee n)\right]^{1/2}\left[b^{1/2}+\left(n b\right)^{-1/2}\right]\right),
\]
which is the same as Lemma B.1 in \cite{BLLW23} and comparable to the rate derived by \cite{KK16} for univariate high-frequency data. Furthermore, setting $b=n^{-1/2}$ and assuming that $p$ diverges at a polynomial rate of $n$, we obtain
\[
\max_{1\leq i\leq p}\max_{1\leq j\leq N}\left\vert \widetilde{X}_{i,t_j}-X_{i,t_j}\right\vert=O_P\left(\left(\log n\right)^{1/2}n^{-1/4}\right).
\]

\end{remark}

\smallskip

We next derive the uniform convergence property of the estimated idiosyncratic spot volatility matrix $\widehat\Sigma_U(\cdot)$ in both the max and spectral norms. Since $\widetilde\Sigma_U(\cdot)$ is equivalent to $\widehat\Sigma_U(\cdot)$, the following theorem continues to hold for $\widetilde\Sigma_U(\cdot)$.

\smallskip

\renewcommand{\thetheorem}{3.\arabic{theorem}}\setcounter{theorem}{0}

\begin{theorem}\label{thm:3.1}

Suppose that Assumptions \ref{ass:1}--\ref{ass:4} are satisfied. Then we have
\begin{equation}\label{eq3.13}
\sup_{h\leq \tau\leq T-h}\left\Vert \widehat\Sigma_U(\tau)-\Sigma_U(\tau)\right\Vert_{\max}=O_P\left(\zeta(p,b,h,{\mathbf n},N)\right),
\end{equation}
and
\begin{equation}\label{eq3.14}
\sup_{h\leq \tau\leq T-h}\left\Vert \widehat\Sigma_U(\tau)-\Sigma_U(\tau)\right\Vert_s=O_P\left(\varpi_p\left[\zeta\left(p,b,h,{\mathbf n},N\right)\right]^{1-q}\right),
\end{equation}
where 
\[\zeta(p,b,h,{\mathbf n},N)=\zeta_1(p,b,{\mathbf n},N)+\zeta_2(p,h,N)\]
with $\zeta_1(p,b,{\mathbf n},N)$ and $\zeta_2(p,h,N)$ defined in Assumption \ref{ass:4}(iii).

\end{theorem}

\smallskip

\begin{remark}\label{re:3.3}

(i)\ The order $\zeta_1(p,b,{\mathbf n},N)$ in the uniform convergence rates is due to the pre-averaging approximation error of the latent prices (e.g., Proposition \ref{prop:3.1} and Remark \ref{re:3.2}), whereas $\zeta_2(p,h,N)$ is the estimation error of the infeasible local PCA directly with latent prices. With an extra term $\left[\varpi_p/(ph)\right]^{1/2}$ in $\zeta_2(p,h,N)$, our uniform convergence rates are a bit slower than that in Theorem 2 of \cite{BLLW23}. If, in addition, assuming $\varpi_p=o\left([p\log (p\vee N)/N]^{1/2}\right)$ and $p\geq N$, we may simplify $\zeta_2(p,h,N)$ to
\begin{equation}\label{eq3.15}
\zeta_2^\star(p,h,N)=\left(\frac{\log(p\vee N)}{Nh}\right)^{1/2}+h^{\delta},
\end{equation}
where the two rates are due to the kernel estimation variance and bias, respectively. In particular, treating $Nh$ as the ``effective pseudo sample size" in kernel-weighted estimation, the first term on the right side of (\ref{eq3.15}) is an optimal minimax rate for large volatility matrix estimation \citep[e.g.,][]{CZ12}.

(ii)\ We next derive an explicit uniform convergence rate for $\widehat\Sigma_U(\cdot)$. For simplicity, we replace $\zeta_2(p,h,N)$ by $\zeta_2^\star(p,h,N)$ defined in (\ref{eq3.15}) in the following discussion, and assume that $q=0$, $\varpi_p$ is bounded, $n_1=\cdots=n_p=n$ and $\beta_1=\cdots=\beta_p=\beta$. We next consider two scenarios: (a) $n^{1-2\beta}b\rightarrow\infty$, and (b) $n^{1-2\beta}b\rightarrow0$. 

\begin{itemize}

\item Scenario (a) often occurs when $\beta$ is close to zero. In this case, $(n b)^{-1}$ is dominated by $(n^{2\beta+1} b)^{-1/2}$ and thus 
\[
\zeta_1(p,b, {\mathbf n}, N)=\left[N\log(p\vee n)\right]^{1/2}\left[(n^{2\beta+1} b)^{-1/2}+b^{1/2}\right].
\] 
Setting $b=n^{-(2\beta+1)/2}$ and $h=N^{-1/(2\delta+1)}$ with $N=n^{(2\beta+1)(2\delta+1)/[2(4\delta+1)]}$, the rate $\zeta(p,b,h,{\mathbf n},N)$ in (\ref{eq3.13}) and (\ref{eq3.14}) becomes $n^{-(2\beta+1)\delta/[2(4\delta+1)]}[\log(p\vee n)]^{1/2}$. Furthermore, when $\beta=0$ and $\delta$ approaches $1/2$, this rate would be close to $n^{-1/12}$ (ignoring the logarithmic rate), comparable to those derived by \cite{ZB14} and \cite{KK16}
in the univariate high-frequency data setting. In fact, this uniform convergence rate is also comparable to that obtained by \cite{CMZ20}. When the microstructure noise is shrinking, i.e., $\beta>0$, our uniform convergence rates would be faster. 

\item Scenario (b) often occurs when $\beta$ is close to $1/2$. In this case, $(n^{2\beta+1} b)^{-1/2}$ is dominated by $(n b)^{-1}$ and thus 
\[
\zeta_1(p,b, {\mathbf n}, N)=\left[N\log(p\vee n)\right]^{1/2}\left[(nb)^{-1}+b^{1/2}\right].
\] 
Setting $b=n^{-2/3}$ and $h=N^{-1/(2\delta+1)}$ with $N=n^{2(2\delta+1)/[3(4\delta+1)]}$, we may obtain that $\zeta(p,b,h,{\mathbf n},N)=n^{-2\delta/[3(4\delta+1)]}[\log(p\vee n)]^{1/2}$ in (\ref{eq3.13}) and (\ref{eq3.14}) which does not rely on $\beta$. In particular, when $\delta$ approaches $1/2$, this rate would be close to $n^{-1/9}$ (ignoring the logarithmic rate).

\end{itemize}

\end{remark}

\smallskip

We next explore the uniform convergence property of $\widehat\Sigma_X(\cdot)$. Due to the time-varying factor model structure (\ref{eq1.2}) with the condition (\ref{eq3.3}), the largest $k$ eigenvalues are spiked, diverging at a rate of $p$. Consequently, the spot volatility matrix $\Sigma_X(\cdot)$ cannot be consistently estimated in the absolute term. Motivated by \cite{FLM13}, we measure the matrix estimate $\widehat\Sigma_X(\tau)$ in the following (time-varying) relative error:
\[\left\Vert \widehat\Sigma_X(\tau)-\Sigma_X(\tau)\right\Vert_{\Sigma_X(\tau)}=\frac{1}{\sqrt{p}}\left\Vert \Sigma_X^{-1/2}(\tau)\widehat\Sigma_X(\tau) \Sigma_X^{-1/2}(\tau)-I_{p} \right\Vert_F.\] 

\smallskip

\begin{theorem}\label{thm:3.2} 

Suppose that Assumptions \ref{ass:1}--\ref{ass:4} are satisfied. Then we have
\begin{equation}\label{eq3.16}
\sup_{h\leq \tau\leq T-h}\left\Vert \widehat\Sigma_X(\tau)-\Sigma_X(\tau)\right\Vert_{\max}=O_P\left(\zeta(p,b,h,{\mathbf n},N)\right),
\end{equation}
and
{\small\begin{equation}\label{eq3.17}
\sup_{h\leq \tau\leq T-h}\left\Vert \widehat\Sigma_X(\tau)-\Sigma_X(\tau)\right\Vert_{\Sigma_X(\tau)}=O_P\left(p^{1/2}\left[\zeta(p,b,h,{\mathbf n},N)\right]^2+\varpi_p\left[\zeta(p,b,h,{\mathbf n},N)\right]^{1-q}\right),
\end{equation}}
where $\zeta(p,b,h,{\mathbf n},N)$ is defined in Theorem \ref{thm:3.1}.

\end{theorem}

\begin{remark}\label{re:3.4}

(i) The uniform convergence rate in the elementwise max norm is the same as that in (\ref{eq3.13}). To guarantee uniform consistency in the relative error of matrix estimation, we need to assume that $p^{1/2}[\zeta(p,b,h,{\mathbf n},N)]^2=o(1)$ and $\varpi_p[\zeta(p,b,h,{\mathbf n},N)]^{1-q}=o(1)$, which restrict the divergence rate of the asset number $p$, see also the discussions in \cite{FLM13} and \cite{WPLL21}.  

(ii) The spot precision matrix, the inverse of the spot volatility matrix, plays an important role in optimal portfolio selection. It is natural to take the inverse of $\widehat\Sigma_X(\cdot)$ as the estimate of $\Sigma_X(\cdot)$. By (\ref{eq2.17}) and the Sherman-Morrison-Woodbury formula, we may write
\[
\widehat\Sigma_X^{-1}(\tau)=\widehat\Sigma_U^{-1}(\tau)-\widehat\Sigma_U^{-1}(\tau)\widetilde\Lambda(\tau)\left[I_k+\widetilde\Lambda(\tau)^{^\intercal}\widehat\Sigma_U^{-1}(\tau)\widetilde\Lambda(\tau)\right]^{-1}\widetilde\Lambda(\tau)^{^\intercal}\widehat\Sigma_U^{-1}(\tau).
\]
Following the standard arguments \citep[e.g.,][]{FLM13, CMZ20, WPLL21} and using Theorem \ref{thm:3.1} and some technical lemmas available in Appendix B of \cite{LLZ2024}, we may show that
\[
\sup_{h\leq \tau\leq T-h}\left\Vert \widehat\Sigma_X^{-1}(\tau)-\Sigma_X^{-1}(\tau)\right\Vert_s=O_P\left(\varpi_p\left[\zeta\left(p,b,h,{\mathbf n},N\right)\right]^{1-q}\right).
\]

(iii) \cite{CMZ20} derive the point-wise convergence rates for $\Sigma_X(\tau)$ (in the elementwise max form) and $\Sigma_U(\tau)$ (in the spectral norm). Theorems \ref{thm:3.1} and \ref{thm:3.2} strengthen their results to the uniform convergence ones (via different estimation method and technique in proofs). Theorem \ref{thm:3.2} also establishes the uniform convergence property for $\widehat\Sigma_X(\cdot)$ in the relative error measurement.

\end{remark}

\section{Monte-Carlo simulation studies}\label{sec4}
\renewcommand{\theequation}{4.\arabic{equation}}
\setcounter{equation}{0}

\subsection{Data generating process}\label{sec4.1}

We generate the latent price process $X_t$ from a drift-free time-varying factor model:
\begin{equation}\label{eq4.1}
d{X}_t= {\Lambda}({t}) d{W}_t^F + {\sigma}^U_td{W}_t^U,
\end{equation}
where ${\Lambda}(t)$ is a $p\times k$ time-varying factor loading process, ${\sigma}^U_t$ is a $p\times p$ (diagonal) idiosyncratic volatility process, ${W}_t^F$ is a $k$-dimensional standard Brownian motion and ${W}_t^U$ is a $p$-dimensional Brownian motion with covariance matrix $\Sigma_\rho$. The number of latent factors is $k=3$. The time-varying factor loading matrix ${\Lambda}(t)=\left[\Lambda_{i,l}(t)\right]_{p\times k}$ is generated from 
\[
 \Lambda^2_{i,l}(t)=b_{i,l}\left[a_{i,l}-\Lambda^2_{i,l}(t)\right] d t+\sigma_{i,l}^{0} \Lambda_{i,l}(t) d W_{i,t}^F, \quad i=1,\cdots,p,\ \ l=1, 2, 3,
\]
where $a_{i,1}=0.01+i/p, a_{i,2}=0.0115+i/p, a_{i,3}=0.0105+i/p,  b_{i,1}=0.006+i/(100 p), b_{i,2}=0.007+i/(100 p), b_{i,3}=0.008+i/(100 p), \sigma_{i,1}^{0}=0.3+i/(5p), \sigma_{i,2}^{0}=\sigma_{i,3}^{0}=0.4+i/(5 p)$, and $W_{i,t}^F$ is the $i$-th element of $W_t^F$. Define $\sigma^U_t = \mathsf{diag}(\sigma^U_{1,t},\cdots,\sigma^U_{p,t})$ with
\[
  (\sigma_{i,t}^{U})^2=\left[0.00053+i /(100p)\right]\left[0.0017+i / p- (\sigma_{i,t}^{U})^2\right] d t+\left[0.0013+i/(10 p)\right] \sigma_{i,t}^{U} d W_{i,t}^U, 
\]
where $W_{i,t}^U$ is the $i$-th element of $W_t^U$. Choose $\Sigma_\rho=(\rho_{ij})_{p\times p}$ as a correlation matrix satisfying the banded structure \citep[e.g.,][]{GA2017}:
\begin{equation*}
\rho_{ij} =
\left\{
\begin{array}{ll}
\rho^{\left|i-j \right|} \cdot \mathsf {I}(\left|i-j \right| \leq 3), & \ i\neq j, \\
1,  & \  i = j,
\end{array}
\right.
\end{equation*}
where $\mathsf {I}(\cdot)$ is an indicator function and $\rho \sim U(0,0.5)$. Consequently, the generated spot volatility structure has the following low-rank plus sparse decomposition:
\begin{equation}\label{eq4.2}
\Sigma_X(\tau)=\Lambda(\tau) \Lambda(\tau)^{^\intercal}+\sigma_\tau^U \Sigma_\rho\sigma_\tau^U=\Sigma_C(\tau)+\Sigma_U(\tau) , \ \  0<\tau<T.
\end{equation}
The number of assets is set to be $p=100,300,500$ and the simulation is repeated for $100$ times. The sampling frequency  $\Delta$ is set to be $1/(6.5\times 60 \times 6)$, which is equivalent to sampling in every $10$ seconds, setting $T$ as one trading day and assuming there are $6.5$ trading hours for each trading day.

\smallskip

In practice, different asset prices are often non-synchronized and contaminated by the micro-structure noise. We generate the contaminated data via model (\ref{eq2.1}). Similar to \cite{FK19}, we generate $\varepsilon_{i,t}$ independently (over $i$ and $t$) from $\sigma_{\epsilon}\cdot {\mathsf N} (0, \sigma_{ii}(t))$, where $\sigma_{ii}(t)$ is the $i$-th diagonal element of the true spot volatility matrix $\Sigma_X(\cdot)$ at time point $t$ and $\sigma_{\epsilon}$ is the noise-to-signal ratio. The noise is thus correlated with the latent prices. We consider $\sigma_\epsilon=0.05, 0.1$ and $0.2$. To simulate the asynchronous data, we use $p$ independent Poisson processes to generate the true data collection time points $t_j^i$ \citep[e.g.,][]{BN2011}. Specifically, for the $i$-th asset, we control $\{t_j^i: j=1,\cdots,n_i\}$ by a Poisson process with mean $\lambda^i \sim U(1,3)$, indicating that asset prices are randomly collected every 10 to 30 seconds and we expect to have $2340/\lambda^i$ observations on average for the $i$-th asset.

\subsection{Spot volatility matrix estimates}\label{sec4.2}

Throughout the simulation studies, we only use the kernel POET estimation method since it is equivalent to the local PCA. To assess the impacts of micro-structure noise and asynchronicity, we consider the following two spot volatility matrix estimators. 

\begin{itemize}

\item Contaminated but synchronised spot volatility estimate $\overline{\Sigma}_X^\mathrm{s}(\tau)$. We consider the synchronised data collected at every $10$ seconds and then apply the developed kernel POET to the contaminated prices.
    
\item Contaminated and asynchronised spot volatility estimate $\overline{\Sigma}^\mathrm{a}_X(\tau)$. We consider the asynchronous asset prices that are randomly collected every 10 to 30 seconds and then apply the kernel POET. It is expected that the resulting finite-sample performance would be worse than the synchronised spot volatility matrix estimate $\overline{\Sigma}_X^\mathrm{s}(\tau)$.
     
\end{itemize}
We employ the following four shrinkage techniques in the spot idiosyncratic volatility matrix estimation:  Smoothly Clipped Absolute Deviation (SCAD), adaptive-lasso (A-lasso), soft thresholding (Soft) and hard thresholding (Hard). In addition, to demonstrate effectiveness of the generalised shrinkage, we also consider the naive estimator as a benchmark that does not apply shrinkage to the spot idiosyncratic volatility matrix estimation. 

\smallskip

For each $\tau$, we use the eigenvalue-ratio criterion \citep[e.g.,][]{AH2013} to determine the number of latent factors. Let $\overline{\Sigma}_{X}^{(r)}(\cdot)$ and $\overline{\Sigma}_{U}^{(r)}(\cdot)$ be generic notation for estimates of ${\Sigma}_{X}^{(r)}(\cdot)$ and ${\Sigma}_{U}^{(r)}(\cdot)$, respectively, for the $r$-th replication. To measure the distance between the true spot volatility matrices and the estimated ones, we compute the following two measurements: the mean spectral norm deviation ($\mathsf{MSN}_X$) and the mean relative norm deviation ($\mathsf{MRN}_X$) defined by
\begin{eqnarray}
\mathsf{MSN}_X &=& \frac{1}{100} \sum_{r=1}^{100} \left( \frac{1}{10} \sum_{j=1}^{10} \left \| \overline{\Sigma}_{X}^{(r)}(\tau_j) - {\Sigma}_{X}^{(r)}(\tau_j)\right \|_{s} \right), \notag\\
\mathsf{MRN}_X &=& \frac{1}{100} \sum_{r=1}^{100} \left( \frac{1}{10} \sum_{j=1}^{10} \left \| \overline{\Sigma}_{X}^{(r)}(\tau_j) - {\Sigma}_{X}^{(r)}(\tau_j)\right \|_{\Sigma_X^{(r)}(\tau_j)} \right),\notag
\end{eqnarray}
where $\| \overline{\Sigma}_{X}^{(r)}(\tau) - {\Sigma}_{X}^{(r)}(\tau) \|_{\Sigma_X^{(r)}(\tau)}$ is defined as in Theorem \ref{thm:3.2} for the $r$-th replication, and $\tau_j, j=1,2,\cdots,10$, are equidistant time points. To compare the performance of various shrinkage approaches in estimating idiosyncratic spot volatility matrices, we compute the following mean spectral norm for the idiosyncratic spot volatility matrices:
\[
\mathsf{MSN}_U =\frac{1}{100} \sum_{r=1}^{100} \left( \frac{1}{10} \sum_{j=1}^{10} \left \| \overline{\Sigma}_{U}^{(r)}(\tau_j)-\Sigma_U^{(r)}(\tau_j) \right \|_{s} \right). 
\]

\subsection{Choice of the tuning parameters}\label{sec4.3}

It is well known that the nonparametric kernel-based estimation is sensitive to the bandwidth selection. In the developed estimation procedure, we need to carefully select two bandwidths: $b$ and $h$, which are involved in the kernel-weighted pre-averaging (\ref{eq2.7}) and realised volatility matrix (\ref{eq2.8}), respectively. 

\smallskip

As recommended by \cite{KK16}, we apply the cross-validation to select the optimal bandwidth in the local pre-averaging for each asset, i.e., for the $i$-th asset, obtain
\[
b^i_\mathrm{opt}= \argmin_{b>0} \mathsf{CV}_i(b),\ \ \mathsf{CV}_i(b)=\sum_{j=1}^{n_i} \left(Y_{i,t^i_j} - \widetilde{X}_{i,-t_i^j}\right)^{2}\mathsf{I}\left\{T_{l}^i \leq t^i_{j} \leq T_{u}^i\right\}, 
\]
where the truncation mechanism is used to avoid the kernel estimation boundary effect, $T_{l}^i=0.05\times T_i$, $T_{u}^i=0.95\times T_i$, and $\widetilde{X}_{i,-t_i^j}$ is the leave-one-out kernel filter defined as in (\ref{eq2.7}) but removing the observation at $t_i^j$. 

\smallskip

In the low-dimensional data setting, we may select the optimal bandwidth entry by entry in the realised volatility matrix estimation. For example, as in \cite{K10} and \cite{KK16}, for $1\leq i_1,i_2\leq p$, we obtain
\[
h^{i_1i_2}_\mathrm{opt}= \argmin_{h>0} \mathsf{CV}_{i_1i_2}(h),\ \ \mathsf{CV}_{i_1i_2}(h)=\sum_{\tau_j} \left[ \left(\Delta\widetilde X_{\tau_j} \Delta\widetilde X_{\tau_j}^\prime / \Delta  \right)_{i_1i_2} -\widetilde{\sigma}_{X,i_1i_2,-\tau_j}\right]^{2}, 
\]
where $\widetilde{\sigma}_{X,i_1i_2,-\tau_j}$ is the ($i_1,i_2$)-entry of leave-one-out kernel-weighted realised co-volatility estimate. However, this bandwidth selection criterion is rather time consuming when the dimension $p$ is large. Hence, we adopt the following criterion:
\[
h_\mathrm{opt}= \argmin_{h>0} \mathsf{CV}(h),\ \ \mathsf{CV}(h)=\sum_{i=1}^{p}\sum_{\tau_j}\left[\left(\Delta\widetilde X_{\tau_j} \Delta\widetilde X_{\tau_j}^\prime / \Delta  \right)_{ii}-\widetilde{\sigma}_{X,ii,-\tau_j}\right]^{2},
\]
where we only evaluate the quadratic loss for the diagonal elements. To further speed up the computation, as recommended by \cite{KK16}, we may repeat the bandwidth selection processes over a small number of replications (say, $5$) and then fix the optimal bandwidth as the average of the selected bandwidth values. 

\smallskip

When applying the generalised shrinkage in the kernel POET, we need to choose an appropriate tuning parameter to control the shrinkage level. As recommended by \cite{FLM13} and \cite{WPLL21} and discussed in Remark \ref{re:3.1}(iv), we set
\[
\rho_{ij}(\tau)=c_\rho\left[{\widetilde{\sigma}_{U,ii}(\tau){\widetilde{\sigma}_{U,jj}(\tau)}}\right]^{1/2}, \ \ i,j=1,\cdots,p
\]
where $c_\rho$ is the minimum positive number guaranteeing positive definiteness of the estimated spot idiosyncratic volatility matrix. Specifically, $c_\rho$ is chosen such that
\[
c_\rho=\inf \left\{c>0:\ \lambda_{\min}\left\{\left(\overline\Sigma_U^c(\tau)\right)\right\}>0,\ \ \overline\Sigma_U^c(\tau)=\left[\overline{\sigma}_{U,ij}^c(\tau)\right]_{p\times p} \right\},
\]
where $\overline{\sigma}_{U,ij}^c(\tau)$ is defined as in (\ref{eq2.10}) using $\rho_{ij}(\tau)=c[{\widetilde{\sigma}_{U,ii}(\tau){\widetilde{\sigma}_{U,jj}(\tau)}}]^{1/2}$.

\subsection{Simulation results}\label{sec4.4}

Tables \ref{table:1} and \ref{table:2} report the $\mathsf{MSN}_{U}$ values of the spot idiosyncratic volatility matrix estimation for the noise-contaminated synchronised and asynchronised data, respectively. In general, the use of shrinkage results in more accurate estimation than the naive one without shrinkage. The improvement by using the shrinkage becomes more significant as the dimension $p$ increases. The SCAD and adaptive-lasso perform somewhat better than the other two shrinkage methods when the high-frequency data are synchronised (see Table \ref{table:1}), and the adaptive-lasso outperforms the other shrinkage methods when the data are asynchronised (see Table \ref{table:2}). It is unsurprising that the $\mathsf{MSN}_{U}$ values increase as the dimension or the noise-to-signal ratio $\sigma_\epsilon$ increases.  

\smallskip

Tables \ref{table:3}--\ref{table:6} present the estimation results for $\Sigma_X$. Tables \ref{table:3} and \ref{table:4} report the $\mathsf{MSN}_{X}$ measurements whereas Tables \ref{table:5} and \ref{table:6} report the $\mathsf{MRN}_{X}$ measurements. As in Tables \ref{table:1} and \ref{table:2}, the spot volatility matrix estimation with the generalised shrinkage outperforms the naive estimation without shrinkage. The difference between the shrinkage estimation and the naive one is more significant in terms of $\mathsf{MRN}_{X}$. Both $\mathsf{MSN}_{X}$ and $\mathsf{MRN}_{X}$ increase as the dimension $p$ grows. In particular, this divergence pattern is more significant for $\mathsf{MSN}_{X}$, which is not uncommon as the true spot volatility matrix is spiked with the largest $k$ eigenvalues diverging at the same rate of $p$. Hence, as recommended in \cite{FLM13}, it is more sensible to use the $\mathsf{MRN}_{X}$ to access the factor-based spot volatility matrix estimation. Among the four shrinkage techniques, the SCAD and adaptive-lasso shrinkage slightly outperform the soft and hard thresholding methods, and the hard thresholding often has the poorest performance (although it is still superior to the naive estimator). Due to the data asynchronisation, the 
$\mathsf{MSN}_{X}$ and $\mathsf{MRN}_{X}$ values in Tables \ref{table:4} and \ref{table:6} are larger than those provided in Tables \ref{table:3} and \ref{table:5}. The patterns of the estimation performance and comparison are similar over different noise-to-signal ratios. In addition, both the $\mathsf{MSN}_{X}$ and $\mathsf{MRN}_{X}$ measurements increase as $\sigma_\epsilon$ increases from $0.05$ to $0.2$, which is expected as ``larger" noise results in ``larger" approximation errors of the first-step pre-averaging.

\begin{table}[H]
\caption{${\sf MSN}_U$ measurements of $\Sigma_U$ for synchronised data} 
\centering 
\setlength\tabcolsep{0.7pt}
\begin{tabular}{c c c c c c c} 
\hline\hline 
$p$\ \  & $\ \ \sigma_{\epsilon}$\ \ \ &\ \ \ Naive\ \ \ &\ \ \ SCAD\ \ \ &\ \ \ A-Lasso\ \ \ & \ \ \ \ Soft\ \ \ \ \ &\ \ \ \ \ Hard\ \ \ \ \\ [1ex] 
\hline
$p=100$ &\ \ \ \ $0.05$ \ \  & 0.0270 &0.0248&0.0267&0.0260 &0.0293 \\
&\ \ \ \ $0.1$ \ \  &  0.0288 &0.0266&0.0267&0.0280 &0.0293 \\
&\ \ \ \ $0.2$ \ \  & 0.0346 &0.0300&0.0298&0.0326&0.0310 \\
\hline
$p=300$ &\ \ \ \  $0.05$\ \ & 0.0432 &0.0291&0.0296& 0.0297&0.0357\\
&\ \ \ \  $0.1$\ \ & 0.0479 &0.0284&0.0285& 0.0287&0.0371\\
&\ \ \ \  $0.2$\ \ & 0.0578 &0.0365&0.0365& 0.0362&0.0441\\
\hline
$p=500$ &\ \ \ \ $0.05$\ \ & 0.0752&0.0298&0.030&0.030&0.0473 \\
&\ \ \ \ $0.1$\ \ & 0.0781&0.0308&0.0316&0.0310&0.0506 \\
&\ \ \ \ $0.2$\ \ & 0.0875&0.0378&0.0377&0.0379&0.0587 \\
\hline\hline 
\end{tabular}
\label{table:1} 
\end{table}

\begin{table}[H]
\caption{${\sf MSN}_U$ measurements of $\Sigma_U$ for asynchronised data} 
\centering 
\setlength\tabcolsep{0.7pt}
\begin{tabular}{c c c c c c c} 
\hline\hline 
$p$\ \  & $\ \ \sigma_{\epsilon}$\ \ \ &\ \ \ Naive\ \ \ &\ \ \ SCAD\ \ \ &\ \ \ A-Lasso\ \ \ & \ \ \ \ Soft\ \ \ \ \ &\ \ \ \ \ Hard\ \ \ \ \\ [1ex] 
\hline
$p=100$ &\ \ \ \ $0.05$ \ \  & 0.0409 &0.0358&0.0357&0.0354 &0.0374 \\
&\ \ \ \ $0.1$ \ \  & 0.0418 &0.0366&0.0363&0.0361 &0.0381 \\
 &\ \ \ \ $0.2$ \ \  & 0.0425 &0.0359&0.0355&0.0353 &0.0380 \\
\hline
$p=300$ &\ \ \ \  $0.05$\ \ & 0.0800 &0.0455&0.0446& 0.0454&0.0610\\
&\ \ \ \  $0.1$\ \ & 0.0802 &0.0461&0.0451& 0.0459&0.0614\\
&\ \ \ \ $0.2$\ \ & 0.0872 &0.0461&0.0450& 0.0460&0.0648\\
\hline
$p=500$ &\ \ \ \ $0.05$\ \ & 0.1327&0.0599&0.0586&0.0600&0.0941 \\
&\ \ \ \ $0.1$\ \ & 0.1340&0.0605&0.0592&0.0605&0.0947 \\
&\ \ \ \ $0.2$\ \ & 0.1345&0.0599&0.0867&0.0600&0.0950 \\
\hline\hline 
\end{tabular}\label{table:2} 
\end{table}

\begin{table}[H]
\caption{${\sf MSN}_X$ measurements of $\Sigma_X$ for synchronised data} 
\centering 
\setlength\tabcolsep{0.7pt}
\begin{tabular}{c c c c c c c } 
\hline\hline 
$p$\ \  & $\ \ \sigma_{\epsilon}$\ \ \ &\ \ \ Naive\ \ \ &\ \ \ SCAD\ \ \ &\ \ \ A-Lasso\ \ \ & \ \ \ \ Soft\ \ \ \ \ &\ \ \ \ \ Hard\ \ \ \ \\ [1ex] 
\hline
$p=100$ &\ \ \ \  $0.05$\ \ &  1.7758 &1.7728 &1.7724&1.7750&1.7740\\
&\ \ \ \  $0.1$ \ \ & 1.8359 &1.8323 &1.8317&1.8356&1.8320\\
&\ \ \ \ $0.2$ \ \ &  2.1769&2.1723&2.1715&2.1756&2.1720\\
\hline
$p=300$ &\ \ \ \ $0.05$ \ \ & 5.2782& 5.2734 &5.2732  &5.2738& 5.2755\\
&\ \ \ \ $0.1$ \ \ &5.4809&5.4721&5.4713&5.4731&5.4751\\
&\ \ \ \ $0.2$ \ \ & 6.5930&6.5833&6.5823&6.5837&6.5865\\
\hline
$p=500$ &\ \ \ \ $0.05$ \ \ &8.645& 8.6364  & 8.6365 &8.6366&8.6388 \\
&\ \ \ \ $0.1$ \ \ &8.9720&8.9622&8.9618&8.9627&8.9660\\
&\ \ \ \ $0.2$ \ \ &10.7411&10.7303  &10.7294  &10.7305&10.7343 \\
\hline\hline
\end{tabular}\label{table:3} 
\end{table}

\begin{table}[H]
\caption{${\sf MSN}_X$ measurements of $\Sigma_X$ for asynchronised data} 
\centering 
\setlength\tabcolsep{0.7pt}
\begin{tabular}{c c c c c c c} 
\hline\hline 
$p$\ \  & $\ \ \sigma_{\epsilon}$\ \ \ &\ \ \ Naive\ \ \ &\ \ \ SCAD\ \ \ &\ \ \ A-Lasso\ \ \ & \ \ \ \ Soft\ \ \ \ \ &\ \ \ \ \ Hard\ \ \ \ \\ [1ex] 
\hline
$p=100$ &\ \ \ \  $0.05$\ \ &  2.3825& 2.3810  &2.3807  &2.3808& 2.3812 \\
&\ \ \ \  $0.1$ \ \ &2.3855& 2.3841  &2.3839  &2.3840& 2.3845\\
&\ \ \ \ $0.2$ \ \ & 2.3885&2.3869&2.3867&2.3870&2.3873\\
\hline
$p=300$ &\ \ \ \ $0.05$ \ \ &7.1376& 7.1357  &7.1357  &7.1357& 7.1367\\
&\ \ \ \ $0.1$ \ \ &7.1412&7.1396&7.1396&7.1397&7.1405\\
&\ \ \ \ $0.2$ \ \ & 7.1415&7.1399&7.1399&7.1400&7.1408\\
\hline
$p=500$ &\ \ \ \ $0.05$ \ \ &11.6455  &11.6437 &11.6439&11.6440&11.6450\\
&\ \ \ \ $0.1$ \ \ &11.6468&11.6453&11.6459&11.6454&11.6463\\
&\ \ \ \ $0.2$ \ \ &11.6470&11.6454   & 11.6456 &11.6454&11.6464\\
\hline\hline
\end{tabular}\label{table:4} 
\end{table}

\begin{table}[H]
\caption{${\sf MRN}_X$ measurements of $\Sigma_X$ for synchronised data} 
\centering 
\setlength\tabcolsep{0.7pt}
\begin{tabular}{c c c c c c c} 
\hline\hline 
$p$\ \  & $\ \ \sigma_{\epsilon}$\ \ \ &\ \ \ Naive\ \ \ &\ \ \ SCAD\ \ \ &\ \ \ A-Lasso\ \ \ & \ \ \ \ Soft\ \ \ \ \ &\ \ \ \ \ Hard\ \ \ \ \\ [1ex] 
\hline
$p=100$ &\ \ \ \  $0.05$\ \ & 1.6260 &1.4179 &1.4078&1.5433&1.4910\\
&\ \ \ \  $0.1$ \ \ &2.2902 &2.1172 &2.1117&2.2680&2.2000\\
&\ \ \ \ $0.2$ \ \ & 2.8643&2.5876&2.7683&2.2904&2.6516\\
\hline
$p=300$ &\ \ \ \ $0.05$ \ \ & 2.9659& 2.2503  &2.2019  &2.2904& 2.5560\\
&\ \ \ \ $0.1$ \ \ &3.4108&2.7957&2.7707&2.8200&3.0466\\
&\ \ \ \ $0.2$ \ \ & 4.0430&3.3056&3.0117&3.0812&3.4895\\
\hline
$p=500$ &\ \ \ \ $0.05$ \ \ &3.9195& 2.9387  & 2.8699 &2.9416&3.3836 \\
&\ \ \ \ $0.1$ \ \ &4.5601&3.5998&3.5648&3.6178&4.0260\\
&\ \ \ \ $0.2$ \ \ &5.3357&4.2404   &4.2161  &4.2471&4.7475 \\
\hline\hline
\end{tabular}\label{table:5} 
\end{table}

\begin{table}[H]
\caption{${\sf MRN}_X$ measurements of $\Sigma_X$ for asynchronised data} 
\centering 
\setlength\tabcolsep{0.7pt}
\begin{tabular}{c c c c c c c} 
\hline\hline 
$p$\ \  & $\ \ \sigma_{\epsilon}$\ \ \ &\ \ \ Naive\ \ \ &\ \ \ SCAD\ \ \ &\ \ \ A-Lasso\ \ \ & \ \ \ \ Soft\ \ \ \ \ &\ \ \ \ \ Hard\ \ \ \ \\ [1ex] 
\hline
$p=100$ &\ \ \ \  $0.05$\ \ &  2.7097& 2.4809  &2.4691  &2.4650& 2.5816 \\
&\ \ \ \  $0.1$ \ \ &2.7484& 2.5469  &2.5359  &2.5354& 2.6380\\
&\ \ \ \ $0.2$ \ \ & 3.0464&2.8049&2.7930&2.7900&2.9100\\
\hline
$p=300$ &\ \ \ \ $0.05$ \ \ &4.0469& 4.4397  &3.4174  &3.4388& 3.7649\\
&\ \ \ \ $0.1$ \ \ &4.0880&4.4732&3.4504&3.4722&3.8022\\
&\ \ \ \ $0.2$ \ \ & 4.5535&3.8059&3.7795&3.8054&4.2053\\
\hline
$p=500$ &\ \ \ \ $0.05$ \ \ &6.2918   &5.1528  &5.1268&5.1525&5.7603\\
&\ \ \ \ $0.1$ \ \ &6.3458&5.2133&5.1874&5.2130&5.8156\\
&\ \ \ \ $0.2$ \ \ &6.6000&5.4200   & 5.3920 &5.4186&6.0456\\
\hline\hline
\end{tabular}\label{table:6} 
\end{table}

\section{An empirical application}\label{sec5}
\renewcommand{\theequation}{5.\arabic{equation}}
\setcounter{equation}{0}

We next apply the proposed approach to the one-min intraday log-price of S$\&$P 500 index constituents. The log-prices are extracted from Bloomberg, ranging from 29 March to 30 June 2021. There are 66 trading days and 505 stocks in total. The 66 trading days are divided into 11 equal-length time intervals, with 6 trading days in each time interval. In the preliminary data analysis, we remove the illiquid stocks that were not traded for more than half of one-minute trading periods on each trading day, resulting in $353$ stocks in total. We also exclude stock prices that are outside the trading hours of 9:30 am (EST) to 4 pm (EST). A recent paper by \cite{LCL2023} estimates the latent factor structure for the same data set, allowing the existence of microstructure noises. In this section, we aim to further explore the spot volatility structure and study its time-varying pattern. For each equal-length time interval, we first apply the kernel-weighted pre-averaging method in (\ref{eq2.7}) to the (noisy) log-price and then construct the kernel-weighted spot volatility matrix estimates in (\ref{eq2.8}). Due to a possible low-rank plus sparse structure, we further implement the kernel POET estimation as in (\ref{eq2.11}). Specifically, we estimate $36$ evenly separated spot volatility matrices in each time interval (i.e., 6 spot volatility matrix estimates for each trading day). As recommended in the simulation, we adopt the adaptive-lasso in the generalised shrinkage with the tuning parameter chosen according to the criteria discussed in Section \ref{sec4.3}. 

\smallskip 

\begin{figure}[ht]
    \centering
    \includegraphics[width=.45\textwidth]{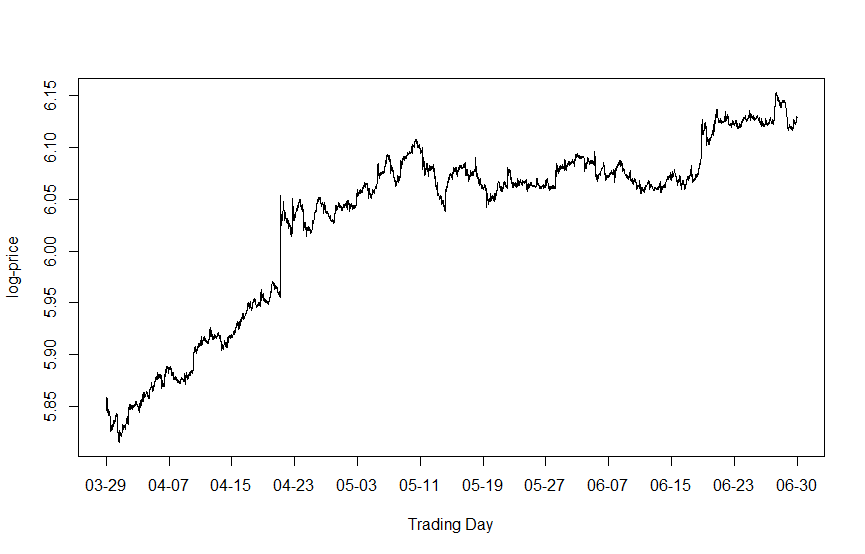}\hfill
    \includegraphics[width=.45\textwidth]{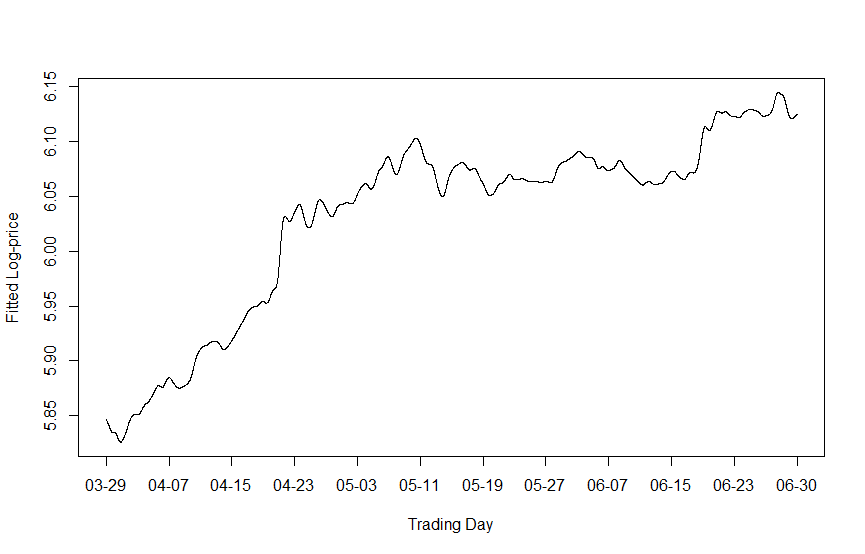}\hfill
    \caption {\emph{The left panel is the plot of the POOL.OQ stock (log) price and the right panel is the plot of the fitted log-price via pre-averaging.}} \label{fig1}
\end{figure}

The left panel of Figure \ref{fig1} depicts the log-price of the POOL.OQ stock over the sampling period whereas the right panel depicts the fitted log-price via the kernel-weighted pre-averaging. We use the Gaussian kernel function in (\ref{eq2.7}) and determine the optimal bandwidth via the cross-validation discussed in Section \ref{sec4.3}. It follows from Figure \ref{fig1} that the fitted log-prices are generally close to the original ones which may be noise contaminated, capturing most of the dynamic patterns. Due to the kernel-weighted smoothing, the fitted prices are slightly smoother than the observed ones. 

\smallskip 

\begin{figure}[ht]
    \centering
    \includegraphics[width=.45\textwidth]{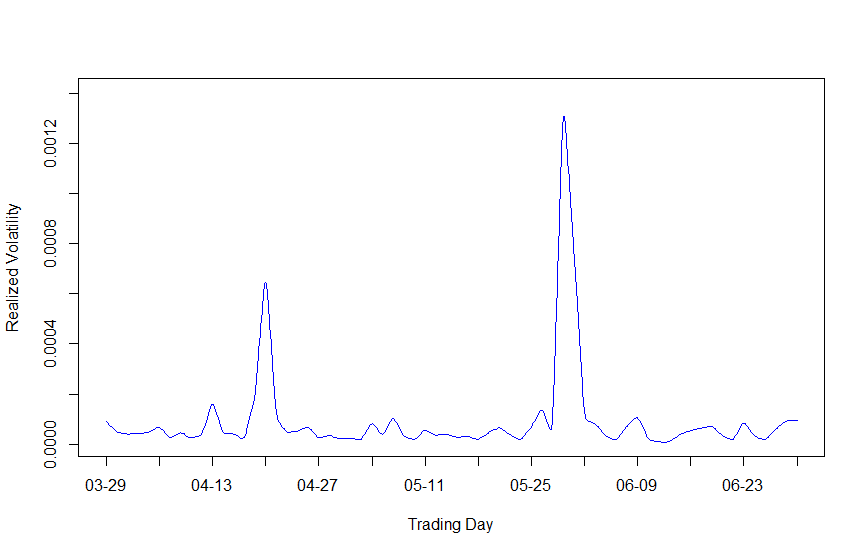}\hfill
    \includegraphics[width=.45\textwidth]{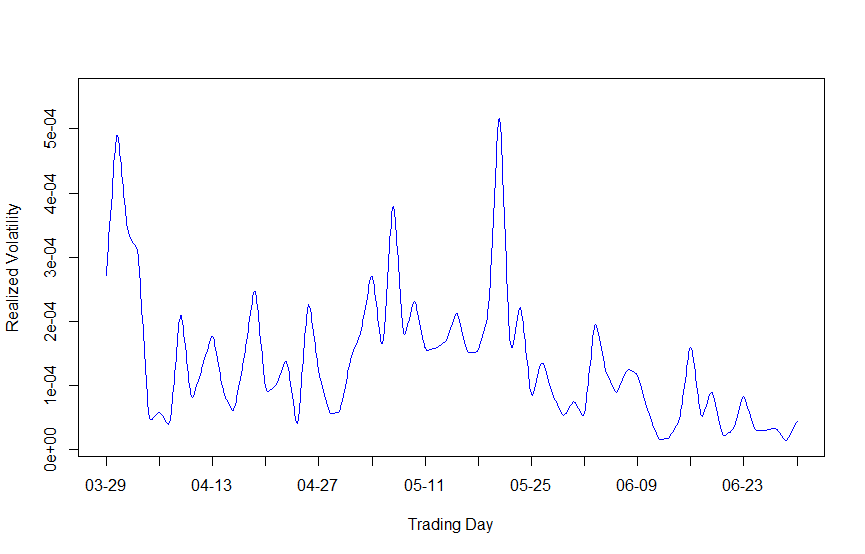}\hfill
    \\
    \includegraphics[width=.45\textwidth]{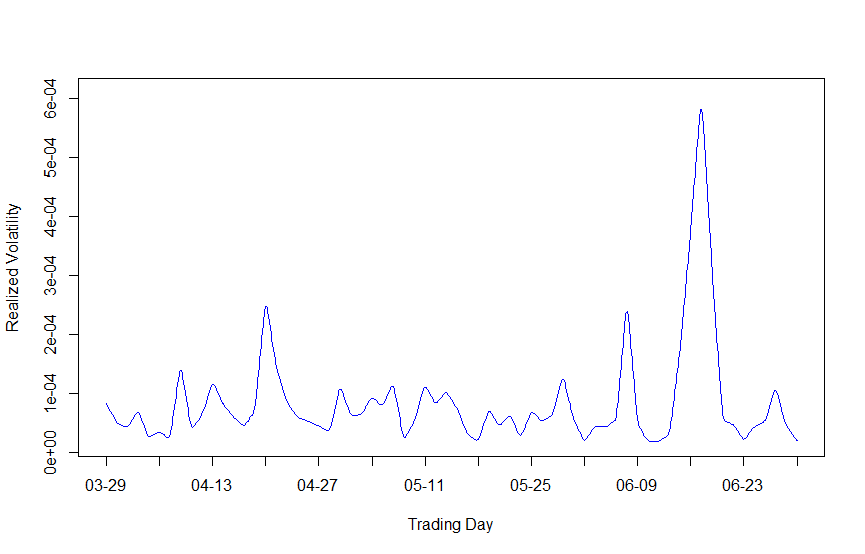}\hfill
    \includegraphics[width=.45\textwidth]{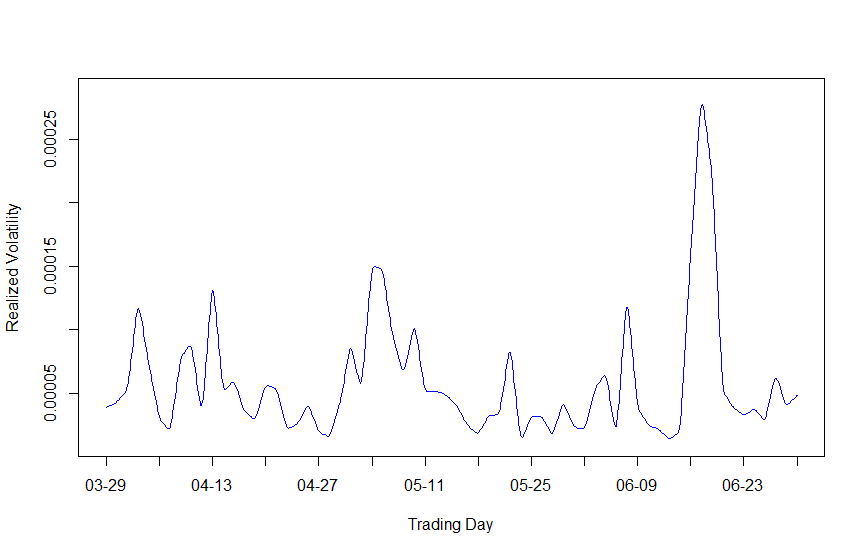}\hfill
    
    \caption {\emph{Plots of the estimated spot volatility of 20\%-th, 40\%-th, 60\%-th and 80\%-th quantile stocks. Upper-left panel: plot of the estimated spot volatility of ABT.N stock (20\%-th quantile); upper-right panel: plot of the estimated spot volatility of SNPS.OQ stock (40\%-th quantile); lower-left panel: plot of the estimated spot volatility of MET.N stock (60\%-th quantile); lower-right panel: plot of the estimated spot volatility of EMR.N stock (80\%-th quantile).}} \label{fig2}
    
\end{figure}

\begin{figure}[ht]
    \centering
    \includegraphics[width=.45\textwidth]{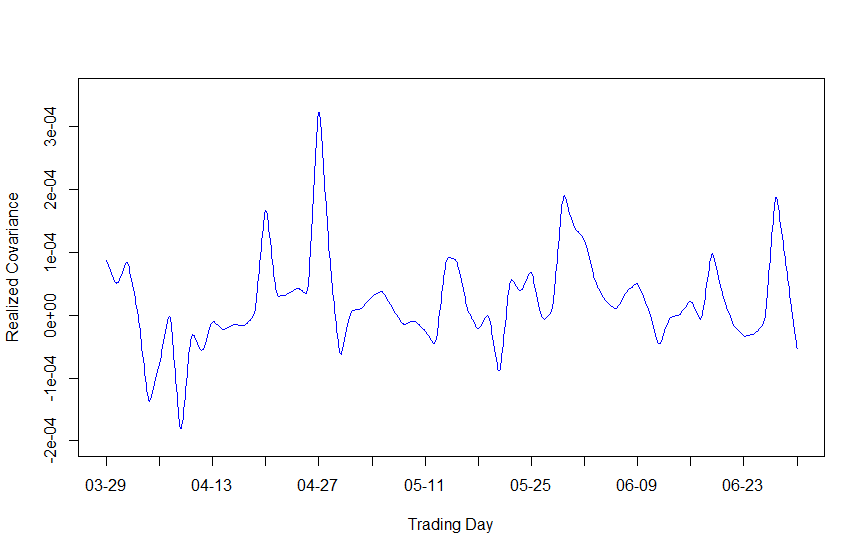}\hfill
    \includegraphics[width=.45\textwidth]{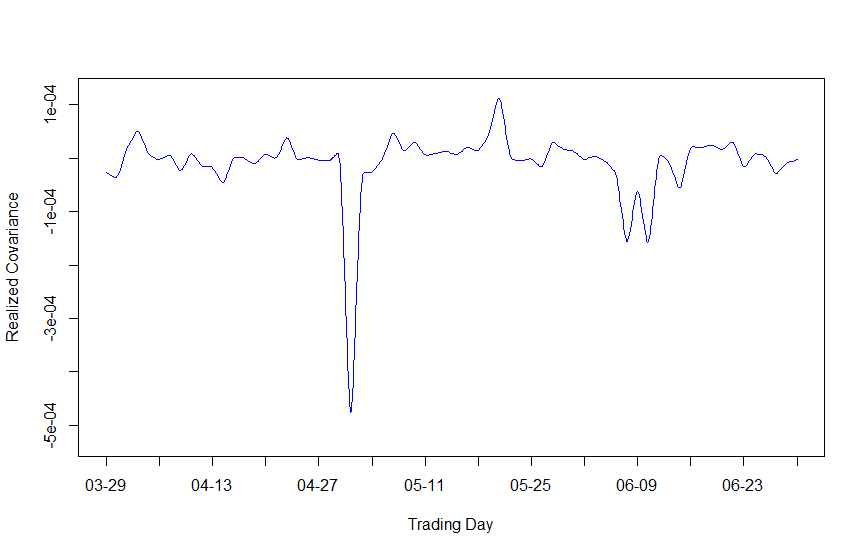}\hfill
    \\
    \includegraphics[width=.45\textwidth]{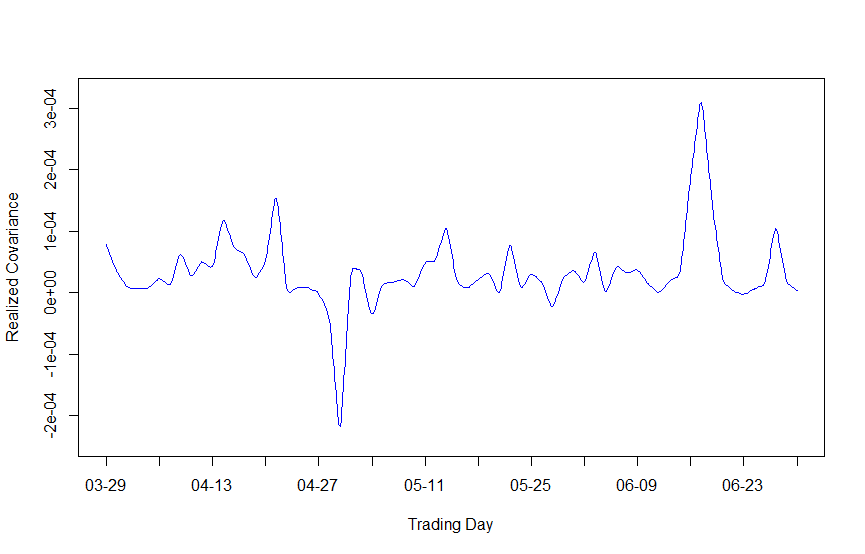}\hfill
    \includegraphics[width=.45\textwidth]{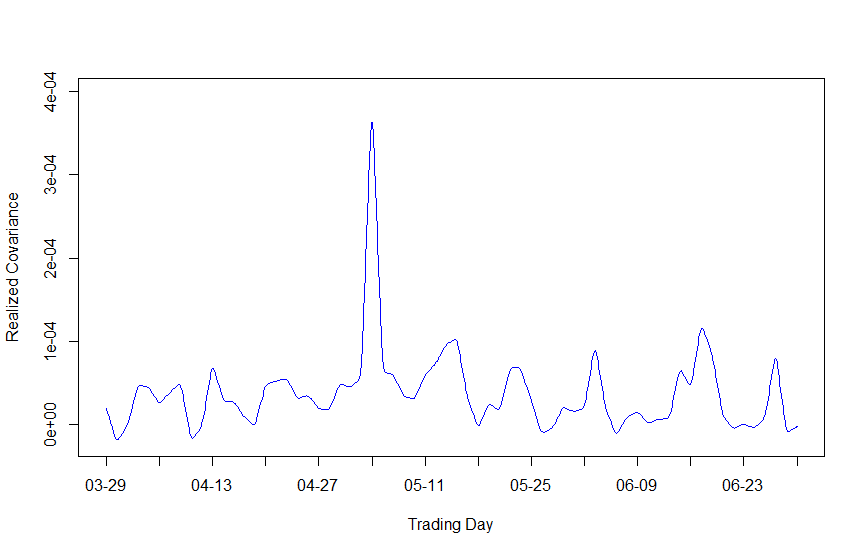}\hfill
    
    \caption {\emph{Plots of the estimated spot covariances of the 20\%-th, 40\%-th, 60\%-th and 80\%-th quantile stock pairs. Upper-left panel: plot of the estimated spot covariance between APA.OQ and MA.N stocks (20\%-th quantile); upper-right panel: plot of the estimated spot covariance between VRTX.OQ and FTV.N stocks (40\%-th quantile); lower-left panel: plot of the estimated spot covariance between BWA.N and CB.N stocks (60\%-th quantile); lower-right panel: plot of the estimated spot covariance between CTVA.N and D.N stocks (80\%-th quantile).}} \label{fig3}
    
\end{figure}

With the kernel POET estimates in (\ref{eq2.11}), we are able to collect the estimated spot volatilities (the diagonal entries of the estimated spot volatility matrix) for all the stocks and the estimated spot covariances (the off-diagonal entries of the estimated spot volatility matrix) between every pair of stocks across all the trading days. We further average the estimated volatilities and covariances within each trading day over the sampling period, and rank them (in decreasing order) according to variances of all the realised volatility and covariance over time. The estimated volatilities and covariances of the 20\%, 40\%, 60\% and 80\% quantile stocks (or stock pairs) across all the trading days are plotted in Figures \ref{fig2} and \ref{fig3}, respectively. It follows from the plots that both the spot volatilities and covariances exhibit significant dynamic changes over time, demonstrating that it is imperative to recover the spot volatility structure over time for high-frequency financial data.

\smallskip 

\begin{figure} 
    \centering
    \includegraphics[width=.4\textwidth]{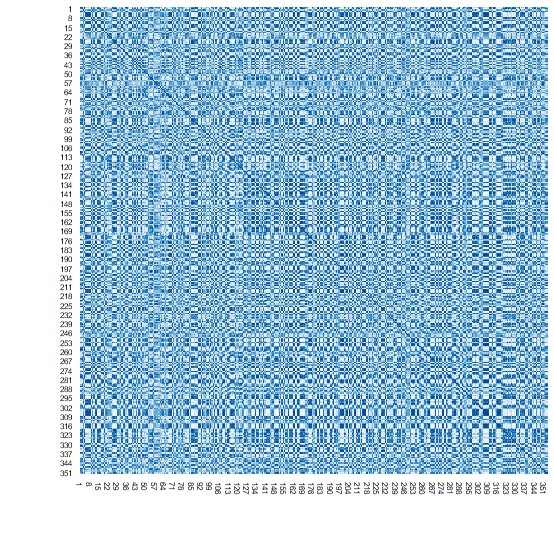}\hfill
    \includegraphics[width=.4\textwidth]{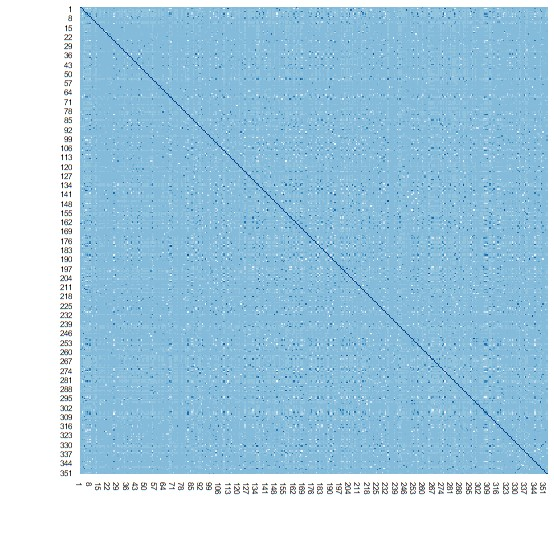}\hfill
     \\[\smallskipamount]
    \includegraphics[width=.4\textwidth]{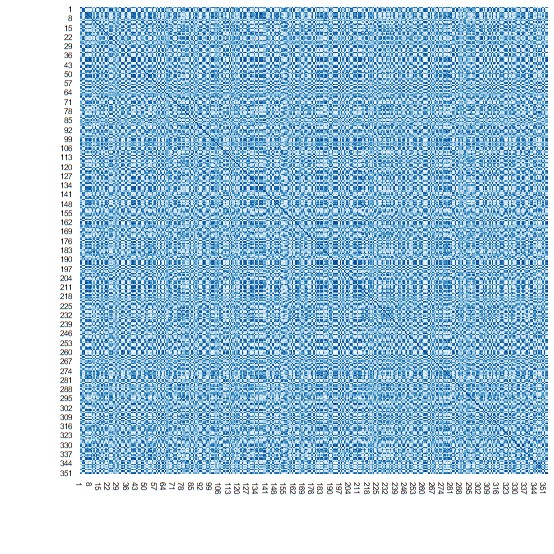}\hfill
    \includegraphics[width=.4\textwidth]{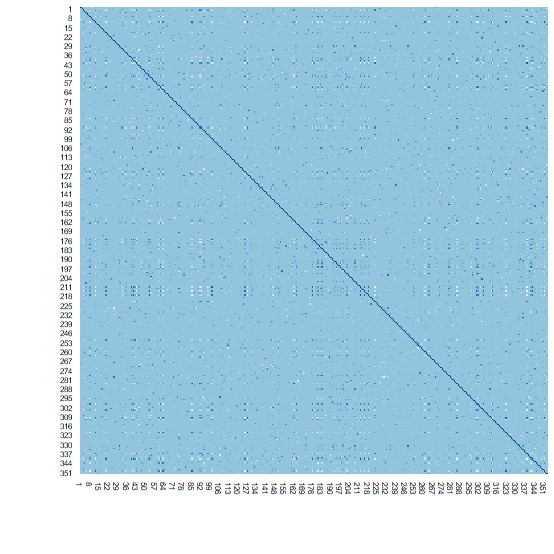}\hfill
    \caption{\emph{The upper panel is the heat maps of the estimated spot volatility matrices for the log prices (left) and idiosyncratic components (right) on 7 June, and the lower panel is the heat maps of the estimated spot volatility matrices for the log prices (left) and idiosyncratic components (right) on 19 May. The shrinkage technique is used to estimated the spot idiosyncratic volatility matrices.}}
    \label{fig4}
\end{figure}

Figure \ref{fig4} plots the heat maps for the estimated spot correlation matrices on two typical trading days: 19 May and 7 June. The heat maps in the upper panel are obtained for 7 June, which represents a typical period of stable volatility with nine estimated common factors. In contrast, those in the lower panel are for 19 May, which corresponds to the period during the crypto-currency market collapse \citep[e.g.,][]{P21} with only two estimated common factors. We note that the upper-left heat map (with nine estimated factors) is slightly denser than the lower-left one (with two estimated factors), which may be partly explained by the magnitude of the market disruption. During the period of the crypto-currency market collapse, its disruption is so large that it nearly dominates the entire stock market. Consequently, the two estimated factors can explain majority of the market volatility. In contrast, the nine estimated factors during the stable period may only explain a limited portion of the market volatility. Therefore, it is reasonable to expect that the spot idiosyncratic correlation matrix tends to be more sparse (after removing the latent factors) when the market collapses. In fact, by applying the shrinkage, the sparse structure of both the heat maps in the right panel is apparent, in which majority of the off-diagonal entries are either zero or very close to zero. This supports the low-rank plus sparse assumption on the spot volatility structure.

\smallskip

\begin{figure}[htbp]
    \centering
    \includegraphics[width=.45\textwidth]{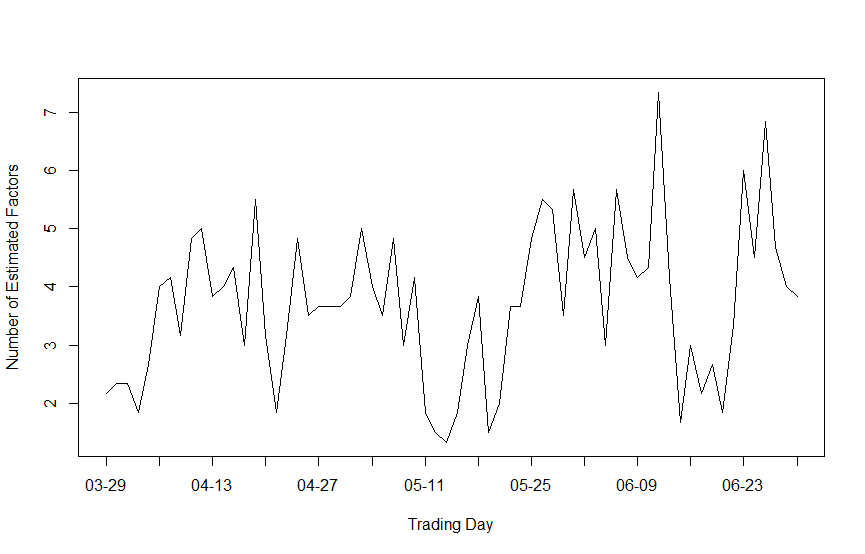}
    \includegraphics[width=.45\textwidth]{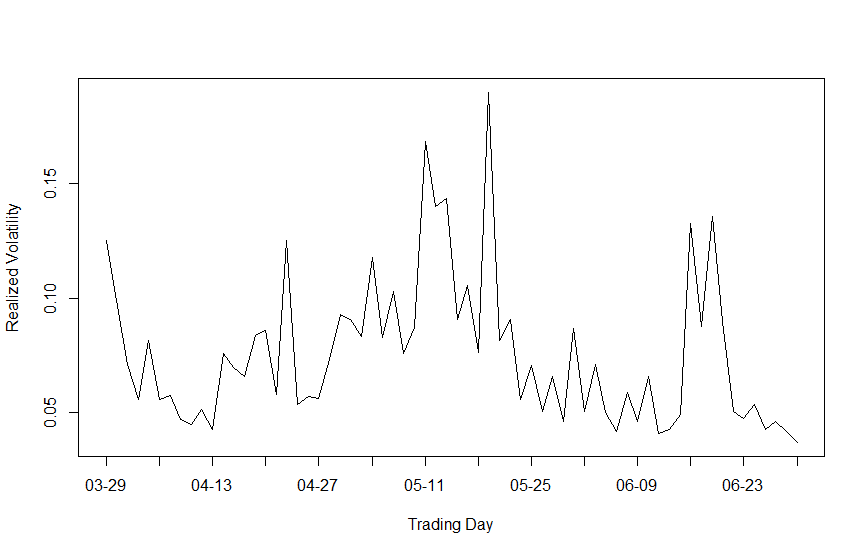}\hfill 
    \caption {\emph{The left panel is the plot of the (time-varying) estimated number of factors (averaged over each trading day), and the right panel is the plot of the realised volatility based on 1-min intraday stock price of S\&P 500 index over the sampling period.}} \label{fig5}
    
\end{figure}

We next further study the time-varying pattern of the factor number estimation and its connection to the market volatility. The left panel in Figure \ref{fig5} plots the averages of the estimated factor numbers over the six pre-determined time intervals in each trading day, and the right panel depicts the realised volatility of the S\&P 500 index intraday prices across trading days. It is well known that the S\&P 500 index is a market value-weighted index and is thus regarded as an indicator of the entire stock market in US. A virtual comparison between the two plots in Figure \ref{fig5} reveals that the averaged factor number tends to be low when the realised volatility of S\&P 500 index is high, whereas the estimated factor number is relatively high when the market realised volatility is low. This pattern is not uncommon in the financial market as the extreme fluctuation in the financial market is often closely correlated with some unexpected financial events or economic disruption. Consequently, a small number of factors tend to dominate the market during the period of market collapse. For example, in May 2021, the crypto-currency market collapsed for the first time, and this shock rapidly spread to the entire stock market. This adequately explains the peak of the realised volatility and the low estimated factor number in the midst of May which can be observed in Figure \ref{fig5}. In contrast, when the financial market is relatively stable, more factors tend to affect the market simultaneously, which is exactly what occurred at the end of May and the beginning of June in 2021.

\section{Conclusion}\label{sec6}

In this paper we propose a new nonparametric methodology and theory for estimating the low-rank plus sparse spot volatility structure of the noise-contaminated and asynchronous high-frequency data with large dimension. The microstructure noises are allowed to be auto-correlated, nonlinear heteroskedastic, asymptotic vanishing and endogenous, and the latent prices satisfy the time-varying continuous-time factor model. Imposing the sparsity restriction on the spot idiosyncratic volatility matrix, we combine the kernel-weighted POET (or local PCA) and pre-averaging techniques to develop the main estimation method. Under some regularity conditions, we derive the uniform convergence property for the estimated spot volatility matrix in various matrix norms and obtain some explicit convergence rates for different scenarios. The Monte-Carlo simulation study shows that the developed estimation method performs well in finite samples. In particular, the empirical application to the S\&P 500 index reveals time-varying patterns of the spot volatility matrix and latent factor number and confirms rationality of assuming the low-rank plus sparse structure.

\section*{Acknowledgement}

The first author's research was partly supported by the Leverhulme Research Fellowship (RF-2023-396) and BA/Leverhulme Small Research Grant (SRG1920/100603).

\section*{Supplementary material}

The supplement contains the proofs of the main theoretical results (in Appendix A) and some technical lemmas with proofs (in Appendix B).

\newpage

\begin{center}

{\Large Supplementary Material to ``Estimating Factor-Based Spot Volatility Matrices with Noisy and Asynchronous High-Frequency Data"}

\end{center}

\bigskip

\appendix

\section{Proofs of the main theoretical results}\label{app:A}
\renewcommand{\theequation}{A.\arabic{equation}}
\setcounter{equation}{0}

Unless explicitly stated differently, all the equations, propositions, remarks, theorems and lemmas mentioned in this appendix refer to those presented in the main text. In this appendix, we provide the detailed proofs of the main theoretical results stated in Section 3. As discussed in Remark 3.1, the local boundedness conditions on $\mu_{i,t}^F$, $\mu_{i,t}^U$, $\sigma_{i_1i_2,t}^U$, $\sigma_{U,i_1i_2}(t)$ and $\Lambda_{i}(t)$ in Assumption 1 can be strengthened to the uniform boundedness conditions over the entire time interval, i.e., (3.8)--(3.10) in main text hold. Throughout the proofs, we let $C$ be a generic positive constant whose value may change from line to line.

\smallskip

\subsection{\textbf{Proof of Proposition 3.1}}\ \ By model (2.1) and the definition of $\widetilde{X}_{i,t}$ in (2.7), we obtain the following decomposition:
\begin{equation}\label{eqA.1}
\widetilde{X}_{i,t_l}-X_{i,t_l}=\Pi_{i,l}(1)+\Pi_{i,l}(2)+\Pi_{i,l}(3)+\Pi_{i,l}(4),
\end{equation}
where 
\begin{eqnarray}
\Pi_{i,l}(1)&=&\sum_{j=1}^{n_i} \left(t_j^i-t_{j-1}^i\right)L_b(t_j^i-t_l)\varepsilon_{i,j},\notag\\
\Pi_{i,l}(2)&=&\sum_{j=1}^{n_i} L_b(t_j^i-t_l)\int_{t_{j-1}^i}^{t_j^i}(X_{i,t_j^i}-X_{i,s})ds,\notag\\
\Pi_{i,l}(3)&=&\sum_{j=1}^{n_i} \int_{t_{j-1}^i}^{t_j^i}\left[L_b(t_j^i-t_l)-L_b(s-t_l)\right]X_{i,s}ds,\notag\\
\Pi_{i,l}(4)&=&\int_{0}^{T_i}L_b(s-t_l)X_{i,s}ds-X_{i,t_l},\ \ T_i=t_{n_i}^i.\notag
\end{eqnarray}

We first use a truncation technique and the Bernstein-type inequality to show
\begin{equation}\label{eqA.2}
{\mathsf P}\left(\bigcup_{i=1}^p\left\{\max_{1\leq l\leq N}\left|  \Pi_{i,l}(1)\right|> \frac{c_\dagger}{4}\nu_1(p,b, n_i)\right\}\right)\rightarrow0,\ \ {\rm as}\ p,\underline{n}\rightarrow\infty,
\end{equation}
where $c_\dagger$ is a sufficiently large positive constant and 
\[\nu_1(p,b, n_i)=\left[\log (p\vee n_i)/(n_i^{2\beta_i+1} b)\right]^{1/2}.\] 
Define $\varepsilon_{i,j}^{\star}=\varepsilon_{i,j}^\ast {\sf I}(  |\varepsilon_{i,j}^\ast|\leq n_{i}^{\kappa_1})$ and $\varepsilon_{i,j}^{\diamond}=\varepsilon_{i,j}^\ast {\sf I}(  |\varepsilon_{i,j}^\ast|> n_{i}^{\kappa_1})$, where $\kappa_1$ is defined in Assumption 3(ii). Note that
\begin{eqnarray}
\Pi_{i,l}(1)&=&n_i^{-\beta_i}\sum_{j=1}^{n_i} L_b(t_j^i-t_l)\chi_{i}(t_j^i)\left(t_j^i-t_{j-1}^i\right)\left[\varepsilon_{i,j}^\star-{\mathsf E}(\varepsilon_{i,j}^\star)\right]+\nonumber\\
&&n_i^{-\beta_i}\sum_{j=1}^{n_i} L_b(t_j^i-t_l)\chi_{i}(t_j^i)\left(t_j^i-t_{j-1}^i\right)\left[\varepsilon_{i,j}^\diamond-{\mathsf E}(\varepsilon_{i,j}^\diamond)\right].\nonumber
\end{eqnarray}
By Assumption 2, we can prove that
\begin{equation}\label{eqA.3}
\max_{1\leq j\leq n_i}\mathsf{E}\left(  \left|\chi_{i}(t_j^i) \varepsilon_{i,j}^\diamond\right|\right) =O\left(n_i^{-C_{\circ}}\right)=o\left(\nu_1(p,b, n_i)\right),
\end{equation}
where $C_\circ$ is an arbitrarily large positive number. As
\begin{equation}\label{eqA.4}
\max_{1\leq i\leq p}\max_{1\leq l\leq N}\sum_{j=1}^{n_i} L_b(t_j^i-t_l)\left(t_j^i-t_{j-1}^i\right)=O(1),
\end{equation}
using (\ref{eqA.3}), (3.4) in Assumption 2(i), (3.5) in Assumption 3(ii) and the Bonferroni and Markov inequalities, we have

\begin{eqnarray}
&&\mathsf{P}\left(\bigcup_{i=1}^p \left\{
\max_{1\leq l\leq N}\left| \sum_{j=1}^{n_i} L_b(t_j^i-t_l)\chi_{i}(t_j^i)\left(t_j^i-t_{j-1}^i\right)\left[\varepsilon_{i,j}^\diamond-{\mathsf E}(\varepsilon_{i,j}^\diamond)\right] \right| >\frac{c_\dagger}{8}n_i^{\beta_i}\nu_1(p,b, n_i)\right\}\right) \nonumber\\ 
&&\leq\mathsf{P}\left( \bigcup_{i=1}^p\left\{\max_{1\leq l\leq N}\left| \sum_{j=1}^{n_i} L_b(t_j^i-t_l)\chi_{i}(t_j^i)\left(t_j^i-t_{j-1}^i\right)\varepsilon_{i,j}^\diamond \right| >\frac{c_\dagger}{9}\nu_\circ(p,b, n_i)\right\}\right) \nonumber\\
&&\leq\mathsf{P}\left(\max_{1\leq i\leq p}\max_{1\leq j\leq n_i}\left| \varepsilon_{i,j}^\diamond\right|  >0\right)\leq\mathsf{P}\left( \bigcup_{i=1}^p\left\{\max_{1\leq j\leq n_i} \left|\varepsilon_{i,j}^\ast\right|  >n_i^{\kappa_1}\right\}\right) \nonumber \\ 
&&\leq\sum_{i=1}^p\sum_{j=1}^{n_i} \mathsf{P}\left(  \left|\varepsilon_{i,j}^\ast\right|  >n_i^{\kappa_1}\right)=O\left(\sum_{i=1}^pn_i \exp\left\{-s_0n_i^{\kappa_1}\right\}\right)=o(1),\label{eqA.5}
\end{eqnarray}
where $s_{0}$ is defined in (3.4) and 
\[\nu_\circ(p,b, n_i)=\left[\frac{\log (p\vee n_i)}{n_i b}\right]^{1/2}.\]  On the other hand, by Assumptions 2 and 4(i) as well as the Bernstein inequality in Lemma \ref{le:B.1}, we may show that
\begin{eqnarray}
&&\mathsf{P}\left(\bigcup_{i=1}^p\left\{\max_{1\leq l\leq N}\left| \sum_{j=1}^{n_i} L_b(t_j^i-t_l)\chi_{i}(t_j^i)\left(t_j^i-t_{j-1}^i\right)\left[\varepsilon_{i,j}^\star-{\mathsf E}(\varepsilon_{i,j}^\star)\right] \right| >\frac{c_\dagger}{8}n_i^{\beta_i}\nu_1(p,b, n_i)\right\}\right) \nonumber\\
&&\leq  \ \sum_{i=1}^{p}\sum_{l=1}^{N}\mathsf{P}\left( \left| \sum_{j=1}^{n_i} L_b(t_j^i-t_l)\chi_{i}(t_j^i)\left(t_j^i-t_{j-1}^i\right)\left[\varepsilon_{i,j}^\star-{\mathsf E}(\varepsilon_{i,j}^\star)\right] \right| >\frac{c_\dagger}{8}\nu_\circ(p,b, n_i)\right) \nonumber\\
&&=O\left(  N\sum_{i=1}^p \exp\left\{ -C_1c_\dagger^2\log(p\vee n_i)\right\} + N\sum_{i=1}^p\phi_i\exp\left\{\frac{\log \gamma_0}{n_i^{\kappa_1}\cdot\nu_\circ(p,b, n_i)}\right\}\right),\label{eqA.6}
\end{eqnarray}
where $C_1$ is a positive and bounded constant, $\gamma_0\in(0,1)$ is defined in Assumption 2(i) and $\phi_i=(n_i/b)^{3/4}n_i^{3\kappa_1/2}\left[\log (p\vee n_i)\right]^{1/4}$. By Assumption 3(i), letting $c_\dagger>0$ be sufficiently large in (\ref{eqA.6}), we have
\begin{equation}\label{eqA.7}
N\sum_{i=1}^p \exp\left\{ -C_1c_\dagger^2\log(p\vee n_i)\right\}=o(1).
\end{equation}
Since $\left[\nu_\circ(p,b, n_i)n_i^{2\kappa_1}\right]^{-1}\rightarrow\infty$ for any $i$ by the first condition in (3.5), and $N\phi_i$ diverges to infinity at a polynomial rate of $n_i$, we may show that
\begin{equation}\label{eqA.8}
N\sum_{i=1}^p\phi_i\exp\left\{\frac{\log \gamma_0}{n_i^{\kappa_1}\nu_\circ(p,b, n_i)}\right\}=o\left(\sum_{i=1}^pn_i \exp\left\{-s_0n_i^{\kappa_1}\right\}\right)=o(1).
\end{equation}
By (\ref{eqA.6})--(\ref{eqA.8}), we have
\begin{equation}\label{eqA.9}
\mathsf{P}\left(\bigcup_{i=1}^p\left\{\max_{1\leq l\leq N}\left| \sum_{j=1}^{n_i} L_b(t_j^i-t_l)\chi_{i}(t_j^i)\left(t_j^i-t_{j-1}^i\right)\left[\varepsilon_{i,j}^\star-{\mathsf E}(\varepsilon_{i,j}^\star)\right] \right| >\frac{c_\dagger}{8}n_i^{\beta_i}\nu_1(p,b, n_i)\right\}\right) =o(1), 
\end{equation}
as $p$ and $\underline {n}$ jointly diverge to infinity. By (\ref{eqA.5}) and (\ref{eqA.9}), we prove (\ref{eqA.2}).

Letting 
\[\nu_2(p,b,n_i)=\sqrt{b\log(p\vee n_i)},\] 
we next show that
\begin{equation}\label{eqA.10}
\mathsf{P}\left(\bigcup_{i=1}^p\left\{\max_{1\leq l\leq n}\left|  \Pi_{i,l}(2)\right|> \frac{c_\dagger}{4}\nu_2(p,b,n_i)\right\}\right)\rightarrow0,\ \ {\rm as}\ p,\underline{n}\rightarrow\infty.
\end{equation}
By (1.2), (2.3) and (2.4), we decompose $\Pi_{i,l}(2)$ as 
\begin{eqnarray}
\Pi_{i,l}(2)&=&\sum_{j=1}^{n_i} L_b(t_k-t_l)\int_{t_{j-1}^i}^{t_j^i}\left[\int_{s}^{t_j^i}\left(\Lambda_{i}(t)^{^\intercal}\mu_{t}^F+\mu_{i,t}^U\right)dt\right]ds+ \nonumber\\
&&\sum_{j=1}^{n_i} L_b(t_k-t_l)\int_{t_{j-1}^i}^{t_j^i}\left[\int_{s}^{t_j^i} \left(\Lambda_{i}(t)^{^\intercal}d W_t^F+\left(\sigma_{i\bullet,t}^U\right)^{^\intercal}d W_t^U\right)\right]ds\nonumber\\
&=:&\Pi_{i,l}(2,1)+\Pi_{i,l}(2,2),\nonumber
\end{eqnarray}
where $\sigma_{i\bullet,t}^U=\left(\sigma_{i1,t}^U,\cdots,\sigma_{ip,t}^U\right)^{^\intercal}$. Let ${\mathcal B}$ denote the event that (3.8)--(3.10) hold, and ${\mathcal B}^c$ be the complement of ${\mathcal B}$. It follows from Assumption 4(i) that 
\[\left\{\max_{1\leq l\leq n}\left|  \Pi_{i,l}(2,1)\right|> \frac{c_\dagger}{8}\nu_2(p,b,n_i)\right\}\] 
is empty for all $i=1,\cdots,p$, conditional on ${\mathcal B}$. As $\mathsf{P}({\mathcal B}^c)\rightarrow0$, we readily have that
\begin{equation}\label{eqA.11}
\mathsf{P}\left(\bigcup_{i=1}^p\left\{\max_{1\leq l\leq N}\left|  \Pi_{i,l}(2,1)\right|> \frac{c_\dagger}{8}\nu_2(p,b,n_i)\right\}\right)\rightarrow0.
\end{equation}
By the Bonferroni inequality, we may show that, for any $\epsilon>0$,
\begin{eqnarray}
&&\mathsf{P}\left(\bigcup_{i=1}^p\left\{\max_{1\leq j\leq n_i}\sup_{t_{j-1}^i\leq s\leq t_j^i}\left\vert\int_{s}^{t_j^i} \left(\Lambda_{i}(t)^{^\intercal}d W_t^F+\left(\sigma_{i\bullet,t}^U\right)^{^\intercal}d W_t^U\right)\right\vert>\epsilon\nu_2(p,b,n_i)\right\}\right)\nonumber\\ 
&&\leq\sum_{i=1}^p\sum_{j=1}^{n_i}\mathsf{P}\left(\sup_{t_{j-1}^i\leq s\leq t_j^i}\left\vert\int_{s}^{t_j^i} \left(\Lambda_{i}(t)^{^\intercal}d W_t^F+\left(\sigma_{i\bullet,t}^U\right)^{^\intercal}d W_t^U\right)\right\vert>\epsilon\nu_2(p,b,n_i)\right)\nonumber\\ 
&&\leq\sum_{i=1}^p\sum_{j=1}^{n_i}\mathsf{P}\left(\sup_{t_{j-1}^i\leq s\leq t_j^i}\left\vert\int_{t_{j-1}^i}^{s} \left(\Lambda_{i}(t)^{^\intercal}d W_t^F+\left(\sigma_{i\bullet,t}^U\right)^{^\intercal}d W_t^U\right)\right\vert>\frac{\epsilon}{2}\nu_2(p,b,n_i)\right) .\label{eqA.12}
\end{eqnarray}
Note that 
\[\left\{\left\vert\int_{t_{j-1}^i}^{s} \left(\Lambda_{i}(t)^{^\intercal}d W_t^F+\left(\sigma_{i\bullet,t}^U\right)^{^\intercal}d W_t^U\right)\right\vert: s\geq t_{j-1}^i\right\}\]
and 
\[
\left\{\exp\left(\psi\left\vert\int_{t_{j-1}^i}^{s} \left(\Lambda_{i}(t)^{^\intercal}d W_t^F+\left(\sigma_{i\bullet,t}^U\right)^{^\intercal}d W_t^U\right)\right\vert\right): s\geq t_{j-1}^i\right\}\ \ {\rm with}\ \psi>0\] 
are sub-martingales, which can be verified via the conditional Jensen inequality. Using the moment generating function for the folded normal random variable and Assumptions 1(ii)(iii) and 3(i), we can prove that
\begin{equation}\label{eqA.13}
\mathsf{E}\left[\exp\left(\psi\left\vert\int_{t_{j-1}^i}^{t_j^i} \left(\Lambda_{i}(t)^{^\intercal}d W_t^F+\left(\sigma_{i\bullet,t}^U\right)^{^\intercal}d W_t^U\right)\right\vert\right)\right]\leq \exp\left\{\frac{\psi^2c_{j}^i C_2}{2n_i}\right\},
\end{equation}
where 
\[C_2=\max_{1\leq i\leq p}\sup_{0\leq t\leq T_i}\left[\Lambda_{i}(t)^{^\intercal}\Lambda_{i}(t)+ \sigma_{U,ii}(t)\right]<\infty.\]
By (\ref{eqA.13}) and Doob's inequality for sub-martingales, we have
\begin{eqnarray}
&&\mathsf{P}\left(\sup_{t_{j-1}^i\leq s\leq t_j^i}\left\vert\int_{t_{j-1}^i}^{s} \left(\Lambda_{i}(t)^{^\intercal}d W_t^F+\left(\sigma_{i\bullet,t}^U\right)^{^\intercal}d W_t^U\right)\right\vert>\frac{\epsilon}{2}\nu_2(p,b,n_i)\right)\nonumber\\
&&=\mathsf{P}\left(\sup_{t_{j-1}^i\leq s\leq t_j^i}\exp\left\{\psi\left\vert\int_{t_{j-1}^i}^{s} \left(\Lambda_{i}(t)^{^\intercal}d W_t^F+\left(\sigma_{i\bullet,t}^U\right)^{^\intercal}d W_t^U\right)\right\vert\right\}>\exp\left\{\frac{\psi\epsilon\nu_2(p,b,n_i)}{2}\right\}\right)\nonumber\\
&&\leq\exp\left(-\frac{\psi\epsilon\nu_2(p,b,n_i)}{2}\right)\mathsf {E}\left[\exp\left(\psi\left\vert\int_{t_{j-1}^i}^{t_j^i} \left(\Lambda_{i}(t)^{^\intercal}d W_t^F+\left(\sigma_{i\bullet,t}^U\right)^{^\intercal}d W_t^U\right)\right\vert\right)\right]\nonumber\\ 
&&\leq\exp\left\{\frac{\psi^2c_{j}^i C_2}{2n_i}-\frac{\psi\epsilon\nu_2(p,b,n_i)}{2}\right\}.\label{eqA.14}
\end{eqnarray}
Then, choosing $\psi=\psi_j^i=\epsilon\nu_2(p,b,n_i)n_i/(2c_j^i C_2)$ in (\ref{eqA.14}), by (\ref{eqA.12}), (\ref{eqA.14}) and Assumption 3, we have
\begin{eqnarray}
&&\mathsf{P}\left(\bigcup_{i=1}^p\left\{\max_{1\leq j\leq n_i}\sup_{t_{j-1}^i\leq s\leq t_j^i}\left\vert\int_{s}^{t_j^i} \left(\Lambda_{i}(t)^{^\intercal}d W_t^F+\left(\sigma_{i\bullet,t}^U\right)^{^\intercal}d W_t^U\right)\right\vert>\epsilon\nu_2(p,b,n_i)\right\}\right)\nonumber\\ 
&&\leq\sum_{i=1}^pn_i\exp\left\{-\frac{\epsilon^2}{8 \overline{c}C_2} (n_ib)\log(p\vee n_i)\right\}=o(1),\label{eqA.15}
\end{eqnarray}
where $\overline{c}$ is defined in Assumption 3(i). This, together with (\ref{eqA.4}), indicates that
\begin{equation}\label{eqA.16}
\mathsf{P}\left(\bigcup_{i=1}^p\left\{\max_{1\leq l\leq N}\left|  \Pi_{i,l}(2,2)\right|> \frac{c_\dagger}{8}\nu_2(p,b,n_i)\right\}\right)\rightarrow0,\ \ {\rm as}\ p,\underline{n}\rightarrow\infty.
\end{equation}
Combining (\ref{eqA.11}) and (\ref{eqA.16}), we prove (\ref{eqA.10}).

We next prove that
\begin{equation}\label{eqA.17}
\mathsf{P}\left(\bigcup_{i=1}^p\left\{\max_{1\leq l\leq N}\left|  \Pi_{i,l}(3)\right|> \frac{c_\dagger}{4}\nu_3(p,b,n_i)\right\}\right)\rightarrow0,\ \ {\rm as}\ p,\underline{n}\rightarrow\infty,
\end{equation}
where 
\[\nu_3(p,b,n_i)=(n_ib)^{-1}\sqrt{\log(p\vee n_i)}.\] Note that 
\[
\vert \Pi_{i,l}(3)\vert\leq\sup_{0\leq t\leq T_i}|X_{i,t}|\cdot \sum_{j=1}^{n_i} \int_{t_{j-1}^i}^{t_j^i}\left\vert L_b(t_j^i-t_l)-L_b(s-t_l)\right\vert ds,
\]
where $T_i=t_{n_i}^i$ is fixed. By the conditions on the kernel function $L(\cdot)$ in Assumption 4(i), we have 
\begin{equation}\label{eqA.18}
\max_{1\leq l\leq n}\sum_{j=1}^{n_i} \int_{t_{j-1}^i}^{t_j^i}\left\vert L_b(t_j^i-t_l)-L_b(s-t_l)\right\vert ds=O\left((n_i b)^{-1}\right).
\end{equation}
On the other hand, by (3.8) and (3.10),
\[
\sup_{0\leq t\leq T_i}|X_{i,t}|\leq\sup_{0\leq t\leq T_i} \left\vert\int_{0}^t \left[\Lambda_{i}(s)^{^\intercal}d W_s^F+\left(\sigma_{i\bullet,s}^U\right)^{^\intercal}d W_s^U\right]\right\vert+\left(C_\mu+k^{1/2}C_\mu C_\Lambda\right)T.
\]
Following the proof of (\ref{eqA.15}), we may show that, for $c_\ddagger>0$ large enough,
{\small\[
{\mathsf{P}\left(\bigcup_{i=1}^p\left\{\sup_{0\leq t\leq T_i} \left\vert\int_{0}^t \left[\Lambda_{i}(s)^{^\intercal}d W_s^F+\left(\sigma_{i\bullet,s}^U\right)^{^\intercal}d W_s^U\right]\right\vert>c_\ddagger\sqrt{\log (p\vee n_i)}\right\}\right)\rightarrow0,\ \ {\rm as}\ p,\underline{n}\rightarrow\infty,}
\]}
indicating that
\begin{equation}\label{eqA.19}
\mathsf{P}\left(\bigcup_{i=1}^p\left\{\sup_{0\leq t\leq T_i}|X_{i,t}|>2c_\ddagger\sqrt{\log (p\vee n_i)}\right\}\right)\rightarrow0,\ \ {\rm as}\ p,\underline{n}\rightarrow\infty.
\end{equation}
By virtue of (\ref{eqA.18}) and (\ref{eqA.19}), we complete the proof of (\ref{eqA.17}) when $c_\dagger$ is sufficiently large.

Finally, we consider $\Pi_{i,l}(4)$. Note that
\begin{eqnarray}
\Pi_{i,l}(4)&=&\left\{\int_{0}^{T_i}L_b(s-t_l)\left[\int_0^s \left(\Lambda_{i}(t)^{^\intercal}\mu_{t}^F+\mu_{i,t}^U\right) dt\right] ds-\int_0^{t_l}\left(\Lambda_{i}(t)^{^\intercal}\mu_{i,t}^F+\mu_{i,t}^U\right) dt\right\}+\nonumber\\
&&\left\{\int_{0}^{T_i}L_b(s-t_l)\left[\int_0^s \left(\Lambda_{i}(t)^{^\intercal}d W_t^F+\left(\sigma_{i\bullet,t}^U\right)^{^\intercal}d W_t^U\right)\right]ds-\right.\nonumber\\
&&\left.\int_0^{t_l}\left(\Lambda_{i}(t)^{^\intercal}d W_t^F+\left(\sigma_{i\bullet,t}^U\right)^{^\intercal}d W_t^U\right)\right\}\nonumber\\
&=:&\Pi_{i,l}(4,1)+\Pi_{i,l}(4,2).\nonumber
\end{eqnarray}
With Assumption 4(i), (3.8) and (3.10), we have 
\begin{equation}\label{eqA.20}
\max_{1\leq i\leq p}\max_{1\leq l\leq N}\left|  \Pi_{i,l}(4,1)\right|=O_P\left(b\right).
\end{equation}
Following the proof of (\ref{eqA.15}), we may show that 
\[\mathsf{P}\left(\bigcup_{i=1}^p\left\{\max_{1\leq l\leq N}\sup_{t_l\leq s\leq t_l+b}\left\vert\int_{t_l}^{s}\left(\Lambda_{i}(t)^{^\intercal}d W_t^F+\left(\sigma_{i\bullet,t}^U\right)^{^\intercal}d W_t^U\right)\right\vert>\frac{c_\dagger}{6} \nu_2(p,b,n_i)\right\}\right)\rightarrow0
\]
and
\[\mathsf{P}\left(\bigcup_{i=1}^p\left\{\max_{1\leq l\leq N}\sup_{t_l-b\leq s\leq t_l}\left\vert\int_{s}^{t_l}\left(\Lambda_{i}(t)^{^\intercal}d W_t^F+\left(\sigma_{i\bullet,t}^U\right)^{^\intercal}d W_t^U\right)\right\vert>\frac{c_\dagger}{6}\nu_2(p,b,n_i)\right\}\right)\rightarrow0
\]
as $p,\underline n\rightarrow\infty$ and $c_\dagger>0$ is large enough. Consequently, we have
\begin{equation}\label{eqA.21}
\mathsf{P}\left(\bigcup_{i=1}^p\left\{\max_{1\leq l\leq N}\left|  \Pi_{i,l}(4,2)\right|> \frac{c_\dagger}{3}\nu_2(p,b,n_i)\right\}\right)\rightarrow0.
\end{equation}
By virtue of (\ref{eqA.20}) and (\ref{eqA.21}), noting that $b/\nu_2(p,b,n_i)\rightarrow0$, we can prove that
\begin{equation}\label{eqA.22}
\mathsf{P}\left(\bigcup_{i=1}^p\left\{\max_{1\leq l\leq N}\left|  \Pi_{i,l}(4)\right|> \frac{c_\dagger}{2}\nu_2(p,b,n_i)\right\}\right)\rightarrow0,\ \ {\rm as}\ p,\underline n\rightarrow\infty.
\end{equation}
The proof of Proposition 3.1 is completed with (\ref{eqA.2}), (\ref{eqA.10}), (\ref{eqA.17}) and (\ref{eqA.22}).\hfill$\blacksquare$

\medskip
\subsection{\textbf{Proof of Theorem 3.1}}\ \ By the property of the shrinkage function, we have
\begin{eqnarray}
&&\sup_{h\leq \tau\leq T-h}\left\Vert \widehat\Sigma_U(\tau)-\Sigma_U(\tau)\right\Vert_{\max}\nonumber\\
&&=\max_{1\leq i_1\leq p}\max_{1\leq i_2\leq p}\sup_{h\leq\tau\leq T-h} \left\vert \widehat{\sigma}_{U,i_1i_2}(\tau)-\sigma_{U,i_1i_2}(\tau)\right\vert\notag\\
&&\leq\max_{1\leq i_1\leq p}\max_{1\leq i_2\leq p}\sup_{h\leq\tau\leq T-h} \left\vert \check{\sigma}_{U,i_1i_2}(\tau)-\sigma_{U,i_1i_2}(\tau)\right\vert+\sup_{0\leq\tau\leq T}\rho(\tau).\notag
\end{eqnarray}
By Assumption 4(iii), in order to prove (3.13), it is sufficient to show that 
\begin{equation}\label{eqA.23}
\max_{1\leq i_1\leq p}\max_{1\leq i_2\leq p}\sup_{h\leq\tau\leq T-h} \left\vert \check{\sigma}_{U,i_1i_2}(\tau)-\sigma_{U,i_1i_2}(\tau)\right\vert=O_P\left(\zeta(p,b,h,{\mathbf n},N)\right).
\end{equation}
Write $U_j(\tau)=\left[U_{1,j}(\tau),\cdots,U_{p,j}(\tau)\right]^{^\intercal}=\Delta U_j K_h^{1/2}(t_j-\tau)$. Note that 
{\small \begin{eqnarray}
\check{\sigma}_{U,i_1i_2}(\tau)&=&\sum_{j=1}^N\widetilde U_{i_1,j}(\tau)\widetilde U_{i_2,j}(\tau)\nonumber\\
&=&\sum_{j=1}^N U_{i_1,j}(\tau) U_{i_2,j}(\tau)+\sum_{j=1}^N\left[\widetilde U_{i_1,j}(\tau)-U_{i_1,j}(\tau)\right]U_{i_2,j}(\tau)+\notag\\ 
&& \sum_{j=1}^N U_{i_1,j}(\tau) \left[\widetilde U_{i_2,j}(\tau)-U_{i_2,j}(\tau)\right]+ \sum_{j=1}^N\left[\widetilde U_{i_1,j}(\tau)-U_{i_1,j}(\tau)\right]\left[\widetilde U_{i_2,j}(\tau)-U_{i_2,j}(\tau)\right].\label{eqA.24}
\end{eqnarray}}
Following the proof of Proposition A.1 in \cite{BLLW23}, we may show that 
\begin{equation}\label{eqA.25}
\max_{1\leq i_1\leq p}\max_{1\leq i_2\leq p}\sup_{h\leq\tau\leq T-h} \left\vert \sum_{j=1}^N U_{i_1,j}(\tau) U_{i_2,j}(\tau)-\sigma_{U,i_1i_2}(\tau)\right\vert=O_P\left(\left(\frac{\log(p\vee N)}{Nh}\right)^{1/2}+h^{\delta}\right).
\end{equation}
By (3.2) in Assumption 1(iii), (3.12) in Remark 3.2(i) and Lemmas \ref{le:B.2}--\ref{le:B.4} in Appendix B, we have
\begin{equation}\label{eqA.26}
\max_{1\leq i\leq p}\sup_{h\leq \tau\leq T-h}\sum_{j=1}^N \left[\widetilde U_{ij}(\tau)-U_{ij}(\tau)\right]^2=O_P\left(\left[\zeta(p,b,h,{\mathbf n},N)\right]^2\right).
\end{equation}
Combining (\ref{eqA.25}), (\ref{eqA.26}) and the Cauchy-Schwarz inequality, we can prove that 
\begin{eqnarray}
&&\max_{1\leq i_1\leq p}\max_{1\leq i_2\leq p}\sup_{h\leq\tau\leq T-h} \left\vert \sum_{j=1}^N\left[\widetilde U_{i_1,j}(\tau)-U_{i_1,j}(\tau)\right]U_{i_2,j}(\tau)\right\vert=O_P\left(\zeta(p,b,h,{\mathbf n},N)\right), \label{eqA.27}\\ 
&&\max_{1\leq i_1\leq p}\max_{1\leq i_2\leq p}\sup_{h\leq\tau\leq T-h} \left\vert \sum_{j=1}^N U_{i_1,j}(\tau) \left[\widetilde U_{i_2,j}(\tau)-U_{i_2,j}(\tau)\right]\right\vert=O_P\left(\zeta(p,b,h,{\mathbf n},N)\right), \label{eqA.28}\\
&&\max_{1\leq i_1\leq p}\max_{1\leq i_2\leq p}\sup_{h\leq\tau\leq T-h} \left\vert \sum_{j=1}^N\left[\widetilde U_{i_1,j}(\tau)-U_{i_1,j}(\tau)\right]\left[\widetilde U_{i_2,j}(\tau)-U_{i_2,j}(\tau)\right]\right\vert\notag\\ 
&&=O_P\left(\left[\zeta(p,b,h,{\mathbf n},N)\right]^2\right)=o_P\left(\zeta(p,b,h,{\mathbf n},N)\right).\label{eqA.29}
\end{eqnarray}
By virtue of (\ref{eqA.25}) and (\ref{eqA.27})--(\ref{eqA.29}), we complete the proof of (\ref{eqA.23}).

We next turn to the proof of (3.14). By the definition of shrinkage, we may show that
\begin{eqnarray}
& &\sup_{h\leq \tau\leq T-h}\left\Vert \widehat\Sigma_U(\tau)-\Sigma_U(\tau)\right\Vert_s  \notag \\
&&\le\sup_{h\leq \tau\leq T-h} \max_{1\le i_1\le p}\sum_{i_2=1}^{p} \left\vert \widehat{\sigma}_{U,i_1i_2}(\tau)-\sigma_{U,i_1i_2}(\tau)\right\vert \notag\\
&&=\sup_{h\leq \tau\leq T-h} \max_{1\le i_1\le p}\sum_{i_2=1}^{p}\left|s_{\rho(\tau)}\left(\check{\sigma}_{U,i_1i_2}(\tau)\right)I\left(\left\vert \check{\sigma}_{U,i_1i_2}(\tau)\right\vert>\rho(\tau)\right)-\sigma_{U,i_1i_2}(\tau)\right|  \notag \\
&&\le\sup_{h\leq \tau\leq T-h} \max_{1\le i_1\le p}\sum_{i_2=1}^{p}\left|s_{\rho(\tau)}\left(\check{\sigma}_{U,i_1i_2}(\tau)\right)-\check{\sigma}_{U,i_1i_2}(\tau)\right|I\left(\left\vert \check{\sigma}_{U,i_1i_2}(\tau)\right\vert>\rho(\tau)\right)+  \notag
\\
&&\ \ \ \sup_{h\leq \tau\leq T-h} \max_{1\le i_1\le p}\sum_{i_2=1}^{p}\left| \check{\sigma}_{U,i_1i_2}(\tau)-\sigma_{U,i_1i_2}(\tau)\right|I\left(\left\vert \check{\sigma}_{U,i_1i_2}(\tau)\right\vert>\rho(\tau)\right)+
\notag \\
&&\ \ \ \sup_{h\leq \tau\leq T-h} \max_{1\le i_1\le p}\sum_{i_2=1}^{p}\left|\sigma_{U,i_1i_2}(\tau)\right|
I\left(\left\vert \check\sigma_{U,i_1i_2}(\tau)\right\vert\le\rho(\tau)\right)  \notag
\\ 
&& =:\Xi_1+\Xi_2+\Xi_3.  \label{eqA.30}
\end{eqnarray}

Define the event 
\[
\mathcal{M}(c)=\left\{\max_{1\leq i_1\leq p}\max_{1\leq i_2\leq p}\sup_{h\leq\tau\leq T-h} \left\vert \check{\sigma}_{U,i_1i_2}(\tau)-\sigma_{U,i_1i_2}(\tau)\right\vert\le c\cdot\zeta(p,b,h,{\mathbf n},N)\right\},
\]
where $c$ is a positive constant. For any small $\epsilon>0$, by (\ref{eqA.23}), we may find $c_\epsilon>0$ such that 
\begin{equation}  \label{eqA.31}
\mathcal{P}\left({\mathcal M}(c_\epsilon)\right)\geq 1-\epsilon.
\end{equation}
As $\left\{\Sigma_U(t):\ 0\leq t\leq T\right\}\in \mathcal{S}(q,\varpi_p)$, by Assumption 4(iii) with $\underline{C}_M\geq2c_\epsilon$ and property (iii) of the shrinkage function, we may show that
\begin{eqnarray}
\Xi_1+\Xi_2 &\le&(\overline{C}_M+c_\epsilon)\zeta(p,b,h,{\mathbf n},N)\left[
 \sup_{h\leq \tau\leq T-h} \max_{1\le i_1\le p}\sum_{i_2=1}^{p}I\left(\left\vert \check{\sigma}_{U,i_1i_2}(\tau)\right\vert>\rho(\tau)\right)\right]  \notag \\
&\le& (\overline{C}_M+c_\epsilon)\zeta(p,b,h,{\mathbf n},N)\left[
 \sup_{h\leq \tau\leq T-h} \max_{1\le i_1\le p}\sum_{i_2=1}^{p}I\left(\left\vert \check{\sigma}_{U,i_1i_2}(\tau)\right\vert>\underline C_M \zeta(p,b,h,{\mathbf n},N)\right)\right] \notag \\
&\le& (\overline{C}_M+c_\epsilon)\zeta(p,b,h,{\mathbf n},N)\left[
 \sup_{h\leq \tau\leq T-h} \max_{1\le i_1\le p}\sum_{i_2=1}^{p}I\left(\left\vert \sigma_{U,i_1i_2}(\tau)\right\vert>c_\epsilon \zeta(p,b,h,{\mathbf n},N)\right)\right] \notag \\
&\leq& C\cdot \zeta(p,b,h,{\mathbf n},N)\left[ \sup_{h\leq \tau\leq T-h} \max_{1\le i_1\le p}\sum_{i_2=1}^{p}\frac{|\sigma_{U,i_1i_2}(\tau)|^q}{\left(c_\epsilon\zeta(p,b,h,{\mathbf n},N)\right)^q}\right]  \notag \\
&=&O_P\left(\Psi\varpi_p\zeta(p,b,h,{\mathbf n},N)^{1-q}\right)=O_P\left(\varpi_p\zeta(p,b,h,{\mathbf n},N)^{1-q}\right)  \label{eqA.32}
\end{eqnarray}
on the event ${\mathcal M}(c_\epsilon)$.

On the other hand, we note that the events $\{\vert \check\sigma_{U,i_1i_2}(\tau)\vert\le \rho(\tau)\}$ and ${\mathcal M}(c_\epsilon)$ jointly imply that $\{|\sigma_{U,i_1i_2}(\tau)|\le (\overline{C}_M+c_\epsilon)\zeta(p,b,h,{\mathbf n},N)\}$ by the triangle inequality. Then,
\begin{eqnarray}
\Xi_3 &\le&\sup_{h\leq \tau\leq T-h} \max_{1\le i_1\le p}\sum_{i_2=1}^{p}\left|\sigma_{U,i_1i_2}(\tau)\right|
I\left(\left\vert \sigma_{U,i_1i_2}(\tau)\right\vert\le \left(\overline{C}
_M+c_\epsilon\right)\zeta(p,b,h,{\mathbf n},N)\right) \notag \\
&\le& \left(\overline{C}_M+c_\epsilon\right)^{1-q}\zeta(p,b,h,{\mathbf n},N)^{1-q}\sup_{h\leq \tau\leq T-h} \max_{1\le i_1\le p}\sum_{i_2=1}^{p}|\sigma_{U,i_1i_2}(\tau)|^q 
\notag \\
&=&O_P\left(\Psi\varpi_p\zeta(p,b,h,{\mathbf n},N)^{1-q}\right)\notag\\
&=&O_P\left(\varpi_p\zeta(p,b,h,{\mathbf n},N)^{1-q}\right).  \label{eqA.33}
\end{eqnarray}
By (\ref{eqA.32}) and (\ref{eqA.33}), letting $\epsilon\rightarrow0$ in (\ref{eqA.31}), we complete the proof of (3.14). \hfill$\blacksquare$

\medskip

\subsection{\textbf{Proof of Theorem 3.2}}\ \ Define
{\small \[ \Sigma_{X}^\ast(\tau)=\Lambda(\tau)\Sigma_{F}^\ast(\tau)\Lambda(\tau)^{^\intercal}+\Sigma_U(\tau)\ \ {\rm with}\ \ \Sigma_{F}^\ast(\tau)=\Delta F(\tau)^{^\intercal}\Delta F(\tau)=\sum_{t=1}^{N}\Delta F_t(\tau)\Delta F_t(\tau)^{^\intercal}.
\]} 
Observe that
\begin{eqnarray}
&&\sup_{h\leq \tau\leq T-h}\left\Vert \widehat\Sigma_X(\tau)- \Sigma_X(\tau)\right\Vert_{\max}\notag\\
&&\le
\sup_{h\leq \tau\leq T-h}\left\Vert \widehat\Sigma_X(\tau)- \Sigma_X^\ast(\tau)\right\Vert_{\max}+
\sup_{h\leq \tau\leq T-h}\left\Vert  \Sigma_X^\ast(\tau)- \Sigma_X(\tau)\right\Vert_{\max},\notag
\end{eqnarray}
and
\begin{eqnarray}
&&\sup_{h\leq \tau\leq T-h}\left\Vert \widehat\Sigma_X(\tau)- \Sigma_X^\ast(\tau)\right\Vert_{\max}\notag\\
&&\leq\sup_{h\leq \tau\leq T-h}\left[\left\Vert \widehat\Sigma_U(\tau)- \Sigma_U(\tau)\right\Vert_{\max}+\left\Vert \widetilde\Lambda(\tau)\widetilde\Lambda(\tau)^{^\intercal}-\Lambda(\tau)\Sigma_{F}^\ast(\tau)\Lambda(\tau)^{^\intercal}\right\Vert_{\max}\right].\notag
\end{eqnarray}
Let $\widetilde\Sigma_{F}(\tau)=\sum_{t=1}^{N}\Delta\widetilde F_t(\tau)\Delta \widetilde F_t(\tau)^{^\intercal}$. By Lemmas \ref{le:B.2} and \ref{le:B.4} in Appendix B, we have
\begin{eqnarray} 
& &\sup_{h\leq \tau\leq T-h}\left\Vert \widetilde\Lambda(\tau)\widetilde\Lambda(\tau)^{^\intercal}-\Lambda(\tau)\Sigma_{F}^\ast(\tau)\Lambda(\tau)^{^\intercal}\right\Vert_{\max}\notag\\
&&=\sup_{h\leq \tau\leq T-h}\left\Vert \widetilde\Lambda(\tau)\widetilde\Sigma_{F}(\tau)\widetilde\Lambda(\tau)^{^\intercal}-\Lambda(\tau)\Sigma_{F}^\ast(\tau)\Lambda(\tau)^{^\intercal}\right\Vert_{\max}\notag\\
&&=O_P\left(\zeta(p,b,h,{\mathbf n},N)\right),\notag
\end{eqnarray}
which, together with (3.13) in Theorem 3.1, leads to 
\begin{equation}\label{eqA.34}
\sup_{h\leq \tau\leq T-h}\left\Vert \widehat\Sigma_X(\tau)- \Sigma_X^\ast(\tau)\right\Vert_{\max}=O_P\left(\zeta(p,b,h,{\mathbf n},N)\right).
\end{equation}
Due to the identification condition of $\Sigma_F(\tau)=\sigma_\tau^F(\sigma_\tau^F)^{^\intercal}=I_k$, we may also show that
\begin{equation}\label{eqA.35}
\sup_{h\leq \tau\leq T-h}\left\Vert  \Sigma_X^\ast(\tau)- \Sigma_X(\tau)\right\Vert_{\max}=O_P\left(\left(\frac{\log(p\vee N)}{Nh}\right)^{1/2}+h^{\delta}\right).
\end{equation}
The proof of (3.16) is completed by virtue of (\ref{eqA.34}) and (\ref{eqA.35}).

We next turn to the proof of (3.17). For any $p\times p$ matrix $\Sigma$, we have
\begin{equation}\label{eqA.36}
\Vert \Sigma\Vert_{\Sigma_X(\tau)}^2=\frac{1}{p}\Vert\Sigma^{-1/2}_X(\tau)\Sigma \Sigma^{-1/2}_X(\tau)\Vert_{F}^2\leq \frac{C}{p}\Vert \Sigma\Vert_F^2,
\end{equation}
since all the eigenvalues of $\Sigma_X(\tau)$ are positive and bounded away from $0$. Note that
\begin{eqnarray}
&&\sup_{h\leq \tau\leq T-h}\left\Vert \widehat\Sigma_X(\tau)- \Sigma_X(\tau)\right\Vert_{\Sigma_X(\tau)}^2\notag\\
&&\le
C\sup_{h\leq \tau\leq T-h}\left[\left\Vert \widehat\Sigma_X(\tau)- \Sigma_X^\ast(\tau)\right\Vert_{\Sigma_X(\tau)}^2+\left\Vert  \Sigma_X^\ast(\tau)- \Sigma_X(\tau)\right\Vert_{\Sigma_X(\tau)}^2\right],\notag
\end{eqnarray}
where
\begin{eqnarray}
&&
\sup_{h\leq \tau\leq T-h}\left\Vert \widehat\Sigma_X(\tau)- \Sigma_X^\ast(\tau)\right\Vert_{\Sigma_X(\tau)}^2\notag\\
&&\le C\sup_{h\leq \tau\leq T-h}\left[\left\Vert \widehat\Sigma_U(\tau)- \Sigma_U(\tau)\right\Vert_{\Sigma_X(\tau)}^2+\left\Vert \widetilde\Lambda(\tau)\widetilde\Lambda(\tau)^{^\intercal}-\Lambda(\tau)\Sigma_{F}^\ast(\tau)\Lambda(\tau)^{^\intercal}\right\Vert_{\Sigma_X(\tau)}^2\right].\notag
\end{eqnarray}
By (\ref{eqA.36}) and (3.14) in Theorem 3.1, we can prove that 
\begin{eqnarray}
&&\sup_{h\leq \tau\leq T-h}\left\Vert \widehat\Sigma_U(\tau)- \Sigma_U(\tau)\right\Vert_{\Sigma_X(\tau)}^2\notag\\
&&\leq\frac{C}{p}\sup_{h\leq \tau\leq T-h}\left\Vert \widehat\Sigma_U(\tau)- \Sigma_U(\tau)\right\Vert_{F}^2\notag\\
&&\leq C\sup_{h\leq \tau\leq T-h}\left\Vert \widehat\Sigma_U(\tau)- \Sigma_U(\tau)\right\Vert_s^2\notag\\
&&= O_{P}\left(\varpi_p^2\left[\zeta(p,b,h,{\mathbf n},N)\right]^{2-2q}\right).\label{eqA.37}
\end{eqnarray}

Let $D_\Lambda(\tau)=\widetilde{\Lambda}(\tau)-\Lambda(\tau)H_\ast(\tau)$, where $H_\ast(\tau)=[H(\tau)]^{-1}$ and $H(\tau)$ is a $k\times k$ rotation matrix defined as in Lemma \ref{le:B.2}. Note that 
\begin{eqnarray*}
&&\widetilde\Lambda(\tau)\widetilde\Lambda(\tau)^{^\intercal}-\Lambda(\tau)\Sigma_{F}^\ast(\tau)\Lambda(\tau)^{^\intercal}\\
&&= D_\Lambda(\tau)D_\Lambda(\tau)^{^\intercal}+D_\Lambda(\tau)H_\ast(\tau)^{^\intercal}\Lambda(\tau)^{^\intercal}+\Lambda(\tau)H_\ast(\tau)D_\Lambda(\tau)^{^\intercal}+\notag\\
&&\Lambda(\tau)\left[H_\ast(\tau)H_\ast(\tau)^{^\intercal}-\Sigma_{F}^\ast(\tau)\right]\Lambda(\tau)^{^\intercal}.
\end{eqnarray*}
By (\ref{eqA.36}) and Lemma \ref{le:B.4}, we have
\begin{eqnarray}
\sup_{h\leq \tau\leq T-h}\left\Vert D_\Lambda(\tau)D_\Lambda(\tau)^{^\intercal}\right\Vert_{\Sigma_X(\tau)}^2&\leq& C\sup_{h\leq \tau\leq T-h} \frac{1}{p}\Vert D_\Lambda(\tau)\Vert_F^4\notag\\
&=&O_{P}\left( p\left[\zeta(p,b,h,{\mathbf n},N)\right]^{4}\right).\label{eqA.38}
\end{eqnarray}
By (2.6) and Assumption 1(iii), we may show that  
\begin{equation}\label{eqA.39}
\sup_{h\leq \tau\leq T-h}\left\Vert \Lambda(\tau)^{^\intercal}\Sigma_{X}^{-1}(\tau)\Lambda(\tau)\right\Vert_s=O_P(1),
\end{equation}
whose proof is similar to the proof of Theorem 2 in \cite{FFL08}. Then, by (\ref{eqA.39}), Lemmas \ref{le:B.3} and \ref{le:B.4}, following the proof of Lemma 13 in \cite{FLM13}, we have
\begin{eqnarray}
&&\sup_{h\leq \tau\leq T-h}\left\Vert \Lambda(\tau)H_\ast(\tau)D_\Lambda(\tau)^{^\intercal}\right\Vert_{\Sigma_X(\tau)}^2\notag\\
&&=\frac{1}{p}\sup_{h\leq \tau\leq T-h}\mathsf{ trace}\left\{H_\ast(\tau)D_\Lambda(\tau)^{^\intercal}\Sigma_{X}^{-1}(\tau)D_\Lambda(\tau)H_\ast(\tau)^{^\intercal}\Lambda(\tau)^{^\intercal}\Sigma_{X}^{-1}(\tau)\Lambda(\tau)\right\}\notag\\
&&\le \frac{C}{p}\sup_{h\leq \tau\leq T-h} \left\Vert H_\ast(\tau)\right\Vert_s^2 \left\Vert D_\Lambda(\tau)\right\Vert^2_F\notag\\
&&=O_P\left(\left[\zeta(p,b,h,{\mathbf n},N)\right]^{2}\right).\label{eqA.40}
\end{eqnarray}
Similarly, we can also prove that 
\begin{equation}\label{eqA.41}
\sup_{h\leq \tau\leq T-h}\left\Vert D_\Lambda(\tau)H_\ast(\tau)^{^\intercal}\Lambda(\tau)^{^\intercal}\right\Vert_{\Sigma_X(\tau)}^2=O_P\left(\left[\zeta(p,b,h,{\mathbf n},N)\right]^{2}\right).
\end{equation}
By Assumption 1(iii) and Lemmas \ref{le:B.2} and \ref{le:B.3}, we may show that 
\begin{eqnarray}
& &\sup_{h\leq \tau\leq T-h}\left\Vert \Lambda(\tau)\left[H_\ast(\tau)H_\ast(\tau)^{^\intercal}-\Sigma_{F}^\ast(\tau)\right]\Lambda(\tau)^{^\intercal}\right\Vert_{\Sigma_X(\tau)}^2\notag\\
&&=\sup_{h\leq \tau\leq T-h}\left\Vert \Lambda(\tau)H_\ast(\tau)\left[\sum_{t=1}^{N}\Delta\widetilde F_t(\tau)\Delta \widetilde F_t(\tau)^{^\intercal}-\sum_{t=1}^{N}H(\tau)\Delta F_t(\tau)\Delta F_t(\tau)^{^\intercal}H(\tau)^{^\intercal}\right]H_\ast(\tau)^{^\intercal}\Lambda(\tau)^{^\intercal}\right\Vert_{\Sigma_X(\tau)}^2\notag\\\notag\\
&&\le\frac{C}{p}\sup_{h\leq \tau\leq T-h} \left\Vert \sum_{t=1}^{N}\Delta\widetilde F_t(\tau)\Delta \widetilde F_t(\tau)^{^\intercal}-\sum_{t=1}^{N}H(\tau)\Delta F_t(\tau)\Delta F_t(\tau)^{^\intercal}H(\tau)^{^\intercal}\right\Vert^2_F\notag\\ 
&&=O_{P}\left(\frac{1}{p}\left[\zeta(p,b,h,{\mathbf n},N)\right]^{2}\right). \label{eqA.42}
\end{eqnarray}
With (\ref{eqA.38}), (\ref{eqA.40})--(\ref{eqA.42}), we have 
\begin{eqnarray}
&&\sup_{h\leq \tau\leq T-h}\left\Vert\widetilde\Lambda(\tau)\widetilde\Lambda(\tau)^{^\intercal}-\Lambda(\tau)\Sigma_{F}^\ast(\tau)\Lambda(\tau)^{^\intercal}\right\Vert_{{\boldsymbol\Sigma}_X(z)}^2\notag\\
&&=O_P\left(p\left[\zeta(p,b,h,{\mathbf n},N)\right]^{4}+\left[\zeta(p,b,h,{\mathbf n},N)\right]^2\right).\label{eqA.43}
\end{eqnarray}
Combining (\ref{eqA.37}) and (\ref{eqA.43}), indicates that 
\begin{eqnarray}
&&\sup_{h\leq \tau\leq T-h}\left\Vert \widehat\Sigma_X(\tau)- \Sigma_X^\ast(\tau)\right\Vert_{\Sigma_X(\tau)}^2\notag\\
&&=O_P\left(p\left[\zeta(p,b,h,{\mathbf n},N)\right]^{4}+\varpi_p^2\left[\zeta(p,b,h,{\mathbf n},N)\right]^{2-2p}\right).\label{eqA.44}
\end{eqnarray}

On the other hand, using (\ref{eqA.36}) and following the proof of (\ref{eqA.42}), we have
\begin{eqnarray}
\sup_{h\leq \tau\leq T-h}\left\Vert  \Sigma_X^\ast(\tau)- \Sigma_X(\tau)\right\Vert_{\Sigma_X(\tau)}^2&=&O_P\left(\frac{1}{p}\left(\left(\frac{\log(p\vee N)}{Nh}\right)^{1/2}+h^{\delta}\right)^2\right)\notag\\
&=&o_P\left(\left[\zeta(p,b,h,{\mathbf n},N)\right]^{2}\right).\label{eqA.45}
\end{eqnarray}
By virtue of (\ref{eqA.44}) and (\ref{eqA.45}), we complete the proof of (3.17).\hfill$\blacksquare$


\section{Technical lemmas and proofs}\label{app:B}
\renewcommand{\theequation}{B.\arabic{equation}}
\setcounter{equation}{0}

In this appendix, we provide some technical lemmas together with their proofs. As in Appendix A, we let $C$ be a generic positive constant whose value may change from line to line. The first lemma is the Bernstein inequality for the $\alpha$-mixing sequence \citep[e.g.,][]{B98}. 

\smallskip

\renewcommand{\thelemma}{{B.\arabic{lemma}}}\setcounter{lemma}{0}

\begin{lemma}\label{le:B.1}
Let $\{Z_t, t\geq1\}$ be a zero-mean $\alpha$-mixing process satisfying ${\mathsf{P}}(|Z_t|\leq B)=1$ for all $t\geq1$. Then, for each integer $q_1\in [1,T/2]$ and each $\epsilon>0$, we have
\begin{equation}  \label{eqB.1}
{\mathsf{P}}\left(\left|\sum\limits_{t=1}^T Z_t\right|>T\epsilon\right)\leq 4\exp\left\{-\frac{\epsilon^2q_1}{8\omega(q_1)}\right\}+22\left(1+\frac{4B}{\epsilon}\right)^{1/2}q_1\alpha(\lfloor q_2\rfloor]),
\end{equation}
where $\omega(q_1)=2\sigma^2(q_1)/q_2^2+B\epsilon/2$, $q_2=T/(2q_1)$,
\begin{equation*}
\begin{array}{ll}
\sigma^2(q_1)=\max\limits_{1\leq j\leq2q_1-1}{\mathsf{E}} & \left\{\left(\lfloor jq_2\rfloor+1-jq_2\right)Z_{\lfloor jq_2\rfloor+1}+Z_{\lfloor jq_2\rfloor+2}+\cdots+ Z_{\lfloor(j+1)q_2\rfloor} \right.\\ 
&\left. +((j+1)q_2-\lfloor(j+1)q_2\rfloor)Z_{\lfloor(j+1)q_2\rfloor+1}\right\}^2,%
\end{array}%
\end{equation*}
$\alpha(\cdot)$ is the $\alpha$-mixing coefficient and $\lfloor\cdot\rfloor$ denotes the floor function.

\end{lemma}

\smallskip

The following lemma derives the uniform mean square convergence rate of the local PCA estimates $\Delta\widetilde F_t(\tau)$, $t=1,\cdots,N$, defined in Section 2.3. Define a $k\times k$ rotation matrix:
\[H(\tau)= \left[\widetilde{V}(\tau)\right]^{-1}\left[\sum_{t=1}^{N}\Delta\widetilde F_t(\tau)\Delta F_t(\tau)^{^\intercal}\right]\left[\frac{1}{p}\Lambda(\tau)^{^\intercal}\Lambda(\tau)\right],\]
where $\widetilde{V}(\tau)=\mathsf{diag}\left\{\widetilde v_1(\tau),\cdots, \widetilde v_k(\tau)\right\}$ with $\widetilde v_j(\tau)$ being the $j$-th largest eigenvalue of $\frac{1}{p}\Delta\widetilde X(\tau)^{^\intercal}\Delta\widetilde X(\tau)$.

\smallskip

\begin{lemma}\label{le:B.2}

Suppose that Assumptions 1--3 and 4(i)(ii) are satisfied. Then, we have
\begin{equation}\label{eqB.2}
\sup_{h\leq \tau\leq T-h}\sum_{t=1}^N\left\Vert \Delta\widetilde F_t(\tau)-H(\tau)\Delta F_t(\tau)\right\Vert^2=O_P\left(\left[\zeta_\circ(p,b,h,{\mathbf n},N)\right]^2\right),
\end{equation}
where 
\[\zeta_\circ(p,b,h,{\mathbf n},N)=\zeta_1(p,b,{\mathbf n},N)+\zeta_2^\circ(p,h,N)\]
with $\zeta_1(p,b,{\mathbf n},N)$ defined in Assumption 4(iii) and $\zeta_{2}^\circ(p,h,N)=\left(\frac{\varpi_p}{ph}\right)^{1/2}+(Nh)^{-1/2}+h^{\delta}$.

\end{lemma}

\medskip
\noindent{\bf Proof of Lemma \ref{le:B.2}}.\ \ By the definition of $\Delta\widetilde F(\tau)$, we have
\begin{equation}\label{eqB.3}
\frac{1}{p}\Delta\widetilde X(\tau)^{^\intercal}\Delta\widetilde X(\tau)\Delta\widetilde F(\tau)=\Delta\widetilde F(\tau)\widetilde V(\tau).
\end{equation}
Let $\Delta X_{j}=X_{t_j}-X_{t_{j-1}}$, $\Delta X_{j}(\tau)=\Delta X_{j}K_h^{1/2}(t_j-\tau)$, and $\Delta\widetilde X_j(\tau)$ be the $j$-th column vector of $\Delta\widetilde X(\tau)$. It follows from (\ref{eqB.3}) that, for $j=1,\cdots,N$,
{ \begin{eqnarray}
\widetilde{V}(\tau)\Delta\widetilde F_j(\tau)&=&\frac{1}{p}\sum_{s=1}^{N}\Delta\widetilde F_s(\tau)\Delta\widetilde X_s(\tau)^{^\intercal}\Delta\widetilde X_j(\tau)\notag\\
&=&\frac{1}{p}\sum_{s=1}^{N}\Delta\widetilde F_s(\tau)\Delta X_s(\tau)^{^\intercal}\Delta X_j(\tau)+\notag\\
&&\frac{1}{p}\sum_{s=1}^{N}\Delta\widetilde F_s(\tau)\left[\Delta\widetilde X_s(\tau)-\Delta X_s(\tau)\right]^{^\intercal}\Delta X_j(\tau)+\notag\\
&&\frac{1}{p}\sum_{s=1}^{N}\Delta\widetilde F_s(\tau)\Delta\widetilde X_s(\tau)^{^\intercal}\left[\Delta\widetilde X_j(\tau)-\Delta X_j(\tau)\right]
.\label{eqB.4}
\end{eqnarray}}

By (1.2) and (2.4), we write
\begin{eqnarray}
\Delta X_s(\tau)&=&\left(\int_{t_{s-1}}^{t_s}\Lambda(t)dF_t\right)K_h^{1/2}(t_s-\tau)+\left(\int_{t_{s-1}}^{t_s}d U_t\right)K_h^{1/2}(t_s-\tau)\notag\\
&=&\left(\int_{t_{s-1}}^{t_s}\Lambda(t)dF_t\right)K_h^{1/2}(t_s-\tau)+\left(\int_{t_{s-1}}^{t_s}\mu_t^Udt\right)K_h^{1/2}(t_s-\tau)+\notag\\
&&\left(\int_{t_{s-1}}^{t_s}\sigma_t^Ud W_t^U\right)K_h^{1/2}(t_s-\tau)\notag\\
&=&R_{s,1}(\tau)+R_{s,2}(\tau)+R_{s,3}(\tau),\notag
\end{eqnarray}
indicating that 
\begin{equation}\label{eqB.5}
\frac{1}{p}\sum_{s=1}^{N}\Delta\widetilde F_s(\tau)\Delta X_s(\tau)^{^\intercal}\Delta X_j(\tau)=\frac{1}{p}\sum_{l_1=1}^3\sum_{l_2=1}^3\sum_{s=1}^{N}\Delta\widetilde F_s(\tau)R_{s,l_1}(\tau)^{^\intercal}R_{j,l_2}(\tau).
\end{equation}
By the smoothness restriction in Assumption 1(iii), we may show that 
\begin{eqnarray}
&&\frac{1}{p}\sum_{s=1}^{N}\Delta\widetilde F_s(\tau)R_{s,1}(\tau)^{^\intercal}R_{j,1}(\tau)\notag\\
&&=\left[\sum_{s=1}^{N}\Delta\widetilde F_s(\tau)\Delta F_s(\tau)^{^\intercal}\right]\left[\frac{1}{p}\Lambda(\tau)^{^\intercal}\Lambda(\tau)\right]\Delta F_j(\tau)\left(1+O_P\left(h^{\delta}\right)\right),\label{eqB.6}
\end{eqnarray}
where $\delta$ is defined in Assumption 1(ii). 

Following the argument in the proof of (\ref{eqA.14}) with $\nu_2(p,b,n_i)$ and $t_j^i$ replaced by $[N^{-1}\log (p\vee N)]^{1/2}$ and $t_s$, respectively, we can prove that 
\begin{equation}\label{eqB.7}
\max_{1\leq i\leq p}\sup_{1\leq s\leq n}|\Delta X_{i,s}|=O_P\left(\sqrt{N^{-1}\log (p\vee N)}\right),
\end{equation}
where $\Delta X_{i,s}$ is the $i$-th component of $\Delta X_{s}$. By (B.7), Assumption 1(i) and the Cauchy-Schwarz inequality, and noting that 
\begin{equation}\label{eqB.8}
\sup_{h\leq \tau\leq T-h}\sum_{s=1}^{N}\left\Vert\Delta\widetilde F_s(\tau)\right\Vert^2=O_P(1)
\end{equation}
from the normalisation restriction (2.14), we have
\begin{eqnarray}
&&\sup_{h\leq \tau\leq T-h}\sum_{j=1}^N\left\Vert\frac{1}{p}\sum_{l_2=1}^3\sum_{s=1}^{N}\Delta\widetilde F_s(\tau)R_{s,2}(\tau)^{^\intercal}R_{j,l_2}(\tau)\right\Vert^2\notag\\
&=&\sup_{h\leq \tau\leq T-h}\sum_{j=1}^N\left\Vert\sum_{s=1}^{N}\Delta\widetilde F_s(\tau)\frac{R_{s,2}(\tau)^{^\intercal}\Delta X_j(\tau)}{p}\right\Vert^2\notag\\
&\leq&\sup_{h\leq \tau\leq T-h}\sum_{s=1}^{N}\left\Vert\Delta\widetilde F_s(\tau)\right\Vert^2\sum_{j=1}^N\sum_{s=1}^{N}\left\Vert\frac{1}{p}R_{s,2}(\tau)^{^\intercal}\Delta X_j(\tau)\right\Vert^2\notag\\
&=&O_P\left(\frac{\log (p\vee N)}{N}\right),\label{eqB.9}
\end{eqnarray}
and similarly
\begin{equation}\label{eqB.10}
\sup_{h\leq \tau\leq T-h}\sum_{j=1}^N\left\Vert\frac{1}{p}\sum_{l_1=1}^3\sum_{s=1}^{N}\Delta\widetilde F_s(\tau)R_{s,l_1}(\tau)^{^\intercal}R_{j,2}(\tau)\right\Vert^2=O_P\left(\frac{\log (p\vee N)}{N}\right).
\end{equation}

Write
\begin{eqnarray}
R_{s,1}(\tau)&=&\left[\Lambda(t_{s-1})\Delta F_s\right]K_h^{1/2}(t_s-\tau)+\left\{\int_{t_{s-1}}^{t_s}\left[\Lambda(t)-\Lambda(t_{s-1})\right] dF_t\right\}K_h^{1/2}(t_s-\tau)\notag\\
&=:&R_{s,1}^\star(\tau)+R_{s,1}^\diamond(\tau),\notag
\end{eqnarray}
and thus
\begin{eqnarray}
&&\sup_{h\leq \tau\leq T-h}\sum_{j=1}^N\left\Vert\frac{1}{p}\sum_{s=1}^{N}\Delta\widetilde F_s(\tau)R_{s,1}(\tau)^{^\intercal}R_{j,3}(\tau)\right\Vert^2\notag\\
&&\leq2\sup_{h\leq \tau\leq T-h}\sum_{j=1}^N\left\Vert\frac{1}{p}\sum_{s=1}^{N}\Delta\widetilde F_s(\tau)R_{s,1}^\star(\tau)^{^\intercal}R_{j,3}(\tau)\right\Vert^2+\notag\\
&&\ \ \ 2\sup_{h\leq \tau\leq T-h}\sum_{j=1}^N\left\Vert\frac{1}{p}\sum_{s=1}^{N}\Delta\widetilde F_s(\tau)R_{s,1}^\diamond(\tau)^{^\intercal}R_{j,3}(\tau)\right\Vert^2.\notag
\end{eqnarray}
Write $R_{s,3}(\tau)=u_sK_h^{1/2}(t_s-\tau)$ with $u_s=\left(u_{1,s},\cdots,u_{p,s}\right)^{^\intercal}=\int_{t_{s-1}}^{t_s}\sigma_t^UdW_t^U$. Note that 
\begin{eqnarray}
&&\sum_{j=1}^N\left\Vert\frac{1}{p}\sum_{s=1}^{N}\Delta\widetilde F_s(\tau)R_{s,1}^\star(\tau)^{^\intercal}R_{j,3}(\tau)\right\Vert^2\notag\\
&&=\sum_{j=1}^N\left\Vert\sum_{s=1}^{N}\Delta\widetilde F_s(\tau) \Delta F_s(\tau)^{^\intercal}\left[\frac{1}{p}\Lambda(t_{s-1})^{^\intercal}u_j\right]\right\Vert^2K_h(t_j-\tau)\notag\\ 
&&\leq\left[\sum_{s=1}^{N}\left\Vert\Delta\widetilde F_s(\tau)\right\Vert^2\right]\left[\sum_{s=1}^{N}\left\Vert\Delta F_s(\tau)\right\Vert^2\right]\left[\sum_{j=1}^N K_h(t_j-\tau)\right]\left[\max_{1\leq s,j\leq n}\left\Vert\frac{1}{p}\Lambda(t_{s-1})^{^\intercal}u_j\right\Vert^2\right].\label{eqB.11}
\end{eqnarray}
By Proposition A.1 in \cite{BLLW23}, we have
\begin{equation}\label{eqB.12}
\sup_{h\leq \tau\leq T-h}\sum_{s=1}^{N}\left\Vert\Delta F_s(\tau)\right\Vert^2=O_P(1).
\end{equation}
As in (\ref{eqA.13}), using the moment generating function for the folded normal random variable, (2.6) and Assumption1(iii), for $l=1,\cdots,k$, we have
\[
{\mathsf E}\left[\exp\left(\psi\left\vert \frac{1}{[p\varpi_p]^{1/2}}\sum_{i=1}^p\Lambda_{il}(t_{s-1})u_{i,j}\right\vert\right)\right]\leq \exp\left\{\frac{\psi^2T C_3}{2N}\right\},
\]
where $\psi>0$, $\Lambda_{il}(\cdot)$ is the $(i,l)$-entry of $\Lambda(\cdot)$, and
\[C_3=\left[\max_{1\leq i\leq p}\max_{1\leq l\leq k}\sup_{0\leq t\leq T}|\Lambda_{il}(t)|^2\right]\left[\sup_{0\leq t\leq T}\frac{1}{p\varpi_p}\sum_{1\leq i_1,i_2\leq p}\sigma_{U,i_1i_2}(t)\right]\]
is bounded with probability approaching one. This, together with the Markov inequality, leads to  
\begin{eqnarray}
&&{\mathsf P}\left(\max_{1\leq s\leq n}\max_{1\leq j\leq N}\left\vert\frac{1}{p}\sum_{i=1}^p\Lambda_{il}(t_{s-1})u_{i,j}\right\vert>m\sqrt{\frac{\varpi_p\log N}{pN}}\right)\nonumber\\
&&={\mathsf P}\left(\max_{1\leq s\leq n}\max_{1\leq j\leq N}\left\vert\frac{1}{(p\varpi_p)^{1/2}}\sum_{i=1}^p\Lambda_{il}(t_{s-1})u_{i,j}\right\vert>m\sqrt{\frac{\log N}{N}}\right)\nonumber\\
&&\leq\sum_{s=1}^N\sum_{j=1}^N\exp\left(-\psi m\sqrt{\frac{\log N}{N}}\right){\mathsf E}\left[\exp\left(\psi\left\vert \frac{1}{(p\varpi_p)^{1/2}}\sum_{i=1}^p\Lambda_{il}(t_{s-1})u_{i,j}\right\vert\right)\right]\nonumber\\
&&\leq N^2\exp\left\{\frac{\psi^2T C_3}{2N}-\psi m\sqrt{\frac{\log N}{N}}\right\}.\label{eqB.13}
\end{eqnarray}
Choosing $\psi=(mN\log N)^{1/2}$ and letting $m>0$ be sufficiently large so that $m^{3/2}-(TmC_3)/2>2$ in (\ref{eqB.13}), we have
\[
{\mathsf P}\left(\max_{1\leq s\leq n}\max_{1\leq j\leq N}\left\vert\frac{1}{p}\sum_{i=1}^p\Lambda_{il}(t_{s-1})u_{i,j}\right\vert>m\sqrt{\frac{\varpi_p\log N}{pN}}\right)=o\left(N^2\exp\{-2\log N\}\right)=o(1)
\]
for any $l=1,\cdots,k$, indicating that 
\begin{equation}\label{eqB.14}
\max_{1\leq s\leq n}\max_{1\leq j\leq N}\left\Vert\frac{\Lambda(t_{s-1})^{^\intercal}u_j}{p}\right\Vert^2=O_P\left(\frac{\varpi_p\log N}{pN}\right).
\end{equation}
By (\ref{eqB.8}), (\ref{eqB.11}), (\ref{eqB.12}), (\ref{eqB.14}) and noting that $\frac{1}{N}\sum_{j=1}^N K_h(t_j-\tau)=O(1)$ uniformly over $h\leq \tau\leq T-h$, we can prove that 
\begin{equation}\label{eqB.15}
\sum_{j=1}^N\left\Vert\frac{1}{p}\sum_{s=1}^{N}\Delta\widetilde F_s(\tau)R_{s,1}^\star(\tau)^{^\intercal}R_{j,3}(\tau)\right\Vert^2=O_P\left(\frac{\varpi_p\log N}{p}\right).
\end{equation}
On the other hand, by the Cauchy-Schwarz inequality, we have
\begin{eqnarray}
&&\sum_{j=1}^N\left\Vert\frac{1}{p}\sum_{s=1}^{N}\Delta\widetilde F_s(\tau)R_{s,1}^\diamond(\tau)^{^\intercal}R_{j,3}(\tau)\right\Vert^2\notag\\
&&\leq\left[\sum_{s=1}^{N}\left\Vert\Delta\widetilde F_s(\tau)\right\Vert^2\right]\left[\sum_{s=1}^{N}\frac{1}{p}\left\Vert \int_{t_{s-1}}^{t_s}\left[\Lambda(t)-\Lambda(t_{s-1})\right] dF_t\right\Vert^2K_h(t_s-\tau)\right]\left[\frac{1}{p}\sum_{j=1}^N \left\Vert u_j\right\Vert^2K_h(t_j-\tau)\right].\notag
\end{eqnarray}
By Proposition A.1 in \cite{BLLW23} and the smoothness condition (3.2), we may show that 
\begin{eqnarray}
&&\sup_{h\leq \tau\leq T-h}\frac{1}{p}\sum_{s=1}^{N}\left\Vert \int_{t_{s-1}}^{t_s}\left[\Lambda(t)-\Lambda(t_{s-1})\right] dF_t\right\Vert^2K_h(t_s-\tau)=O_P\left(N^{-2\delta}\right),\label{eqB.16}\\
&&\sup_{h\leq \tau\leq T-h}\frac{1}{p}\sum_{j=1}^N \left\Vert u_j\right\Vert^2K_h(t_j-\tau)=O_P(1).\label{eqB.17}
\end{eqnarray}
By (\ref{eqB.8}), (\ref{eqB.16}) and (\ref{eqB.17}), we have
\begin{equation}\label{eqB.18}
\sup_{h\leq \tau\leq T-h}\sum_{j=1}^N\left\Vert\frac{1}{p}\sum_{s=1}^{N}\Delta\widetilde F_s(\tau)R_{s,1}^\diamond(\tau)^{^\intercal}R_{j,3}(\tau)\right\Vert^2=O_P\left(N^{-2\delta}\right).
\end{equation}
By virtue of (\ref{eqB.15}) and (\ref{eqB.18}), we prove that
\begin{equation}\label{eqB.19}
\sup_{h\leq \tau\leq T-h}\sum_{j=1}^N\left\Vert\frac{1}{p}\sum_{s=1}^{N}\Delta\widetilde F_s(\tau)R_{s,1}(\tau)^{^\intercal}R_{j,3}(\tau)\right\Vert^2=O_P\left(\frac{\varpi_p\log N}{p}+N^{-2\delta}\right),
\end{equation}
and similarly,
\begin{equation}\label{eqB.20}
\sup_{h\leq \tau\leq T-h}\sum_{j=1}^N\left\Vert\frac{1}{p}\sum_{s=1}^{N}\Delta\widetilde F_s(\tau)R_{s,3}(\tau)^{^\intercal}R_{j,1}(\tau)\right\Vert^2=O_P\left(\frac{\varpi_p\log N}{p}+N^{-2\delta}\right).
\end{equation}

Observe that 
\begin{eqnarray}
&&\sum_{j=1}^N\left\Vert\frac{1}{p}\sum_{s=1}^{N}\Delta\widetilde F_s(\tau)R_{s,3}(\tau)^{^\intercal}R_{j,3}(\tau)\right\Vert^2\notag\\
&&\leq2\sum_{j=1}^N\left\Vert\sum_{s=1}^{N}\Delta\widetilde F_s(\tau)\left[\frac{u_s^{^\intercal}u_j}{p}-{\mathsf E}\left(\frac{u_s^{^\intercal}u_j}{p}\right)\right]K_h^{1/2}(t_s-\tau)\right\Vert^2K_h(t_j-\tau)+\notag\\
&&\ \ \ \ 2\sum_{j=1}^N\left\Vert\sum_{s=1}^{N}\Delta\widetilde F_s(\tau){\mathsf E}\left(\frac{u_s^{^\intercal}u_j}{p}\right)K_h^{1/2}(t_s-\tau)\right\Vert^2K_h(t_j-\tau)\notag\\
&&\leq2\left[\sum_{s=1}^{N}\left\Vert\Delta\widetilde F_s(\tau)\right\Vert^2\right]\left[\sum_{s=1}^{N}K_h^2(t_s-\tau)\right]\left[\sum_{s=1}^N\sum_{j=1}^N\left\vert \frac{u_s^{^\intercal}u_j}{p}-{\mathsf E}\left(\frac{u_s^{^\intercal}u_j}{p}\right)\right\vert^4\right]^{1/2}+\notag\\
&&\ \ \ \ 2\left[\sum_{s=1}^{N}\left\Vert\Delta\widetilde F_s(\tau)\right\Vert^2\right]\left[\sum_{s=1}^N\sum_{j=1}^N\left\vert{\mathsf E}\left(\frac{u_s^{^\intercal}u_j}{p}\right)\right\vert^2 K_h(t_s-\tau)K_h(t_j-\tau)\right].
\notag
\end{eqnarray}
By Assumption 1(ii), we may show that 
\begin{eqnarray}
&&\sup_{h\leq \tau\leq T-h}\sum_{s=1}^N\sum_{j=1}^N\left\vert{\mathsf E}\left(\frac{u_s^{^\intercal}u_j}{p}\right)\right\vert^2 K_h(t_s-\tau)K_h(t_j-\tau)\notag\\
&&=\sup_{h\leq \tau\leq T-h}\sum_{s=1}^N\left\vert{\mathsf E}\left(\frac{u_s^{^\intercal}u_s}{p}\right)\right\vert^2 K_h^2(t_s-\tau)=O\left(\frac{1}{Nh}\right),\notag
\end{eqnarray}
which, together with (\ref{eqB.8}), leads to 
{\small \begin{equation}\label{eqB.21}
\sup_{h\leq \tau\leq T-h}\left[\sum_{s=1}^{n}\left\Vert\Delta\widetilde F_s(\tau)\right\Vert^2\right]\left[\sum_{s=1}^N\sum_{j=1}^N\left\vert{\mathsf E}\left(\frac{u_s^{^\intercal}u_j}{p}\right)\right\vert^2 K_h(t_s-\tau)K_h(t_j-\tau)\right]=O_P\left(\frac{1}{Nh}\right).
\end{equation}}
As in the proof of (B.11) in \cite{K18}, by the uniform sparsity assumption (2.6), we have 
\[\sum_{s=1}^N\sum_{j=1}^N{\mathsf E}\left[\left\vert \frac{u_s^{^\intercal}u_j}{p}-{\mathsf E}\left(\frac{u_s^{^\intercal}u_j}{p}\right)\right\vert^4\right]=O\left(\left(\frac{\varpi_p}{Np}\right)^2\right),
\]
which, together with (\ref{eqB.8}) and $\frac{1}{N}\sum_{s=1}^N K_h^2(t_j-\tau)=O\left(h^{-1}\right)$ uniformly over $h\leq \tau\leq T-h$, indicates that
\begin{equation}\label{eqB.22}
\sup_{h\leq \tau\leq T-h}\left[\sum_{s=1}^N\left\Vert\Delta\widetilde F_s(\tau)\right\Vert^2\right]\left[\sum_{s=1}^NK_h^2(t_s-\tau)\right]\left[\sum_{s=1}^N\sum_{j=1}^N\left\vert \frac{u_s^{^\intercal}u_j}{p}-{\mathsf E}\left(\frac{u_s^{^\intercal}u_j}{p}\right)\right\vert^4\right]^{1/2}=O_P\left(\frac{\varpi_p}{ph}\right).
\end{equation}
Then, by (\ref{eqB.21}) and (\ref{eqB.22}), we can prove that 
\begin{equation}\label{eqB.23}
\sup_{h\leq \tau\leq T-h}\sum_{j=1}^N\left\Vert\frac{1}{p}\sum_{s=1}^N\Delta\widetilde F_s(\tau)R_{s,3}(\tau)^{^\intercal}R_{j,3}(\tau)\right\Vert^2=O_P\left(\frac{\varpi_p}{ph}+\frac{1}{Nh}\right).
\end{equation}

By (3.12), (\ref{eqB.8}) and Proposition A.1 in \cite{BLLW23}, we can prove that 
\begin{equation}\label{eqB.24}
\sup_{h\leq \tau\leq T-h}\sum_{j=1}^N\left\Vert\frac{1}{p}\sum_{s=1}^N\Delta\widetilde F_s(\tau)\left[\Delta\widetilde X_s(\tau)-\Delta X_s(\tau)\right]^{^\intercal}\Delta X_j(\tau)\right\Vert^2=O_P\left(\zeta_1^2(p,b,{\mathbf n},N)\right)
\end{equation}
and
\begin{equation}\label{eqB.25}
\sup_{h\leq \tau\leq T-h}\sum_{j=1}^N\left\Vert\frac{1}{p}\sum_{s=1}^N\Delta\widetilde F_s(\tau)\Delta\widetilde X_s(\tau)^{^\intercal}\left[\Delta\widetilde X_j(\tau)-\Delta X_j(\tau)\right]\right\Vert^2=O_P\left(\zeta_1^2(p,b,{\mathbf n},N)\right).
\end{equation}

By virtue of (\ref{eqB.6}), (\ref{eqB.9}), (\ref{eqB.10}), (\ref{eqB.19}), (\ref{eqB.20}), (\ref{eqB.23})--(\ref{eqB.25}), Assumption 4(ii) and (\ref{eqB.26}) in Lemma \ref{le:B.3} below, we complete the proof of (\ref{eqB.2}).\hfill$\blacksquare$

\smallskip

Let $V_0(\tau)={\rm diag}\{v_1(\tau),\cdots,v_k(\tau)\}$ with $v_j(\tau)$ being the $j$-th largest eigenvalue of $\Sigma_\Lambda(\tau)$ defined in (3.3), and $\Delta X(\tau)$ be defined similarly to $\Delta \widetilde X(\tau)$ with $\Delta \widetilde X_i$ replaced by $\Delta X_i$. Let $W_0(\tau)$ be a $k\times k$ matrix consisting of the eigenvectors of $\Sigma_\Lambda(\tau)$. Lemma \ref{le:B.3} below derives the uniform consistency properties of $\widetilde V(\tau)$ and $H(\tau)$ both of which are defined above Lemma \ref{le:B.2}.

\smallskip

\begin{lemma}\label{le:B.3} 

Suppose that Assumptions 1--3 and 4(i)(ii) are satisfied. Then we have
\begin{equation}\label{eqB.26}
\sup_{h\leq \tau\leq T-h}\left\Vert \widetilde V(\tau)-V_0(\tau)\right\Vert_s=o_P(1)
\end{equation}
and
\begin{equation}\label{eqB.27}
\sup_{h\leq \tau\leq T-h}\left\Vert H(\tau) -H_0(\tau)\right\Vert_s=o_P(1),
\end{equation}
where $H_0(\tau)=\left[V_0(\tau)\right]^{-1/2}W_0(\tau)\left[\Sigma_\Lambda(\tau)\right]^{1/2}$.

\end{lemma}

\smallskip

\noindent{\bf Proof of Lemma \ref{le:B.3}}.\ \ We start with the proof of (\ref{eqB.26}). By Weyl's inequality, we have
\begin{eqnarray}
&&\sup_{h\leq\tau\leq T-h}\max_{1\leq l\leq k}\left\vert \widetilde v_l(\tau)-v_l(\tau)\right\vert\notag\\
&&\leq\sup_{h\leq \tau\leq T-h}\left\Vert \frac{1}{p}\Delta\widetilde X(\tau)^{^\intercal}\Delta \widetilde X(\tau)-\frac{1}{p}\Delta X(\tau)^{^\intercal}\Delta X(\tau)\right\Vert_s+\notag\\
&&\ \ \ \sup_{h\leq \tau\leq T-h}\left\Vert \frac{1}{p}\Delta X(\tau)^{^\intercal}\Delta X(\tau)-\Delta F(\tau)\left[\frac{1}{p}\Lambda(\tau)^{^\intercal}\Lambda(\tau)\right]\Delta F(\tau)^{^\intercal}\right\Vert_s+\notag\\
&&\ \ \ \sup_{h\leq \tau\leq T-h}\left\Vert \frac{1}{p}\Lambda(\tau)\left[\Delta F(\tau)^{^\intercal}\Delta F(\tau)-I_k\right] \Lambda(\tau)^{^\intercal}\right\Vert_s+\notag\\
&&\ \ \ \sup_{h\leq \tau\leq T-h}\left\Vert \frac{1}{p}\Lambda(\tau)^{^\intercal}\Lambda(\tau)-\Sigma_\Lambda(\tau)\right\Vert_s.\label{eqB.28}
\end{eqnarray}
By (3.12) in Remark 3.2 and (3.7) in Assumption 4(ii), we may show that
\begin{eqnarray}
&&\sup_{h\leq \tau\leq T-h}\left\Vert \frac{1}{p}\Delta\widetilde X(\tau)^{^\intercal}\Delta \widetilde X(\tau)-\frac{1}{p}\Delta X(\tau)^{^\intercal}\Delta X(\tau)\right\Vert_s\nonumber\\
&&\leq\sup_{h\leq \tau\leq T-h}\left\Vert \frac{1}{p}\left[\Delta\widetilde X(\tau)-\Delta X(\tau)\right]^{^\intercal}\Delta \widetilde X(\tau)\right\Vert_s+\nonumber\\ 
&&\ \ \ \ \sup_{h\leq \tau\leq T-h}\left\Vert \frac{1}{p}\Delta X(\tau)^{^\intercal} \left[\Delta\widetilde X(\tau)-\Delta X(\tau)\right]\right\Vert_s\nonumber\\
&&=O_P\left(\zeta_1(p,b,{\mathbf n}, N)\right)=o_P(1), \label{eqB.29}
\end{eqnarray}
where $\zeta_1(p,b,{\mathbf n}, N)$ is defined in Assumption 4(iii). Using Proposition A.1 in \cite{BLLW23} and noting that $\sigma_t^F=I_k$, we have
\[
\sup_{h\leq\tau\leq T-h} \left\Vert\Delta F(\tau)^{^\intercal}\Delta F(\tau)-I_k\right\Vert_s=o_P(1),
\]
which, together with (3.10), indicates that
\begin{equation}\label{eqB.30}
\sup_{h\leq \tau\leq T-h}\left\Vert \frac{1}{p}\Lambda(\tau)\left[\Delta F(\tau)^{^\intercal}\Delta F(\tau)-I_k\right] \Lambda(\tau)^{^\intercal}\right\Vert_s=o_P(1).
\end{equation}
By (3.3) in Assumption 1(iii), we readily have that
\begin{equation}\label{eqB.31}
\sup_{h\leq \tau\leq T-h}\left\Vert \frac{1}{p}\Lambda(\tau)^{^\intercal}\Lambda(\tau)-\Sigma_\Lambda(\tau)\right\Vert_s=o_P(1).
\end{equation}
By (\ref{eqB.28})--(\ref{eqB.31}), to complete the proof of (\ref{eqB.26}), it remains to show that 
\begin{equation}\label{eqB.32}
\sup_{h\leq \tau\leq T-h}\left\Vert \frac{1}{p}\Delta X(\tau)^{^\intercal}\Delta X(\tau)-\Delta F(\tau)\left[\frac{1}{p}\Lambda(\tau)^{^\intercal}\Lambda(\tau)\right]\Delta F(\tau)^{^\intercal}\right\Vert_s=o_P(1).
\end{equation}

Letting $\Delta X_{i,s}(\tau)$ be the $i$-th element of $\Delta X_{s}(\tau)$ and $\sigma_{i\bullet, t}^U$ the $i$-th row vector of $\sigma_{t}^U$, it follows from (1.2) and (2.4) that
\begin{eqnarray}
\Delta X_{i,s}(\tau)&=&\left\{\left(\int_{t_{s-1}}^{t_s}\Lambda_{i}(t)^{^\intercal}dF_t\right)+\left(\int_{t_{s-1}}^{t_s}\mu_{i,t}^Udt\right)+\left[\int_{t_{s-1}}^{t_s}\left(\sigma_{i\bullet, t}^U\right)^{^\intercal}d W_t^U\right]\right\}K_h^{1/2}(t_s-\tau)\notag\\
&=:&R_{i,s,1}(\tau)+R_{i,s,2}(\tau)+R_{i,s,3}(\tau).\notag
\end{eqnarray}
Then we write
\[R_{i\bullet,l}(\tau)=\left[R_{i,1,l}(\tau),\cdots,R_{i,n,l}(\tau)\right]^{^\intercal},\ \ i=1,\cdots,p,\ \ l=1,2,3,\]
and
\[
\frac{1}{p}\Delta X(\tau)^{^\intercal}\Delta X(\tau)=\sum_{l_1=1}^3\sum_{l_2=1}^3\frac{1}{p}\sum_{i=1}^p R_{i\bullet,l_1}(\tau)R_{i\bullet,l_2}(\tau)^{^\intercal}.
\]
Similarly to the proofs of (\ref{eqB.6}), (\ref{eqB.9}), (\ref{eqB.19}) and (\ref{eqB.23}), we may show that uniformly over $h\leq \tau\leq T-h$,
{\small\begin{eqnarray}
&& \left\Vert\frac{1}{p}\sum_{i=1}^p R_{i\bullet,1}(\tau)R_{i\bullet,1}(\tau)^{^\intercal}-\Delta F(\tau)\left[\frac{1}{p}\Lambda(\tau)^{^\intercal}\Lambda(\tau)\right]\Delta F(\tau)^{^\intercal}\right\Vert_s=O_P\left(h^\delta \log(p\vee N)\right), \label{eqB.33}\\ 
&&\frac{1}{p}\left\Vert\sum_{i=1}^p\sum_{l=1}^3 R_{i\bullet,2}(\tau)R_{i\bullet,l}(\tau)^{^\intercal}+\sum_{i=1}^p\sum_{l=1}^3 R_{i\bullet,l}(\tau)R_{i\bullet,2}(\tau)^{^\intercal}\right\Vert_s=O_P\left(\sqrt{\frac{\log(p\vee N)}{N}}\right),\label{eqB.34}\\ 
&&\frac{1}{p}\left\Vert\sum_{i=1}^p R_{i\bullet,1}(\tau)R_{i\bullet,3}(\tau)^{^\intercal}+\sum_{i=1}^p R_{i\bullet,3}(\tau)R_{i\bullet,1}(\tau)^{^\intercal}\right\Vert_s=O_P\left(\sqrt{\frac{\varpi_p\log^2(p\vee N)}{ph}}+\sqrt{\frac{\log(p\vee N)}{N^{2\delta}}}\right),\label{eqB.35}\\ 
&&\frac{1}{p}\left\Vert\sum_{i=1}^p R_{i\bullet,3}(\tau)R_{i\bullet,3}(\tau)^{^\intercal}\right\Vert_s=O_P\left(\sqrt{\frac{\varpi_p}{ph}}+\frac{1}{Nh}\right).\label{eqB.36}
\end{eqnarray}}
Combining (\ref{eqB.33})--(\ref{eqB.36}) with Assumption 4(ii)(iii), we complete the proof of (\ref{eqB.26}).

We next turn to the proof of (\ref{eqB.27}) which is similar to the proof of (A.10) in \cite{K18}. Let $\widetilde{W}(\tau)=W(\tau)[Q(\tau)]^{-1}$, where
\[W(\tau)= \left[\frac{1}{p}\Lambda(\tau)^{^\intercal}\Lambda(\tau)\right]^{1/2}\left[\Delta F(\tau)^{^\intercal}\Delta\widetilde F(\tau)\right],\ \ Q(\tau)=\left[\mathsf {diag}\left\{W(\tau)^{^\intercal}W(\tau)\right\}\right]^{1/2}.\]
Define
{\small \begin{eqnarray}
M(\tau)&=& \left[\frac{1}{p}\Lambda(\tau)^{^\intercal}\Lambda(\tau)\right]^{1/2}\left[\Delta F(\tau)^{^\intercal}\Delta F(\tau)\right] \left[\frac{1}{p}\Lambda(\tau)^{^\intercal}\Lambda(\tau)\right]^{1/2},\notag\\
D(\tau)&=& \left[\frac{1}{p}\Lambda(\tau)^{^\intercal}\Lambda(\tau)\right]^{1/2} \Delta F(\tau)^{^\intercal}\left[\frac{\Delta \widetilde X(\tau)^{^\intercal}\Delta \widetilde X(\tau)-\Delta F(\tau)\Lambda(\tau)^{^\intercal}\Lambda(\tau)\Delta F(\tau)^{^\intercal}}{p}\right]\Delta \widetilde F(\tau).\notag
\end{eqnarray}}
By the definition of $\Delta \widetilde{F}(\tau)$ in (\ref{eqB.3}), we have
{\[
\left\{M(\tau)+D(\tau)[W(\tau)]^{-1}\right\}\widetilde{W}(\tau)=\widetilde{W}(\tau)\widetilde{V}(\tau),
\]}
indicating that {\footnotesize $\widetilde{W}(\tau)$ is a $k\times k$} matrix consisting of the eigenvectors of $M(\tau)+D(\tau)[W(\tau)]^{-1}$. Write 
\[H(\tau)=\left[\widetilde{V}(\tau)\right]^{-1}\left[Q(\tau)\widetilde{W}(\tau)^{^\intercal}\right]\left[\frac{1}{p}\Lambda(\tau)^{^\intercal}\Lambda(\tau)\right]^{1/2}.\]
Hence, in order to prove (\ref{eqB.27}), we only need to show 
\begin{eqnarray}
\sup_{h\leq \tau\leq T-h}\left\Vert Q^2(\tau)-V_0(\tau)\right\Vert_s=o_P(1),\label{eqB.37}\\
\sup_{h\leq \tau\leq T-h}\left\Vert \widetilde W(\tau)-W_0(\tau)\right\Vert_s=o_P(1).\label{eqB.38}
\end{eqnarray}

Let $W_\ast(\tau)$ be a $k\times k$ matrix consisting of the eigenvectors of $M(\tau)$. By the $\sin \theta$ theorem in \cite{DK70}, we have
\begin{equation}\label{eqB.39}
\left\Vert \widetilde W(\tau)-W_0(\tau)\right\Vert_s\leq C \Vert D(\tau)\Vert_s\cdot \Vert [W(\tau)]^{-1}\Vert_s.
\end{equation}
Using (3.3) in Assumption 1(iii) and Lemma \ref{le:B.2}, and noting that $H(\tau)$ is asymptotically non-singular, we may show that
\begin{equation}\label{eqB.40}
\sup_{h\leq \tau\leq T-h} \Vert [W(\tau)]^{-1}\Vert_s=O_P(1).
\end{equation}
Meanwhile, by (3.3), (\ref{eqB.8}), (\ref{eqB.12}), (\ref{eqB.29}) and (\ref{eqB.32}), we can prove that
\begin{eqnarray}
&&\sup_{h\leq \tau\leq T-h} \Vert D(\tau)\Vert_s\notag\\
&&\leq C\sup_{h\leq \tau\leq T-h}\left\Vert\frac{\Delta \widetilde X(\tau)^{^\intercal}\Delta \widetilde X(\tau)-\Delta F(\tau)\Lambda(\tau)^{^\intercal}\Lambda(\tau)\Delta F(\tau)^{^\intercal}}{p}\right\Vert_s\notag\\
&&\leq C\sup_{h\leq \tau\leq T-h}\left\Vert\frac{\Delta X(\tau)^{^\intercal}\Delta X(\tau)-\Delta F(\tau)\Lambda(\tau)^{^\intercal}\Lambda(\tau)\Delta F(\tau)^{^\intercal}}{p}\right\Vert_s+o_P(1)\notag\\
&&=o_P(1).\label{eqB.41}
\end{eqnarray}
By virtue of (\ref{eqB.39})--(\ref{eqB.41}), we complete the proof of (\ref{eqB.38}).

It remains to prove (\ref{eqB.37}). By the triangular inequality, we have
\[
\left\Vert Q^2(\tau)-V_0(\tau)\right\Vert_s\leq \left\Vert Q^2(\tau)-\widetilde{V}(\tau)\right\Vert_s+\left\Vert \widetilde{V}(\tau)-V_0(\tau)\right\Vert_s.
\]
By (\ref{eqB.3}), (\ref{eqB.29}) and (\ref{eqB.32}), 
\begin{eqnarray}
&&\sup_{h\leq \tau\leq T-h} \Vert Q^2(\tau)-\widetilde{V}(\tau)\Vert_s\notag\\ 
&\leq& C\sup_{h\leq \tau\leq T-h}\left\Vert \Delta \widetilde{F}(\tau)^{^\intercal}\frac{\Delta \widetilde X(\tau)^{^\intercal}\Delta \widetilde X(\tau)-\Delta F(\tau)\Lambda(\tau)^{^\intercal}\Lambda(\tau)\Delta F(\tau)^{^\intercal}}{p}\Delta \widetilde{F}(\tau)\right\Vert_s\notag\\
&\leq&C\sup_{h\leq \tau\leq T-h}\left\Vert\frac{\Delta X(\tau)^{^\intercal}\Delta X(\tau)-\Delta F(\tau)\Lambda(\tau)^{^\intercal}\Lambda(\tau)\Delta F(\tau)^{^\intercal}}{p}\right\Vert_s+o_P(1)\notag\\
&=&o_P(1).\label{eqB.42}
\end{eqnarray}
Then, by (\ref{eqB.26}) and (\ref{eqB.42}), we complete the proof of (\ref{eqB.37}).\hfill$\blacksquare$

\smallskip

Let $H_\ast(\tau)=[H(\tau)]^{-1}$ be the inverse of the rotation matrix $H(\tau)$. The following lemma derives the uniform convergence rate for the estimated factor loading functions.

\smallskip

\begin{lemma}\label{le:B.4}

Suppose that Assumptions 1--3 and 4(i)(ii) are satisfied. Then, 
\begin{equation}\label{eqB.43}
\max_{1\leq i\leq p}\sup_{h\leq \tau\leq T-h}\left\Vert \widetilde\Lambda_i(\tau)-H_\ast(\tau)^{^\intercal}\Lambda_i(\tau)\right\Vert=O_P\left(\zeta_1(p,b,{\mathbf n}, N)+\zeta_{2}(p,h,N)\right)
\end{equation}
where $\zeta_1(p,b,{\mathbf n}, N)$ and $\zeta_{2}(p,h,N)$ are defined in Assumption 4(iii) and $H_\ast(\tau)=H(\tau)^{^\intercal}+o_P(1)$ uniformly over $h\leq \tau\leq T-h$.

\end{lemma}

\smallskip

\noindent{\bf Proof of Lemma \ref{le:B.4}}.\ \ By (1.2) and the definition of $\widetilde\Lambda_i(\tau)$, we have
\begin{eqnarray}
\widetilde\Lambda_i(\tau)&=&\sum_{j=1}^N\Delta\widetilde{X}_{i,j}\Delta\widetilde{F}_j(\tau)K_h^{1/2}(t_j-\tau)\notag\\
&=&\sum_{j=1}^N\left[\int_{t_{j-1}}^{t_j}\Lambda_i(t)^{^\intercal}F_tdt\right]H(\tau)\Delta F_j(\tau)K_h^{1/2}(t_j-\tau)+\sum_{j=1}^N\Delta U_{i,j}(\tau)H(\tau)\Delta F_j(\tau)+\notag\\
&&\sum_{j=1}^N\Delta X_{i,j}\left[\Delta\widetilde{F}_j(\tau)-H(\tau)\Delta F_j(\tau)\right]K_h^{1/2}(t_j-\tau)+\notag\\
&&\sum_{j=1}^N\left(\Delta\widetilde{X}_{i,j}-\Delta X_{i,j}\right)\Delta\widetilde{F}_j(\tau)K_h^{1/2}(t_j-\tau),\notag
\end{eqnarray}
where $\Delta U_{i,j}(\tau)=(U_{i,t_j}-U_{i,t_{j-1}})K_h^{1/2}(t_j-\tau)$.

Using (3.12), (\ref{eqB.8}) and the Cauchy-Schwarz inequality, we may show that 
\begin{eqnarray}
&&\max_{1\leq i\leq p}\sup_{h\leq\tau\leq T-h}\left\Vert \sum_{j=1}^N\left(\Delta\widetilde{X}_{i,j}-\Delta X_{i,j}\right)\Delta\widetilde{F}_j(\tau)K_h^{1/2}(t_j-\tau)\right\Vert \notag\\
&&\leq \max_{1\leq i\leq p}\sup_{h\leq\tau\leq T-h}\left[ \sum_{j=1}^N\left(\Delta\widetilde{X}_{i,j}-\Delta X_{i,j}\right)^2K_h(t_j-\tau)\right]^{1/2} \sup_{h\leq \tau\leq T-h}\left[\sum_{s=1}^N\left\Vert\Delta\widetilde F_s(\tau)\right\Vert^2\right]^{1/2}\notag\\ 
&&=O_P\left(\zeta_1(p,b,{\mathbf n}, N)\right)\cdot O_P(1)=O_P\left(\zeta_1(p,b,{\mathbf n}, N)\right).\label{eqB.44}
\end{eqnarray}
By Lemma \ref{le:B.2}, we have
\begin{eqnarray}
&&\max_{1\leq i\leq p}\sup_{h\leq\tau\leq T-h}\left\Vert \sum_{j=1}^N\Delta X_{i,j}\left[\Delta\widetilde{F}_j(\tau)-H(\tau)\Delta F_j(\tau)\right]K_h^{1/2}(t_j-\tau)\right\Vert\notag\\
&&=O_P\left(\zeta_1(p,b,{\mathbf n}, N)+\zeta_{2}^\circ(p,h,N)\right).\label{eqB.45}
\end{eqnarray}
Following the proof of Proposition A.1 in \cite{BLLW23}, we can prove that
\[
\max_{1\leq i\leq p}\sup_{h\leq\tau\leq T-h}\left\Vert \sum_{j=1}^N\Delta U_{i,j}(\tau)\Delta F_j(\tau)\right\Vert =O_P\left(\sqrt{\frac{\log(p\vee N)}{Nh}}\right),
\]
which, together with the fact that $\sup_{h\leq \tau\leq T-h}\left\Vert H(\tau)\right\Vert_s=O_P(1)$ by Lemma \ref{le:B.3}, leads to
\begin{eqnarray}
&&\max_{1\leq i\leq p}\sup_{h\leq\tau\leq T-h}\left\Vert \sum_{j=1}^N\Delta U_{i,j}(\tau)H(\tau)\Delta F_j(\tau)\right\Vert\notag\\
&&=O_P\left(\sqrt{\frac{\log(p\vee N)}{Nh}}\right)=O_P(\zeta_2(p,h,N)).\label{eqB.46}
\end{eqnarray}
By (\ref{eqB.44})--(\ref{eqB.46}) and noting that $\zeta_2^\circ(p,h,N)=O(\zeta_2(p,h,N))$, we only need to show that
{\small\begin{eqnarray}
&&\max_{1\leq i\leq p}\sup_{h\leq\tau\leq T-h}\left\Vert\sum_{j=1}^N\left[\int_{t_{j-1}}^{t_j}\Lambda_i(t)^{^\intercal}F_tdt\right]H(\tau)\Delta F_j(\tau)K_h^{1/2}(t_j-\tau)-H_\ast(\tau)^{^\intercal}\Lambda_i(\tau)\right\Vert\nonumber\\ 
&&=O_P\left(\zeta_1(p,b,{\mathbf n}, N)+\zeta_{2}(p,h,N)\right).\label{eqB.47}
\end{eqnarray}}

By (\ref{eqB.12}), (\ref{eqB.27}) and the smoothness condition in Assumption 1(iii),  we have
\begin{equation}\label{eqB.48}
\max_{1\leq i\leq p}\sup_{h\leq\tau\leq T-h}\left\Vert\sum_{j=1}^N\left[\int_{t_{j-1}}^{t_j}\Lambda_i(t)^{^\intercal}F_tdt\right]H(\tau)\Delta F_j(\tau)K_h^{1/2}(t_j-\tau)-H(\tau)\Delta F(\tau)^{^\intercal}\Delta F(\tau)\Lambda_i(\tau)\right\Vert=O_P(h^{\delta}).
\end{equation}
On the other hand, by Lemmas \ref{le:B.2} and \ref{le:B.3} as well as Assumption 1(iii), we have 
\begin{eqnarray}
&&H(\tau)\Delta F(\tau)^{^\intercal}\Delta F(\tau)\Lambda_i(\tau)-H_\ast(\tau)^{^\intercal}\Lambda_i(\tau)\notag\\
&=&\left[H(\tau)\Delta F(\tau)^{^\intercal}\Delta F(\tau)H(\tau)^{^\intercal}- \Delta \widetilde F(\tau)^{^\intercal}\Delta \widetilde F(\tau)\right]H_\ast(\tau)^{^\intercal}\Lambda_i(\tau)\notag\\
&\leq&C \left\Vert \Delta F(\tau)H(\tau)^{^\intercal}- \Delta \widetilde F(\tau)\right\Vert\notag\\
&=&O_P\left(\zeta_1(p,b,{\mathbf n}, N)+\zeta_{2}^\circ(p,h,N)\right).\label{eqB.49}
\end{eqnarray}
By virtue of (\ref{eqB.48}) and (\ref{eqB.49}), we prove (\ref{eqB.47}), thus completing the proof of (\ref{eqB.43}).

Finally, by (\ref{eqB.27}) in Lemma \ref{le:B.3}, we readily have that 
\[\sup_{h\leq \tau\leq T-h}\left\Vert H(\tau) H(\tau)^{^\intercal} -I_k\right\Vert_s=o_P(1),
\]
indicating that $H_\ast(\tau)=H(\tau)^{^\intercal}+o_P(1)$ uniformly over $h\leq \tau\leq T-h$. \hfill$\blacksquare$


\end{document}